\RequirePackage{fix-cm}
\RequirePackage{fixltx2e}
\documentclass[12pt,twoside,english,aip, jcp,floatfix]{revtex4-1}

\usepackage[T1]{fontenc}
\usepackage[latin9]{inputenc}
\usepackage{geometry}
\usepackage{xcolor}
\usepackage{babel}
\usepackage{amsmath}
\usepackage{amssymb}
\usepackage{graphicx}
\usepackage[unicode=true,
 bookmarks=false,
 breaklinks=true,pdfborder={0 0 0},pdfborderstyle={},backref=false,colorlinks=true]
 {hyperref}
\hypersetup{pdftitle={The long road to calibrated prediction uncertainty for computational chemistry methods},
 pdfauthor={Pascal PERNOT & Andreas SAVIN},
 citecolor =blue, linkcolor = blue,  urlcolor  = blue}

\makeatletter

\providecommand{\tabularnewline}{\\}

\renewcommand{\textendash}{--}

\makeatother

\begin{document}

\title{The long road to calibrated prediction uncertainty in computational
chemistry}

\author{Pascal PERNOT}

\affiliation{Institut de Chimie Physique, UMR8000 CNRS,~\\
Université Paris-Saclay, 91405, Orsay, France}
\email{pascal.pernot@universite-paris-saclay.fr}

\begin{abstract}
\noindent Uncertainty quantification (UQ) in computational chemistry
(CC) is still in its infancy. Very few CC methods are designed to
provide a confidence level on their predictions, and most users still
rely improperly on the mean absolute error as an accuracy metric.
The development of reliable UQ methods is essential, notably for CC
to be used confidently in industrial processes. A review of the CC-UQ
literature shows that there is no common standard procedure to report
or validate prediction uncertainty. I consider here analysis tools
using concepts (calibration and sharpness) developed in meteorology
and machine learning for the validation of probabilistic forecasters.
These tools are adapted to CC-UQ and applied to datasets of prediction
uncertainties provided by composite methods, Bayesian Ensembles methods,
machine learning and \emph{a posteriori} statistical methods.
\end{abstract}
\maketitle
\tableofcontents{}

\newpage{}

\section{Introduction}

As stated in recent perspective articles \citep{Lejaeghere2020,Rommel2021,Reiher2022},
uncertainty quantification (UQ) in computational chemistry (CC) is
still in its early stages of development. For instance, in electronic
structure theory, at the exception of the BEEF-type methods \citep{Wellendorff2014},
none of the methods implemented in popular computational chemistry
codes provides an uncertainty or a confidence index on the calculated
properties. Confidence is generally based on benchmark studies which
provide performance indices, such as the ubiquitous mean absolute
error (MAE). However, it is well established that, lacking a probabilistic
interpretation, the MAE \emph{should not} be used as an uncertainty
proxy \citep{Ruscic2014,Pernot2015,Pernot2018}. As will be seen below,
prediction uncertainty might be much more complex to estimate than
usual performance indices.

This difficulty in quantifying our confidence in model predictions
has far reaching consequences. It is an obstacle for computational
chemistry to stand on par with, or to replace, physical measurements
\citep{Irikura2004} or to be used in decision making \citep{Wan2021}.
It also has a strong impact on \emph{multi-scale simulation}, where
propagation of uncertainty through the scales is necessary to assess
the reliability of predictions \citep{Gabriel2021,Mao2021,Ye2021},
or in \emph{iterative learning} to minimize the cost of high-level
calculations \citep{Proppe2016,Simm2016,Hie2020}. In \emph{benchmarking},
a very sensible way to select among a set of levels of theory would
be to pick one with a fit-to-purpose prediction uncertainty. It is
also worth to note that uncertainty provides a metric for the comparison
of measurement values, a prerequisite to the production of \emph{reproducible}
results \citep{Volodina2021}. In all such applications, a prediction
uncertainty estimate has to be \emph{fair}: underestimation is potentially
dangerous (overconfidence), and overestimation is wasteful. Reaching
a good balance is the main challenge of UQ. 

For the purpose of the present study, I sorted UQ approaches into
two classes, according to their embedding level within the computational
chemistry method:
\begin{itemize}
\item \emph{Embedded} UQ methods produce prediction uncertainty concurrently
with property predictions. This is an heterogeneous class which presently
encompasses the above-mentioned BEEF or Bayesian Ensemble approach
\citep{Wellendorff2014} and several \emph{bottom-up correction} methods
with detailed uncertainty budgets (the \emph{Type B} methods of Ruscic
\citep{Ruscic2014}), such as the Feller-Peterson-Dixon method \citep{Feller2008,Feller2017}
or the ATOMIC protocol \citep{Bakowies2019,Bakowies2020,Bakowies2021}.
Some machine learning methods, such as Bayesian neural networks which
are designed to provide uncertainty along with prediction, also belong
here \citep{Tran2020,Gawlikowski2021}. These methods are akin to
the ``molecule-specific'' UQ concept \textcolor{violet}{\citep{Reiher2022}}.
\item \emph{A-posteriori} UQ methods use a set of predictions and a reference
dataset to estimate prediction uncertainty, generally after trend
correction \citep{Pernot2011,Lejaeghere2014,Lejaeghere2014a,Lejaeghere2016,DeWaele2016,Pernot2015,Proppe2017,Lejaeghere2020}.
This is sometimes referred to as ``the statistical approach'', albeit
statistical tools are also at the heart of embedded methods. Ruscic
defined it as the \emph{Type A} method \citep{Ruscic2014}. Some examples,
at various sophistication levels, are the correction of harmonic vibrational
frequencies by scaling factors \citep{Scott1996,Pernot2011}, regression
analysis of various properties \citep{Faver2011b,Lejaeghere2014,Lejaeghere2014a,Pernot2015,Proppe2016,Proppe2017,Das2021},
correction by Gaussian Processes \citep{Kennedy2001,Proppe2019a}
and $\Delta$-Machine Learning ($\Delta$-ML) methods \citep{Ramakrishnan2015,Tran2020,Nandi2021,Unzueta2021,Bhattacharjee2021,Hruska2022}. 
\end{itemize}
Very often, the validation of prediction uncertainty values provided
by these CC-UQ methods is based on the qualitative appreciation of
the agreement of prediction uncertainty with the amplitude of errors
\citep{Csontos2010,DeWaele2016}. For the Bayesian ensemble methods,
calibration has been assessed by a visual examination of the normality
of $z$-scores histograms \citep{Mortensen2005}, or by visual evaluation
of the width of error bars \citep{Wellendorff2012,Wellendorff2014}.
When quantitative tools have been used, notably for the validation
of a-posteriori methods, a popular test statistic is the \emph{prediction
interval coverage probability} (PICP) \citep{Shrestha2006,Gawlikowski2021}.
PICP tests have been done either on the full dataset \citep{Proppe2017,Proppe2021,Bakowies2021},
or using a splitting scheme between calibration and validation sets
\citep{Pernot2015,Proppe2021}. Globally, there does not seem to be
(yet) a consensus in the community on the adequate validation vocabulary,
concepts and tools, which would be necessary to compare the merits
of different CC-UQ strategies.

In the past few decades, probabilistic forecasting for meteorology
has been the object of fundamental developments of validation concepts
and methods \citep{Gneiting2007a,Gneiting2014}. Two major concepts
resulting from these studies are \emph{calibration} (\emph{reliability}),
and \emph{sharpness} (\emph{resolution})\emph{. }A probabilistic prediction
method is said to be \emph{calibrated} if the confidence of predictions
matches the probability of being correct for all confidence levels
\citep{Tomani2020}. A \emph{calibrated} method is \emph{sharp} if
it produces the \emph{tightest} possible confidence intervals \citep{Kuleshov2018}.
Sharpness is conditional on calibration (a method cannot be sharp
if it is not calibrated). With the emergence of prediction uncertainty
estimation in Machine Learning, several calibration and sharpness
metrics, and graphical checks are also now used in this field \citep{Kuleshov2018,Tran2020,Lai2021,Gawlikowski2021}. 

Most computational chemistry methods are based on deterministic algorithms,
but our lack of knowledge about their prediction errors links them
to probabilistic forecasters. Part of these validation tools have
recently been introduced to the computational chemistry field by Tran
\emph{et al.} \citep{Tran2020} to compare the calibration and sharpness
of several ML algorithms trained to predict adsorption energies. This
is, to my knowledge the only study of this kind in the application
field of interest. My intent in this study is to adapt and apply calibration/sharpness
validation methods to a wider computational chemistry domain, and
to evaluate their pertinence in various scenarios, with the expectation
that such methods could be more generally used in the community. 

In the Sec.\,\ref{sec:The-CC-UQ-Context}, I review the specifics
of CC-UQ (error sources, UQ methods) and previous validation practices.
I then present (Sec.\,\ref{sec:Methods-for-the}) the panel of validation
tools for probabilistic forecasters (calibration and sharpness metrics
and graphical checks), and I propose adapted methods for the typical
CC-UQ outputs. In Sec.\,\ref{sec:Results}, these tools are applied
to a variety of datasets. The main features of this study are reported
and discussed in the conclusions (Sec.\,\ref{sec:Conclusion}).

\section{The computational chemistry UQ context\label{sec:The-CC-UQ-Context}}

A major obstacle to UQ in computational chemistry is the predominant
role of systematic errors. As uncertainty is a \emph{non-negative}
parameter expected to quantify \emph{unpredictable} errors \citep{GUM},
estimation (and correction) of \emph{predictable} or systematic errors
is a challenging, but necessary preliminary step to CC-UQ. The international
reference guide for metrology (\emph{a.k.a.} 'the GUM' \citep{GUM})
states that ``\emph{It is assumed that the result of a measurement
has been corrected for all recognized significant systematic effects
and that every effort has been made to identify such effects.}''
(GUM, Sect. 3.2.4). As a consequence, the partition between unpredictable
and predictable errors depends on the efforts that a modeler is ready
or able to invest in the analysis. 

I next consider the main sources of errors in computational chemistry
(Sect.\,\ref{subsec:Error-sources-in}) and the existing UQ methods,
based on different approaches to systematic error corrections (Sect.\,\ref{subsec:Main-error-correction}). 

\subsection{Error sources in computational chemistry\label{subsec:Error-sources-in}}

The main error sources in computational chemistry have been reviewed
recently \citep{Lejaeghere2020,Rommel2021}, and can be tagged as
\emph{numerical}, \emph{parametric}, and \emph{model }errors. I come
back briefly on these categories in order to discuss the expected
errors distributions, which is an essential ingredient to define UQ
validation methods.

\subsubsection{Numerical Errors}

These include errors due to finite arithmetics implementation of computational
chemistry codes and, for stochastic methods, to the random errors
resulting from finite sampling. For properly implemented and converged
methods, numerical errors can often be assumed to be well controlled
and negligible against other error sources \citep{Irikura2004} (except
for instances of numerical chaos \citep{Feher2012,Lafage2020}). However,
for most algorithms we are still missing rigorous error bounds \citep{Cances2017,Herbst2020},
and estimation of the amplitude of numerical errors due to finite
arithmetics requires computational approaches. In the Monte Carlo
framework, one observes the effects of tiny perturbations of model
inputs \citep{Irikura2004,Feher2012} and/or arithmetic operations
\citep{Scott2007,Denis2016,Chatelain2019}, without need to alter
the codes. The use of interval arithmetics has also been proposed
\textcolor{blue}{\citep{Janes2011}}, but it might require in-depth
recoding, barely an option for legacy computational chemistry codes.
This is however not a fatality for newly developed codes, as shown
recently by Herbst \emph{et al.} \citep{Herbst2020} with the development
of a DFT code in the Julia language accepting arbitrary floating point
types, including intervals. For some parameters, the amplitude of
numerical errors can also be estimated by systematic convergence studies
\citep{Carbogno2021}.

One can consider numerical errors as random variables, but their distribution
is not necessarily normal. Non-normal distributions occur, for instance,
for rounding errors in floating point arithmetics \citep{Higham2019},
or stochastic errors for Quantities of Interest (QoIs) in molecular
dynamics \citep{Wan2021}.

\subsubsection{Parametric errors}

These occur in methods involving the statistical estimation of parameters
with respect to reference data (\emph{e.g.}, semi-empirical force-fields,
extrapolation schemes in composite methods, semi-empirical methods,
statistical corrections...). An essential property of parametric errors
is that their amplitude should decrease when the size of the reference
dataset increases. The amplitude of parametric errors is typically
estimated by Monte Carlo uncertainty propagation, from a probability
density function of the parameters obtained, for instance, by Bayesian
inference \citep{Cailliez2011,Cailliez2020}. 

The distribution of parametric errors depends on the distribution
of the uncertain parameters and on the functional form of the model
with respect to these parameters. Here again, normality cannot be
assumed as a default distribution feature.

\subsubsection{Model errors}

For computational chemistry, model errors include \emph{level-of-theory}
errors (\emph{e.g.}, the choice of a density functional approximation,
DFA) and \emph{representation} errors (\emph{e.g.}, the choice of
a basis set or discretization grid). They are often the dominant contributions
to a computational chemistry uncertainty budget. Compared to numerical
and parametric errors, which are mostly aleatoric, model errors are
systematic. The estimation and correction of model errors are usually
handled either by bottom-up approaches of physics-based corrections
\citep{Ruscic2014}, or by corrective statistical models (the so-called
\emph{discrepancy functions} \citep{Kennedy2001}) involving the comparison
with a dataset of reference results \citep{Pernot2017}. It is important
to note that, in contrast to parametric errors, model errors \emph{do}
\emph{not} decrease when the size of the reference dataset increases. 

Model errors can present any type of distribution \citep{Pernot2018}.
However, a recent study on the shape of error distributions for different
QoIs and DFAs showed that the distributions are most often unimodal,
with various levels of asymmetry, and that they are in most cases
heavier-tailed than a normal distribution \citep{Pernot2021}.

\subsection{Main error correction and prediction uncertainty estimation methods\label{subsec:Main-error-correction}}

I review here the main methods appearing in the CC-UQ literature to
correct systematic errors and estimate prediction uncertainty, focusing
on the level of information they are able to provide, which goes from
a standard uncertainty to a full distributions of predictions. 

\subsubsection{Bottom-up correction methods\label{subsec:Bottom-up-correction-methods}}

Proceeding from a chosen level-of-theory/representation and applying
successive corrections based on physical criteria is a systematic
way to establish an uncertainty budget: each correction, being imperfect,
coming with some uncertainty. This kind of approach is referred to
as ``\emph{Type B} methods'' by Ruscic \citep{Ruscic2014}, in reference
to the method defined in the GUM \citep{GUM} for the estimation of
uncertainty \emph{in the absence of} a statistical sample (statistical
analysis of data samples defines\emph{ Type A} methods \citep{Ruscic2014,Klippenstein2017a}).
It appears however that some form of Type A analysis lies at the heart
of some bottom-up correction methods. For instance, the Feller-Peterson-Dixon
(FPD) method considers two main uncertainty components: the complete
basis-set (CBS) and the zero-point energy (ZPE) corrections. For the
CBS correction, uncertainty is estimated as the half-spread of the
corrections provided by five different extrapolation models \citep{Feller2008}.
For ZPE, it is taken as half the difference in the last step of a
series of approximations \citep{Feller2012}. I note for further reference
that in the FPD approach uncertainties are combined linearly instead
of quadratically, a worst case scenario ignoring possible error compensations
\citep{Feller2006}. This is expected to provide a ``crude but conservative
uncertainty'', from which the estimation of prediction intervals
is not straightforward and complicates the interpretation of calibration
tests.

In contrast, the ATOMIC composite method aims ``to provide realistic
corrections and uncertainty estimates corresponding to intervals of
95\% confidence...'' \citep{Bakowies2019}. Here, a mixture of linear
and quadratic uncertainty combination is used in deriving the final
uncertainty. Bakowies \citep{Bakowies2020} summarizes the interest
of the bottom-up correction approach:\emph{ ``The proposed model
is a welcome alternative to statistical assessment, first because
it does not depend on comparison with experiment, second because it
recognizes the expected scaling of error with system size, and third
because it provides a detailed account of the importance of various
contributions to overall error and uncertainty.``} 

\subsubsection{A-posteriori prediction uncertainty estimation}

A-posteriori approaches elaborate a statistical model independent
of the computational chemistry method to estimate its prediction uncertainty,
generally after correction of the original model predictions for systematic
errors \citep{Pernot2011,Pernot2015,Vishwakarma2021,Das2021}. As
for bottom-up corrections, different prediction uncertainty estimation
models are necessary for different predicted properties by a given
method. 

The a-posteriori approach requires three elements:
\begin{enumerate}
\item A reference dataset on which to calibrate the statistical model, the
ideal being a large set (at least several hundreds) of high-accuracy
values.
\item A statistical model to correct the (systematic) trends in the model
errors. This optional step might involve from a simple shift to machine
learning models.
\item A statistical method to estimate the prediction uncertainty from the
residual errors, the complexity of which depends on their distribution. 
\end{enumerate}
If a trend correction is applied (step 2), it is important to note
that the a-posteriori approach does not provide a prediction uncertainty
for the original computational chemistry method, but for its \emph{corrected}
version. This has been clearly illustrated for the correction of vibrational
frequencies by scaling factors \citep{Pernot2011}, or for the linear
correction of DFT-predicted Mössbauer isomer shifts \citep{Proppe2017}. 

One can point out two main obstacles to the success of the a-posteriori
approach, affecting the first and third steps: 
\begin{itemize}
\item The reference dataset should contain as many as possible high-accuracy
data, which, depending on the studied property, might be difficult
to achieve, notably for experimental data. For the application of
simple correction models, the errors dataset should also be \emph{homogeneous},
in the sense that it should not contain multiple contradictory trends
preventing the success of step 2. This might be less stringent for
machine learning correction models, which should be able to correct
for multiple trends, provided a relevant set of input features. In
addition, care should be taken that the prediction uncertainty is
corrected from reference data uncertainty \citep{Pernot2015,Klippenstein2017a},
a problem absent from bottom-up correction methods. 
\item The ability to establish reliable prediction uncertainties heavily
relies on the distribution of the corrected errors. Skewed or heteroscedastic
error distributions (\emph{i.e.}, where the variance of the errors
varies along the predictor variable) might be specifically challenging
for the definition of an uncertainty. The difficulty to establish
and communicate asymmetric and/or heteroscedastic uncertainty might
even be considered as a criterion to reject some computational chemistry
methods for prediction uncertainty estimation.
\end{itemize}
A-posteriori methods have been mostly used to estimate standard uncertainties
or prediction intervals. Depending on the approach, a full distribution
might be accessible (\emph{e.g.}, for regression models, Gaussian
processes...).

\subsubsection{Bayesian ensemble methods}

In the DFT framework, the Bayesian Error Estimation density Functional
(BEEF) family of methods has been designed to provide uncertainty
on its predictions \citep{Mortensen2005,Medford2014,Ulissi2017,Houchins2017,Guan2019}. 

In a first step, the parameters of a DFA are calibrated by Bayesian
inference against a reference dataset. However, the resulting parametric
uncertainties are typically insufficient to cover the amplitude of
prediction errors (remember that the amplitude of parametric errors
decreases with the size of the calibration dataset). 

To restore the validity of the statistical model, the variance-covariance
matrix of the reference data is scaled by a \emph{'temperature}' factor
such that the \emph{mean} prediction variance matches the variance
of prediction errors \citep{Wellendorff2012}. The resulting probability
density function (pdf) of the DFA parameters is captured as an ensemble
of parameters values which can then be used to estimate prediction
uncertainty on any relevant property by Monte Carlo sampling. Note
that, as Bayesian ensembles are calibrated to cover the amplitude
of prediction errors, their prediction uncertainty is affected by
reference data uncertainty. When comparing predictions to experimental
values, care should be taken not to count experimental uncertainty
twice.

In the available literature the results of BEEF models are summarized
by a mean value and a standard deviation and the ensembles are not
used for validation.

\subsubsection{Machine learning}

Materials science and catalysis see an intensive development of ML
algorithms to replace DFT calculations \citep{Montavon2013,Ramakrishnan2014,Ramakrishnan2015,Rupp2015,Faber2017,Ulissi2017,Zaspel2019,Tran2020,CesardeAzevedo2021,Duan2021,Nandi2021,Venkatraman2021}.
DFT error correction by ML models is also a pathway actively explored
to obtain high accuracy results from low-level DFT calculations \citep{Ramakrishnan2015,Bhattacharjee2021,Nandi2021,Unzueta2021}.

UQ is central to many automatized applications in screening or design
of efficient compounds, and several ML algorithms are now available
to estimate prediction uncertainty, through either distribution- or
ensemble-based methods. A recent comparison of the calibration and
sharpness of these methods by Tran \emph{et al.} \citep{Tran2020}
reveals a wide spectrum of reliability of ML-UQ methods. 

\section{Methods for the validation of prediction uncertainty\label{sec:Methods-for-the}}

The general framework for the validation of probabilistic predictions
is first presented to define the calibration and sharpness concepts
and the associated statistical test and metrics. The cases with restricted
information, in the shape of an uncertainty (standard or expanded)
instead of a full distribution, are then considered, for which validation
methods are derived from the general framework. The concepts are illustrated
on synthetic data.

\subsection{General case: probabilistic predictions}

A probabilistic prediction provides a distribution over the values
that can be taken by the Quantity of Interest (QoI), $V$. The predicted
cumulative distribution function (CDF) for $V$ is noted $F(V)$ and
the reciprocal quantile function $F^{-1}(p)\equiv F_{p}^{-1}$, where
$p$ is a probability. For the sake of validation, predictions are
made for a series of $M$ test systems for which one has reference
values $\{R_{i}\}_{i=1}^{M}$. For each reference system $i$, one
has thus a predicted CDF $F_{i}(V)$ and the corresponding quantile
function $F_{p,i}^{-1}$.

Few UQ methods provide prediction distributions, and approximations
of CDF functions are often available through\emph{ ensembles} of values,
representative of the predictive distribution. Such ensembles are
generated by ML-UQ methods, Bayesian Ensemble BEEF-type methods or
stochastic algorithms.

\subsubsection{Definitions: Calibration and sharpness}

A method is considered to be calibrated (or reliable) if the confidence
of predictions matches the probability of being correct for all confidence
levels \citep{Tomani2020,Tran2020}. Formally, the empirical and predicted
CDFs should be identical \citep{Kuleshov2018}, \emph{i.e.
\begin{equation}
\lim_{M\rightarrow\infty}\frac{1}{M}\sum_{i=1}^{M}\boldsymbol{1}\left(R_{i}\le F_{p,i}^{-1}\right)=p,\,\forall p\in[0,1]\label{eq:calibration}
\end{equation}
}where $\boldsymbol{1}(x)$ is the \emph{indicator function} for proposition
$x$, taking values 1 when $x$ is true and 0 when $x$ is false.
This equation can be generalized to prediction intervals \citep{Kuleshov2018}\emph{
\begin{equation}
\lim_{M\rightarrow\infty}\frac{1}{M}\sum_{i=1}^{M}\boldsymbol{1}\left(R_{i}\in I_{p,i}\right)=p,\,\forall p\in[0,1]\label{eq:calibration-1}
\end{equation}
}where 
\begin{equation}
I_{p,i}=\left[F_{(1-p)/2,i}^{-1},F_{(1+p)/2,i}^{-1}\right]\label{eq:pred-int}
\end{equation}
 is the 100$p$\,\% prediction interval for the predicted value $V_{i}$. 

However, this calibration condition is not sufficient to ensure that
the prediction uncertainties or confidence intervals are useful. Eqs.\,\ref{eq:calibration}-\ref{eq:calibration-1}
provide an average calibration assessment over the test set, but does
not guarantee that calibration is realized \emph{locally} for each
predicted value. A useful prediction model must also be \emph{sharp},
\emph{i.e.}, prediction intervals should be as tight as possible around
any predicted value \citep{Kuleshov2018}.\bigskip{}

\paragraph*{Example}
\begin{quote}
Let us consider a toy model where the predictive CDF $F(V)$ is learned
from an ensemble of errors $E_{i}=R_{i}-V_{i}$ with uncorrected bias
and without accounting for the linear dependence of $E$ on $V$ (Fig.\,\ref{fig:example-definitions}(a)).
\begin{figure}[t]
\begin{centering}
\begin{tabular}{ll}
\includegraphics[height=6cm]{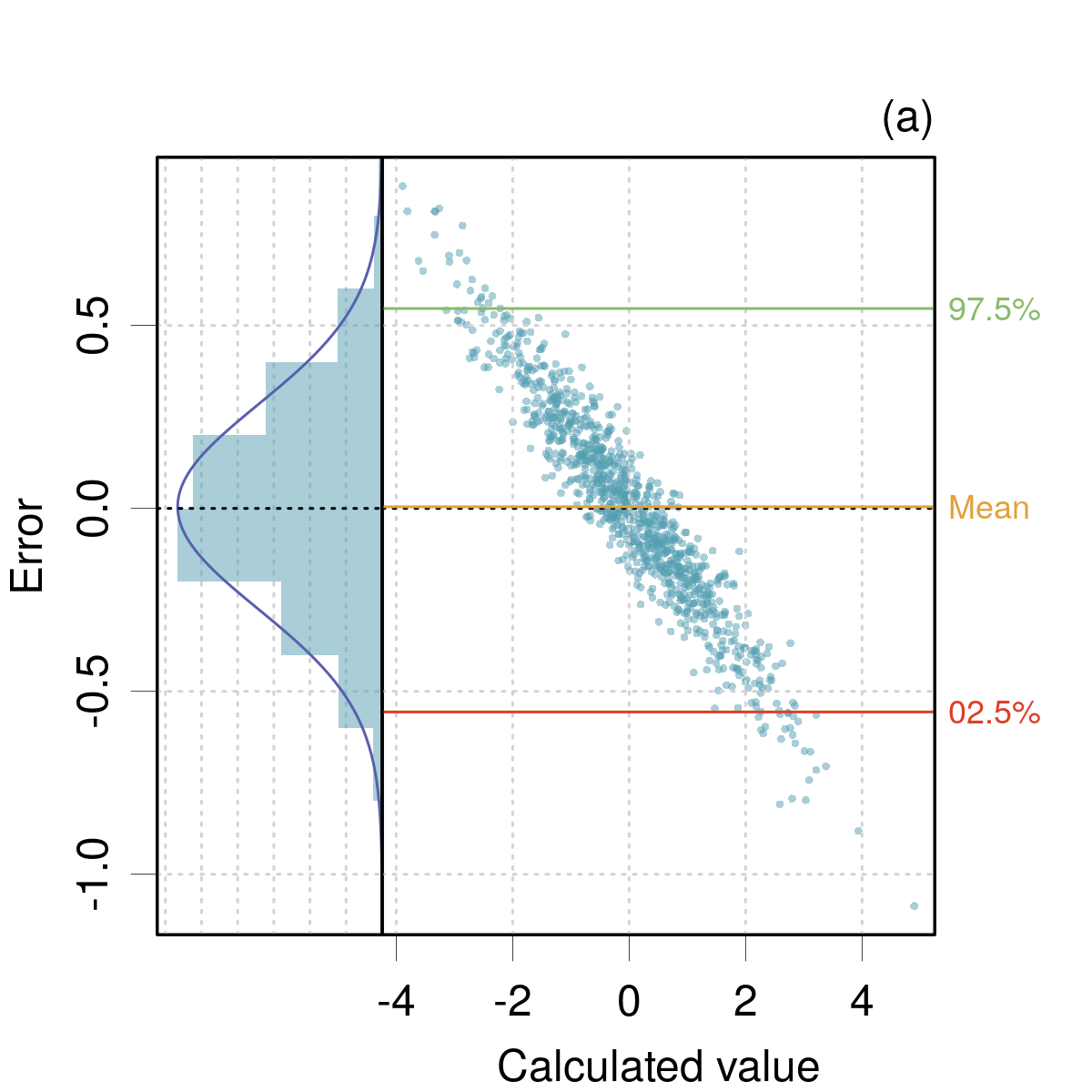} & \includegraphics[height=6cm]{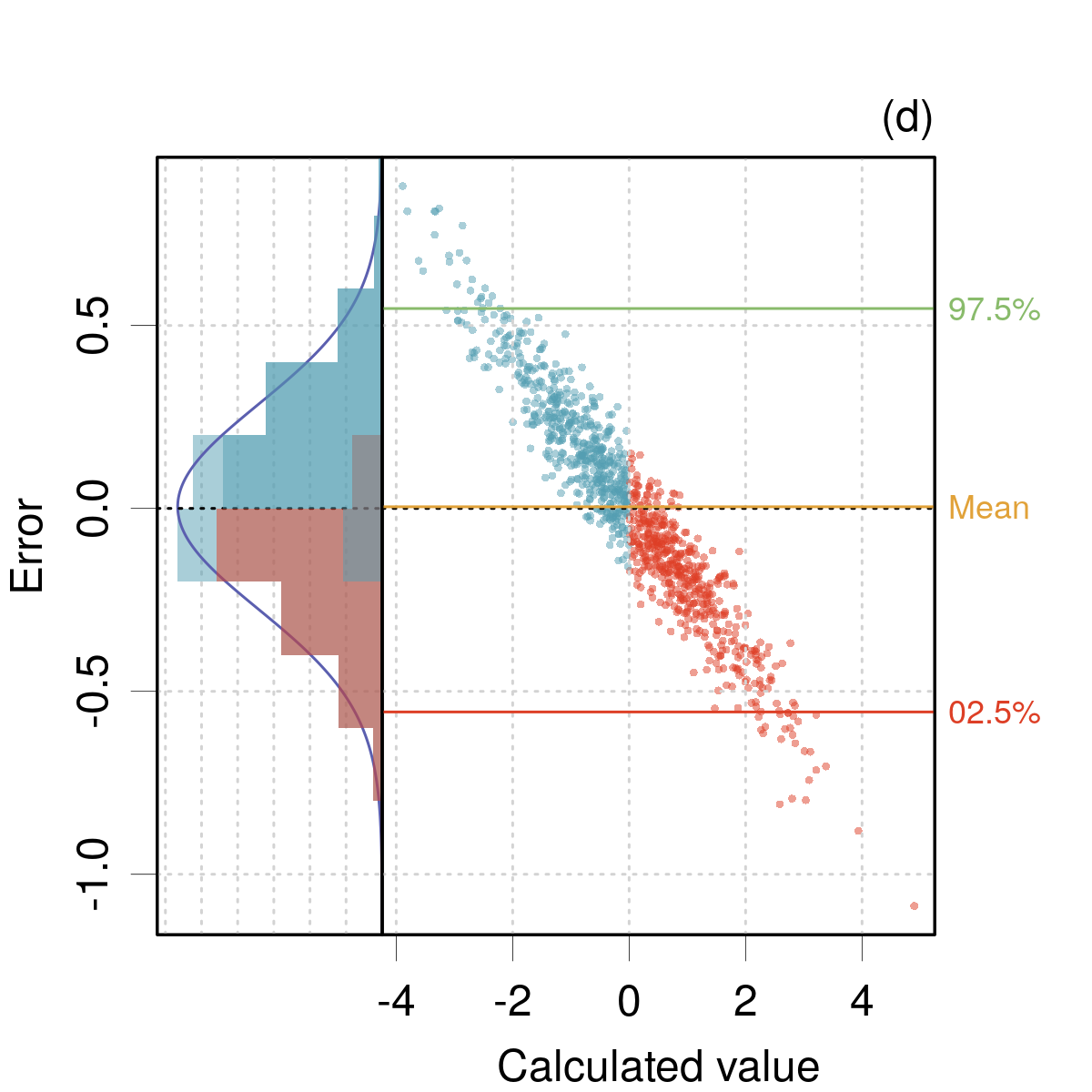}\tabularnewline
\includegraphics[height=6cm]{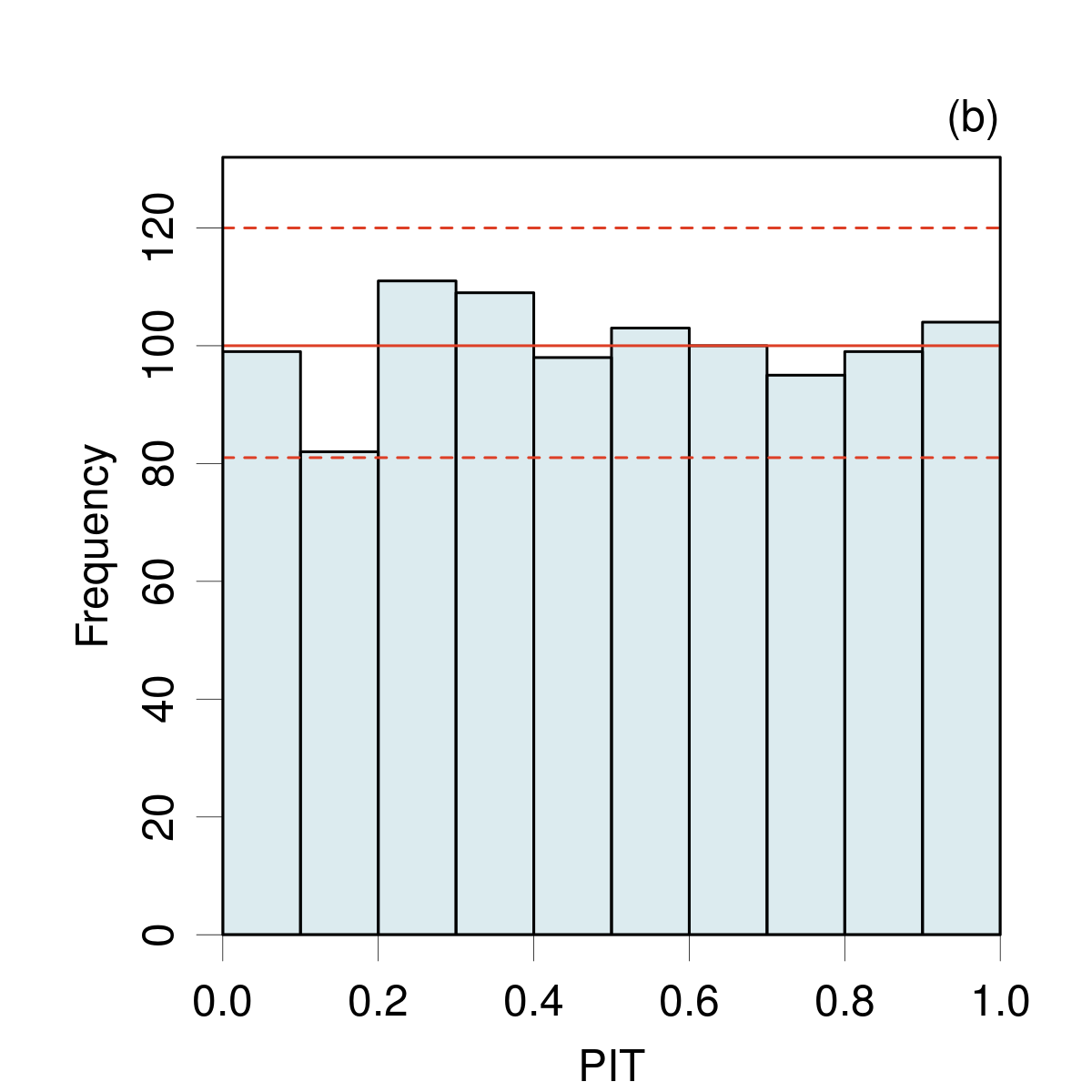} & \includegraphics[height=6cm]{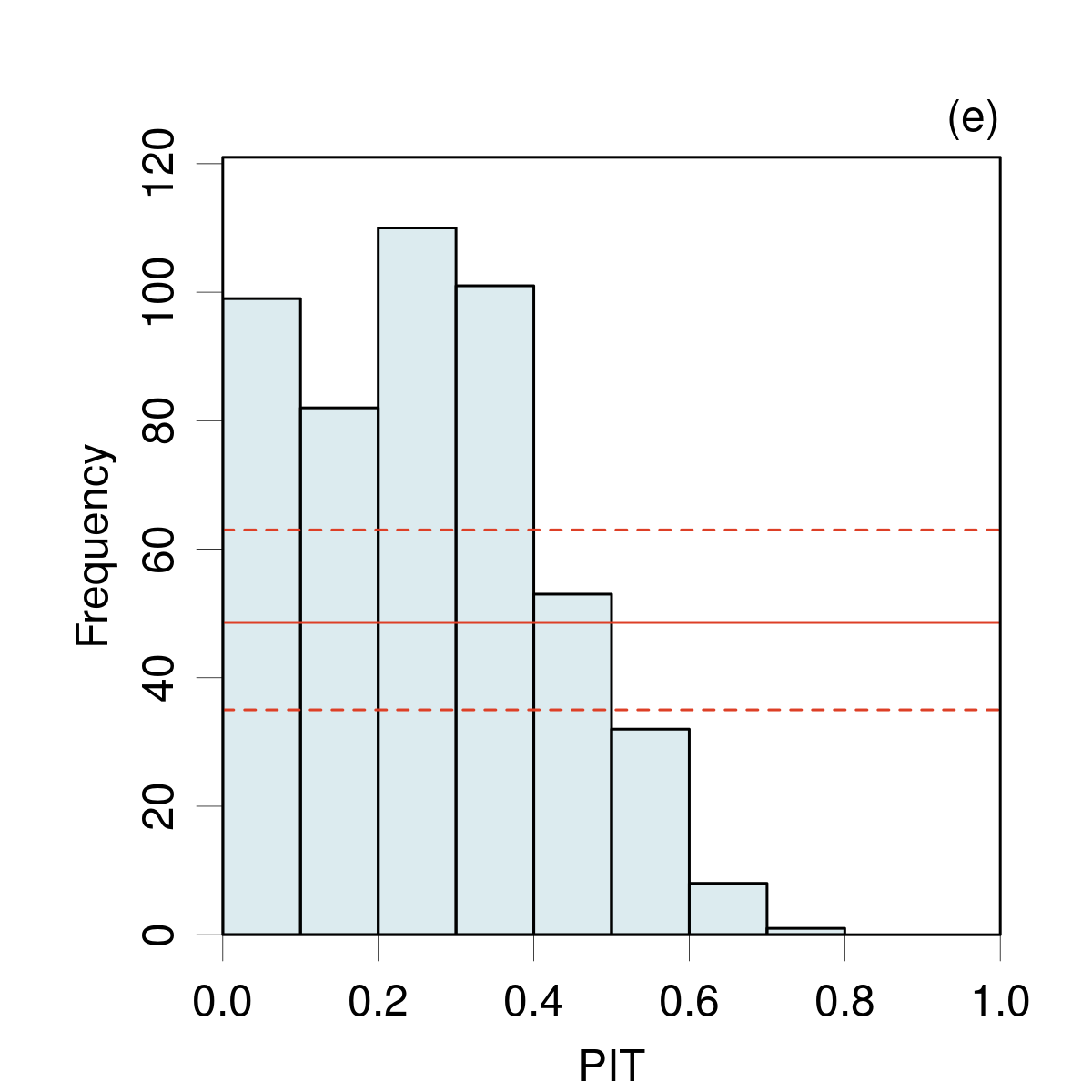}\tabularnewline
\includegraphics[height=6cm]{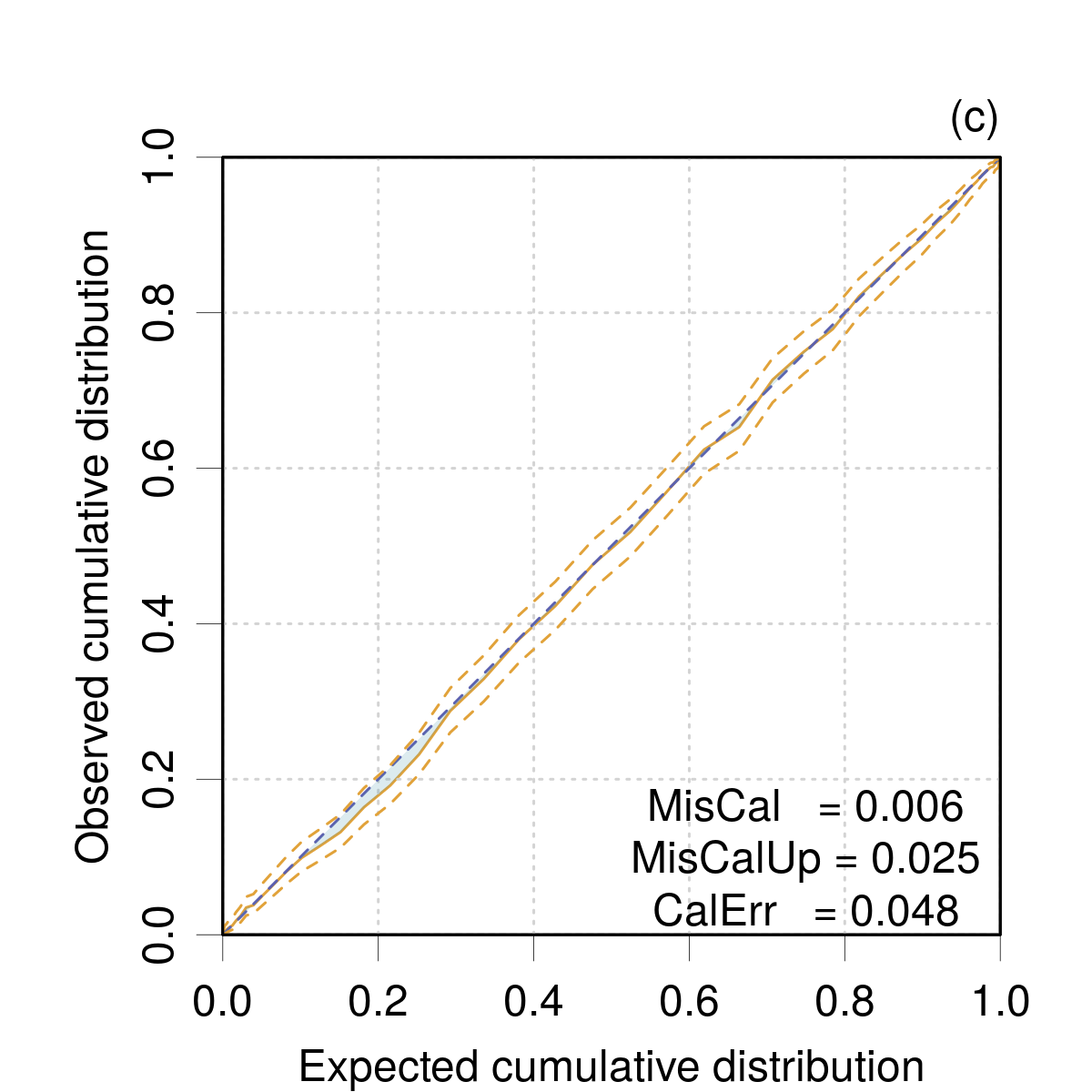} & \includegraphics[height=6cm]{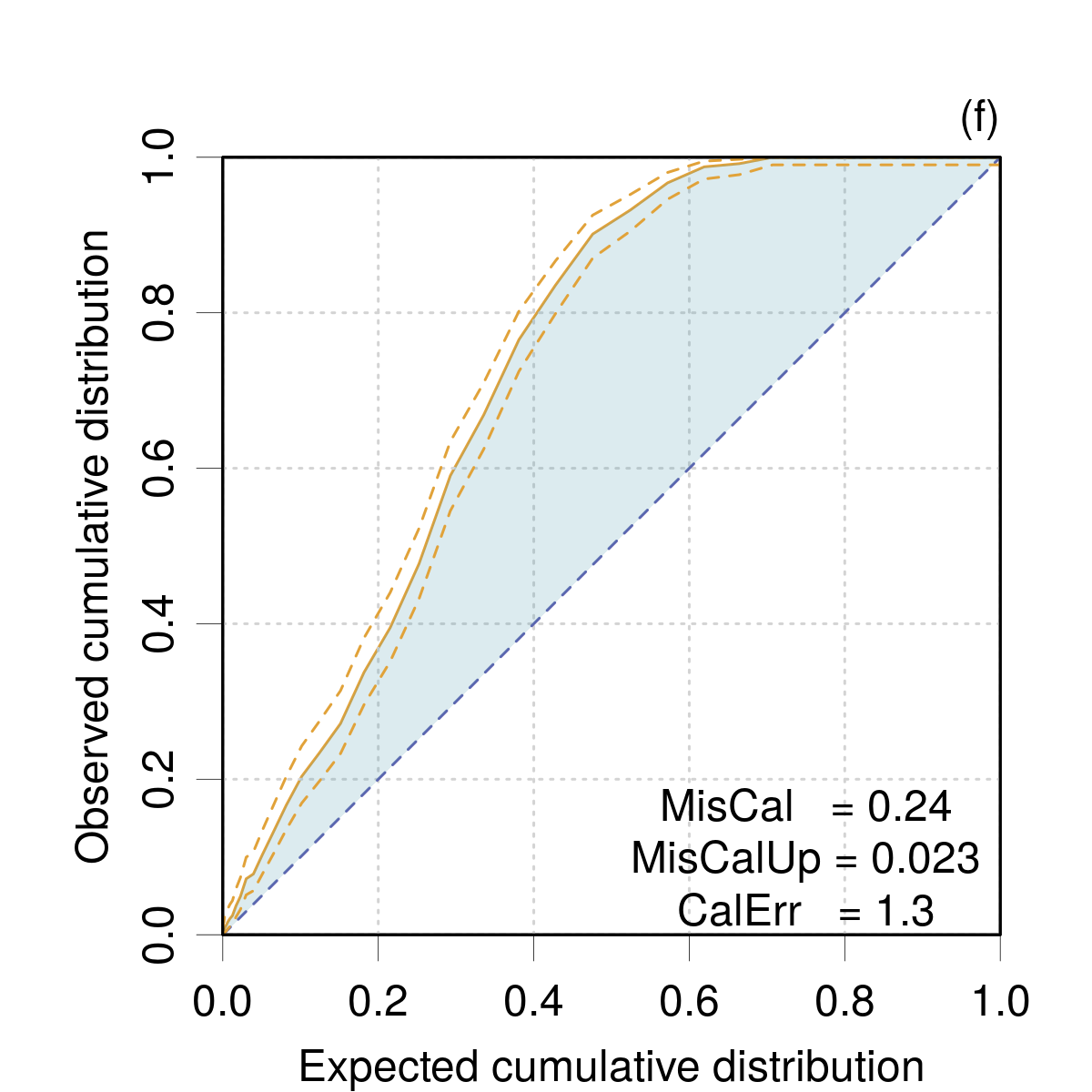}\tabularnewline
\end{tabular}
\par\end{centering}
\caption{\label{fig:example-definitions}Error samples from a normal distribution
with a linear trend: (a,d) Error distribution plots; (b,e) Probability
integral transform (PIT) histograms; (c,f) Calibration curves. In
(d-f), the uncertainty statistics estimated over the whole dataset
are used to validate a subset restricted to the positive calculated
values (red dots).}
\end{figure}
 Assuming that the test set has the same properties as the learning
set, using the system-independent $F_{p}^{-1}$ function will provide
correct prediction intervals over the test set, and Eqs.\,\ref{eq:calibration}-\ref{eq:calibration-1}
will be satisfied. However, if $F_{p}^{-1}$ is used to design prediction
intervals for a local subset of the test set, for instance the positive
values of $V$ ((Fig.\,\ref{fig:example-definitions}(d)), the intervals
will clearly be oversized and ill-centered: sharpness will not be
ensured. In the first case, calibration tests such as the probability
integral transform (PIT) histogram presented below (Sect.\,\ref{par:Probability-integral-transform})
do not detect a problem (Fig.\,\ref{fig:example-definitions}(b)),
while the same test on the local subset of the data is diagnostic
of a calibration problem (Fig.\,\ref{fig:example-definitions}(e)). 
\end{quote}

\subsubsection{Calibration tests and metrics\label{subsec:Diagnostic-checks}}

Several visual checks and statistics have been proposed to estimate
calibration and sharpness of probabilistic predictions \citep{Cook2006,Gneiting2007a,Gneiting2014,Kuleshov2018}.
I review below the most pertinent ones for CC-UQ applications and
extend the toolbox with a graphical representation enabling to assess
simultaneously average and local calibration. Some of these tools
have been used in a ML setup, where statistical uncertainty on the
validation statistics is negligible; this is not the case in a typical
CC-UQ problem, so I added confidence intervals to all estimators.

\paragraph{Probability integral transform histogram.\label{par:Probability-integral-transform}}

The probability integral transform (PIT) is the value that the predictive
CDF attains at a test value $R_{i}$ \citep{Gneiting2014}
\begin{equation}
PIT_{i}=F_{i}(R_{i})
\end{equation}
For a calibrated method, a histogram of the PIT values for the validation
set should be uniform over {[}0,1{]}. The PIT histogram enables a
visual check of Eq.\,\ref{eq:calibration}, which does not require
additional information. 

Due to the finite size $M$ of the validation set, one should not
expect a perfectly uniform histogram. In order to assess significant
deviations from the uniform distribution, a 95\,\% confidence interval
on the bin heights in a uniform histogram is obtained from the quantiles
of the Poisson distribution with rate $M/n_{bin}$ where $n_{bin}$
is the number of bins in the PIT histogram. Significant deformations
of the histogram from a uniform distribution can be used to diagnose
calibration problems (see the Supplementary Material, Sec.\,III).
Alternatively or in complement to this graphical check \citep{Sailynoja2021},
some authors recommend statistical tests for uniformity \citep{Cook2006}.\bigskip{}

\paragraph*{Example (continued)}
\begin{quote}
PIT histograms are shown in Fig.\,\ref{fig:example-definitions}(b,e).
For the first one, the binned PIT values stand in the confidence range
for a uniform histogram, assessing the calibration of the predictions.
A contrario, the second histogram presents significant deviations
from the uniform, with an excess of small values pointing to a negative
bias in the predictions. 
\end{quote}

\paragraph{Calibration curve.}

Calibration can also be checked by comparing the estimated success
rate in the left hand side of Eq.\,\ref{eq:calibration} 
\begin{equation}
\eta_{p}=\frac{1}{M}\sum_{i=1}^{M}\boldsymbol{1}\left(R_{i}\le F_{p,i}^{-1}\right)\label{eq:PICP-1}
\end{equation}
to the target probability $p$ for a series of $p$ values in $[0,1]$.
By plotting $\eta_{p}$ vs $p$, one gets a \emph{calibration curve}
\citep{Kuleshov2018,Tran2020}\footnote{In statistics, this representation is commonly called a \emph{pp-plot},
which is the reciprocal representation of the quantile-quantile plot
(\emph{qq-plot}). }. For a calibrated method, the curve should lie on the identity line.
Note that calibration curves might also be designed from Eq.\,\ref{eq:calibration-1},
but testing small coverage intervals does not seem very pertinent
in a UQ setup. 

To account for finite $M$ values and assess the overlap of the calibration
curve with the identity line, a 95\,\% confidence band is plotted
around the calibration curve (see Sec.\,\ref{subsec:Binomial-proportions-confidence}
for implementation details). As for the PIT histogram, deformations
of the calibration curve from the identity line are diagnostic of
calibration problems (see the Supplementary Material, Sec.\,III). 

\paragraph{Miscalibration area and calibration error.}

The area between the calibration curve and the identity line (\emph{miscalibration
area,} noted MisCal \citep{Tran2020}) can be used as a calibration
metric. To use it for testing, we consider as an upper limit half
of the area of the 95\,\% confidence band around the curve (MisCalUp):
if MisCal~$>$~MisCalUp, the identity line does not lie within the
95\,\% confidence band, and calibration can be questioned. 

A calibration error (CalErr) is also proposed in the literature, as
the sum of squared differences over the probabilities in the calibration
curve \citep{Kuleshov2018}, or its square root \citep{Tran2020}.
The latter is retained here for its homogeneity with an error. As
MisCal, it can be used to compare several methods on their calibration
level.\bigskip{}

\paragraph*{Example (continued)}
\begin{quote}
Fig.\,\ref{fig:example-definitions}(c,f) shows the calibration curves
corresponding to both scenarios in Fig.\,\ref{fig:example-definitions}(a,b).
The one corresponding to the full calibration dataset does not deviate
notably from the identity line, and the MisCal statistic is much smaller
that its upper limit MisCalUp. In the second case, there is no ambiguity
about miscalibration, for either the curve or the statistics. As a
confirmation, the calibration error in the second case (CalErr$=1.3$)
is much larger than in the first case ($0.048$).
\end{quote}

\paragraph{PICP testing for a series of target coverage probabilities.\label{par:PICP-testing-for}}

The effective coverage of a $100p$\,\% prediction interval $I_{p}$
is estimated by its \emph{prediction interval coverage probability}
(PICP) \citep{Shrestha2006,Gawlikowski2021}, as the ratio of $S_{p}$,
the number of successes ($R_{i}\in I_{p,i}$) , to the size $M$ of
the validation set 
\begin{equation}
\nu_{p}=S_{p}/M\label{eq:PICP}
\end{equation}
where
\[
S_{p}=\sum_{i=1}^{M}\boldsymbol{1}\left(R_{i}\in I_{p,i}\right)
\]
To test if a PICP value $\nu_{p}$ is compatible with the target coverage
probability $p$, one checks if $p$ lies within a 95\,\% confidence
interval around $\nu_{p}$ (see Section\,\ref{subsec:Binomial-proportions-confidence}
for implementation details). 

Direct application of Eq.\,\ref{eq:calibration-1} would lead to
test PICP values for a regular set of coverage probabilities in $[0,1]$.
However, one is typically interested in large coverage values, and
there is little practical interest to consider low-probability intervals.
In the following, I will consider $p=0.5$ (inter-quartile range),
$0.75$ and $0.95$. 

\subsubsection{Sharpness metrics\label{subsec:Sharpness-tests-and}}

Several sharpness metrics have been proposed, such as the mean prediction
interval width \citep{Lai2021}, or the mean variance of the prediction
distributions \citep{Kuleshov2018}. Here I retain the definition
of Tran \emph{et al.}~\citep{Tran2020} 
\begin{equation}
sha=\sqrt{\frac{1}{M}\sum_{i=1}^{M}\mathrm{Var}(F_{i})}\label{eq:MPU-1}
\end{equation}
where $\mathrm{Var}(F_{i})$ is the variance associated with the CDF
at point $i$, $F_{i}(V)$. $sha$ has the dimension of an uncertainty
on the QoI and it corresponds to the \emph{mean prediction uncertainty}
(MPU) used in an earlier study by Pernot \emph{et al. }\citep{Pernot2015}.
The mean predictive variance ($sha^{2}$) has also been used as a
model performance metric by Proppe \emph{at al.} \citep{Proppe2017},
and as a calibration statistic for the BEEF methods \citep{Wellendorff2012,Medford2014,Wellendorff2014}.

It is considered that $sha$ should be small, but no threshold value
is available to distinguish between sharp and unsharp methods. $sha$
is therefore a convenient metric to compare several methods, but not
for self-standing sharpness assessment.

\subsubsection{Local calibration\label{subsec:Local-calibration}}

\paragraph{Local Coverage Probability (LCP)\label{par:Local-Coverage-Probability}.}

One can check \emph{local} calibration by performing PICP tests on
subsets of the validation data. The simplest scenario is to split
the dataset into contiguous areas\footnote{I choose ``area'' here in order to avoid confusions arising from
the use of ``interval'' with different meanings. It also generalizes
well to multi-dimensional predictor variables.} of the predictor variable (typically the QoI $V$), but other splitting
schemes can be considered. If prediction uncertainty is not constant,
it might also be interesting to use it as a predictor variable to
check its impact on local PICP estimates.

Two constraints have to be considered: (i) the subsets sizes have
to remain large enough for PICP testing to have some power, and (ii)
using equi-sized subsets could simplify the appreciation of multiple
tests. The first constraint limits the resolution of the LCP analysis,
and for small datasets, rather than splitting the dataset into small
sets, one might use overlapping subsets. Whatever the splitting scheme,
the LCP method proceeds as follows:
\begin{enumerate}
\item Choose a set of equi-sized areas of the predictor variable (contiguous
or overlapping).
\item Repeat for several values of probability $p$:
\begin{enumerate}
\item within each area (indexed by $k$), estimate the local PICP, $\nu_{p,k}=S_{p,k}/M_{k}$ 
\item plot $\nu_{p,k}$ and its 95\,\% confidence interval at the center
of each area. 
\end{enumerate}
\end{enumerate}
The LCP plot enables to check (1) average calibration, from the agreement
between estimated and target coverage probabilities, and (2) local
calibration, from the uniformity of the estimated coverage across
the local areas of the predictor variable(s). 

As for PICP testing, one is mostly interested in large coverage probabilities,
and I propose to check only the 0.5, 0.75 and 0.95 probability levels.
Note that the LCP analysis uses $M_{k}$ values smaller than $M$.
In consequence, tests on local PICPs have larger uncertainty and are
more likely to be permissive, and a false impression of good calibration
might arise. It is essential to keep in mind that (1) calibration
should be estimated based on PICPs for the whole dataset, and (2)
trends in the local PICP values with respect to the target coverage
are important diagnostic features.\bigskip{}

\paragraph*{Example (continued)}
\begin{quote}
Fig.\,\ref{fig:example-definitions-LCP}(a,b) shows the LCP analysis
results for two scenarios. For the full set represented in Fig.\,\ref{fig:example-definitions}(a),
one can see in the margin of Fig.\,\ref{fig:example-definitions-LCP}(a)
the PICP values and their 95\,\% CIs, confirming the good calibration,
in line with the PIT histogram, the calibration curve and statistics.
In this case, the LCP procedure selects $n=\left\lfloor 1000/150\right\rfloor =6$
contiguous areas and estimates the local PICP values, which are reported
at the center of each area. It is clear that the predictions are not
sharp. Fig.\,\ref{fig:example-definitions-LCP}(b) shows the LCP
analysis for the same dataset \emph{corrected from its linear trend}.
In this case, the estimated prediction CDF is both calibrated and
sharp. 
\begin{figure}[t]
\begin{centering}
\begin{tabular}{ll}
\includegraphics[height=6cm]{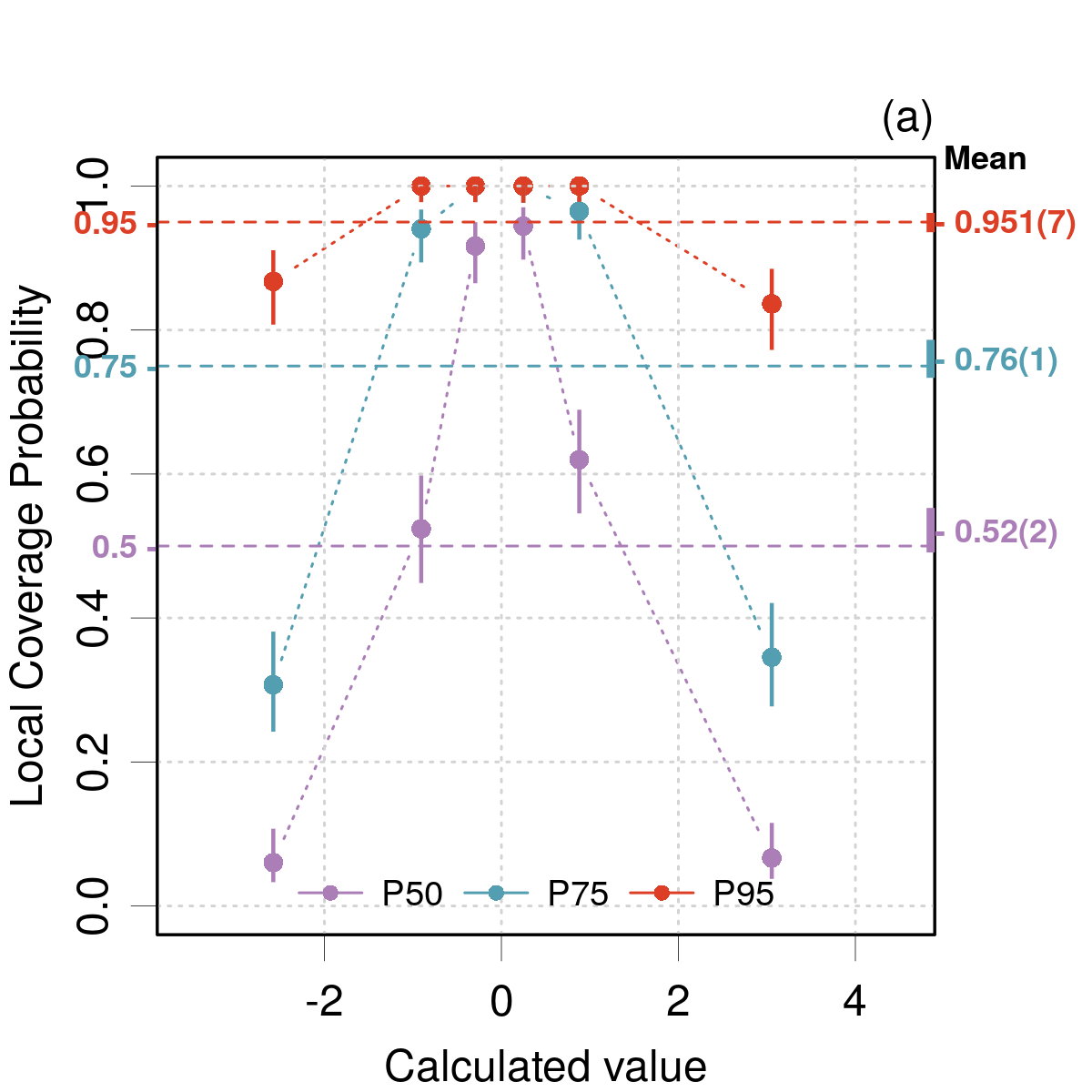} & \includegraphics[height=6cm]{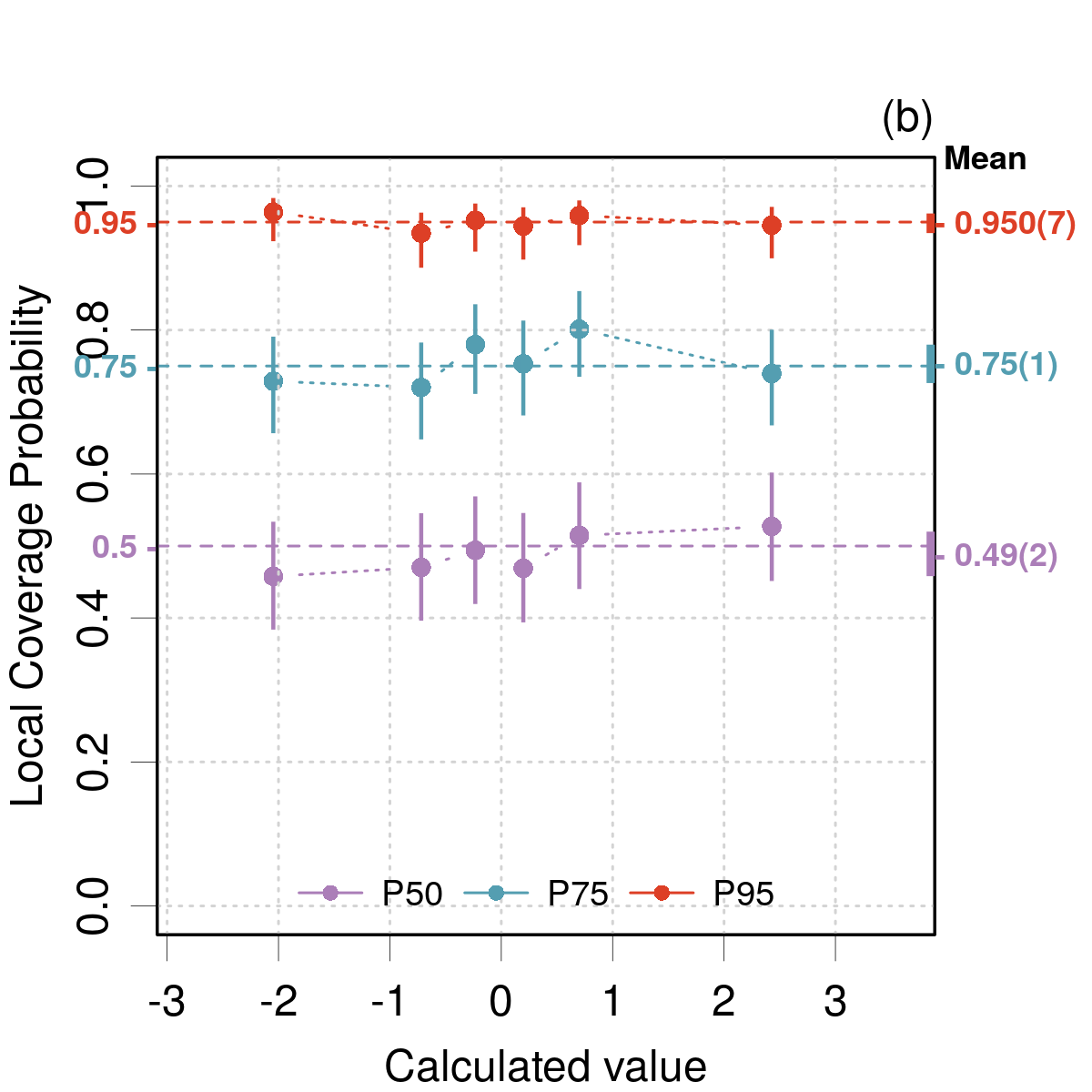}\tabularnewline
\end{tabular}
\par\end{centering}
\caption{\label{fig:example-definitions-LCP}LCP analysis: (a) the same dataset
as in Fig.\,\ref{fig:example-definitions}(a); (b) after linear trend
correction.}
\end{figure}
\end{quote}

\subsection{Validation of expanded uncertainty}

The majority of CC-UQ methods do not provide a prediction CDF ($F_{i}$),
but limited summary statistics, such as sets of predicted values $V_{i}\pm u_{V,i}$.
In order to use validation methods based on Eqs.\,\ref{eq:calibration}-\ref{eq:calibration-1},
one would have to make assumptions on the underlying CDFs. As we have
seen that there is no typical shape for the computational chemistry
errors distributions (Sect.\,\ref{subsec:Error-sources-in}), it
is preferable to avoid such assumptions and derive distribution-free
validation methods. Let us see how the elementary test in Eq.\,\ref{eq:calibration-1}
\begin{align}
R_{i} & \in I_{p,i}\label{eq:elem-test}
\end{align}
can be implemented in such cases. 

Assuming the symmetry of the prediction interval, one can write 
\begin{equation}
I_{p,i}=\left[V_{i}-U_{p,i},V_{i}+U_{p,i}\right]
\end{equation}
where $U_{p,i}$ is the expanded prediction uncertainty at the $p$
level. Eq.\,\ref{eq:elem-test} is therefore equivalent to 
\begin{equation}
|E_{i}|\le U_{p,i}\label{eq:testU}
\end{equation}
where $E_{i}=R_{i}-V_{i}$ is an error. Note the $U_{p}$ should account
for both calculation and reference errors. 

Hence, when an expanded prediction uncertainty for the errors \textendash{}
typically $U_{95,i}$ \textendash{} is available, on can consider
Eq.\,\ref{eq:testU} without hypothesis on the errors distribution
other than symmetry and the ones made to estimate $U_{p,i}$. This
enables us to use PICP testing (Sect.\,\ref{par:PICP-testing-for})
and the LCP analysis for sharpness.

\subsection{Validation of standard uncertainty using $z$-scores}

When a standard uncertainty $u_{E,i}$ is available, Eq.\,\ref{eq:testU}
can be rewritten as
\begin{equation}
|E_{i}|\le f_{p,i}u_{E,i}
\end{equation}
where $f_{p,i}$ is a coverage factor associated with the errors CDF
$F_{i}$, at point $i$. Considering that the errors distribution
strongly depends on the dominant error source in a calculation, it
is often difficult to define the errors CDF on which to estimate coverage
factors. In such cases, it might be interesting to directly test the
consistency of the errors and uncertainties through their ratio, as
shown below.

Let us assume that we have unbiased errors with unknown distribution,
but known standard deviation $\sigma_{i}>0$,
\begin{align}
\mathrm{E}(E_{i}) & =0\\
\mathrm{Var}(E_{i}) & =\sigma_{i}^{2}
\end{align}
Then, the $z$-scores, $z_{i}=E_{i}/\sigma_{i}$, are unit-scaled
and zero-centered variables, \emph{i.e.},
\begin{align}
\mathrm{E}(z_{i}) & =\mathrm{E}(E_{i})/\sigma_{i}=0\\
\mathrm{Var}(z_{i}) & =\mathrm{Var}(E_{i})/\sigma_{i}^{2}=1
\end{align}
Therefore, if the errors are unbiased and the $\sigma_{i}$ values
are correctly estimated by $u_{E,i}$, the distribution of $z$-scores
should be unbiased with unit variance,
\begin{align}
\mathrm{E}(Z) & =0\label{eq:Zvalid}\\
\mathrm{Var}(Z) & =1\label{eq:Zvalid1}
\end{align}

In the absence of information on the $z$-score distribution, testing
the value of $\mathrm{Var}(Z)$ cannot be done using the popular chi-squared
test, which is not robust with respect to deviations from the underlying
normality assumption (Supplementary Material, Sec.\,II). As for the
PICP, the $z$-score variance is tested against a 95\,\% confidence
interval (estimated by bootstrapping; see Sect.\,\ref{subsec:Testing--scores-variance}).

\subsubsection{Local $z$-score variance (LZV) analysis}

Similar to the LCP analysis for coverage probabilities (Sect.\,\ref{subsec:Sharpness-tests-and}),
I propose a local analysis based on the variance of $z$-scores, estimated
in a series of contiguous or overlapping areas {[}Local z-score Variance
(LZV) analysis{]}. For each subset, $\mathrm{Var}(Z)$ and its 95\%
confidence interval are estimated and plotted on a graph. Intervals
that do not cross the unity variance line might be considered as problematic,
but here also, trends are of diagnostic interest.\bigskip{}

\subsubsection*{Example}
\begin{quote}
A dataset of errors has been generated in the hypothesis of a Student's-$t$
errors distribution ($\mathcal{T}(\nu)$) and non-uniform random variances
issued from a chi-squared distribution ($\chi^{2}(\nu)$) scaled to
have unit mean
\begin{align}
E_{i} & =u_{E,i}\sqrt{3/5}\mathcal{T}(\nu=5)\label{eq:genz-1}\\
u_{E,i} & =0.01\sqrt{s_{i}/<s>}\\
s_{i} & \sim\chi^{2}(\nu=4)\label{eq:genz-2}
\end{align}
The errors and $z$-scores distributions for a sample of $N=1000$
$z$-scores are shown in Fig.\,\ref{fig:example-zscores}(a,b). The
$z$-score statistics are $\mathrm{E}(Z)=-0.045(30)$ and $\mathrm{Var}(Z)=0.894(53)$,
with a 95\,\% probability interval equal to $\left[0.8,1.0\right]$
for the variance, which is validated. 
\begin{figure}[!th]
\begin{centering}
\begin{tabular}{ll}
\includegraphics[height=6cm]{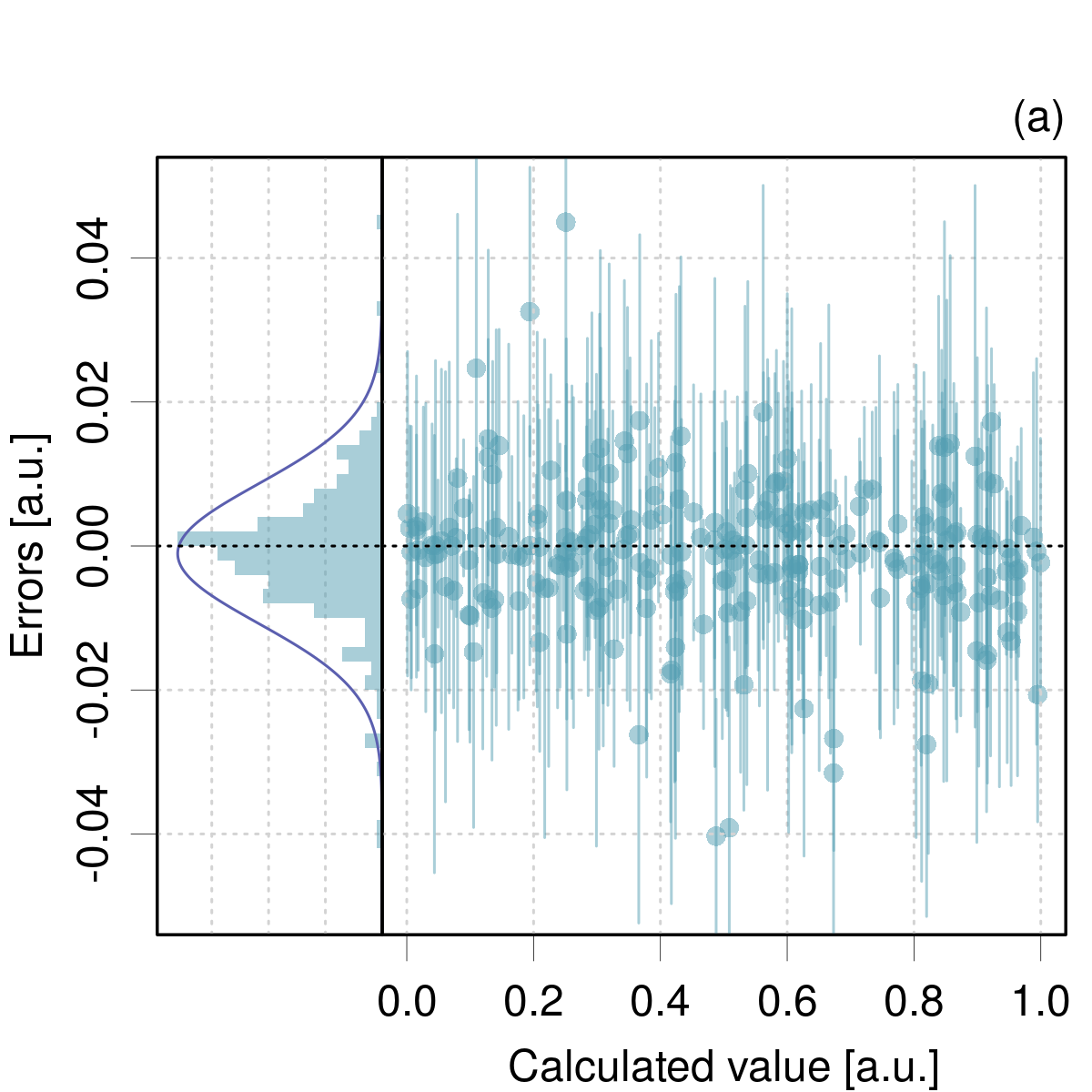} & \includegraphics[height=6cm]{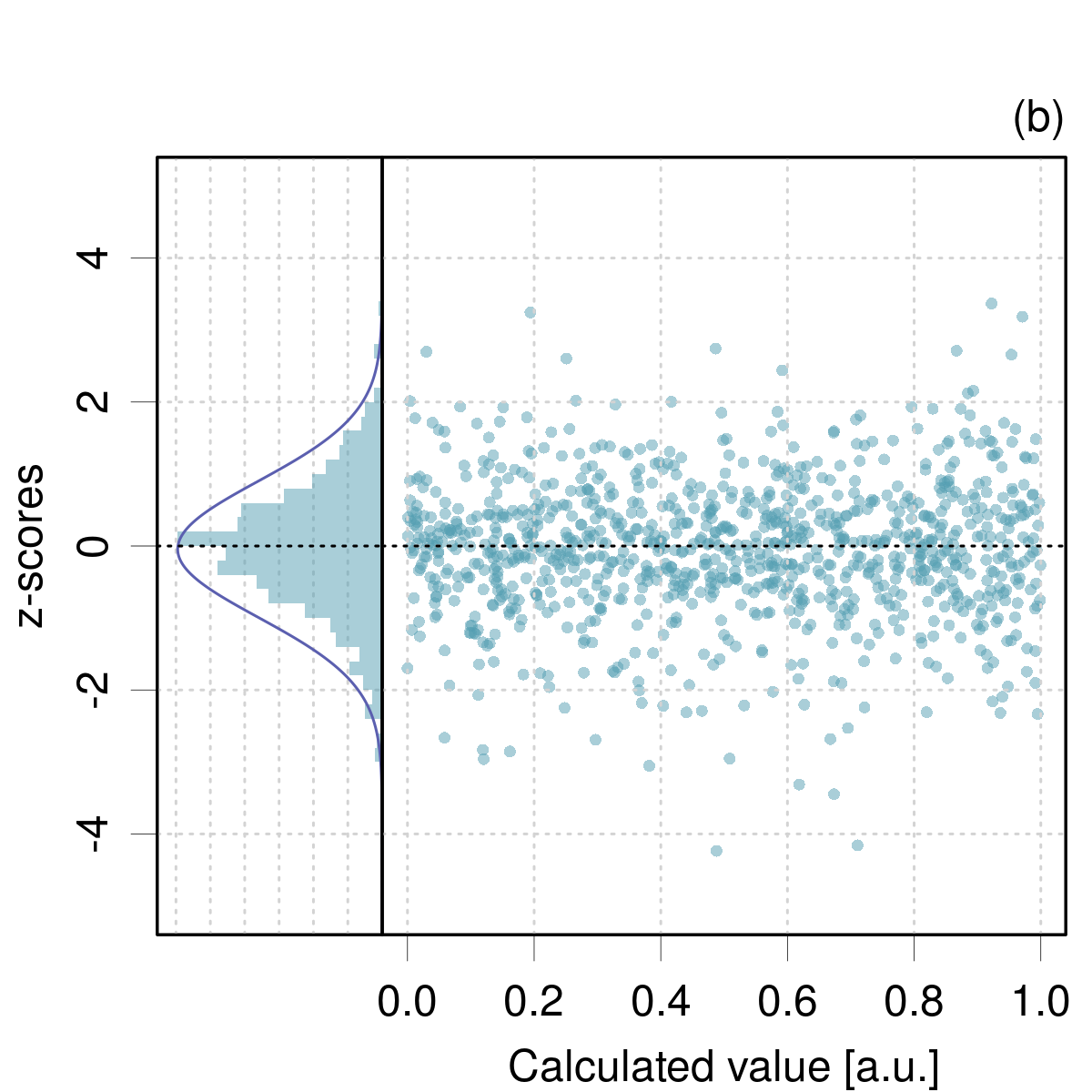}\tabularnewline
\includegraphics[height=6cm]{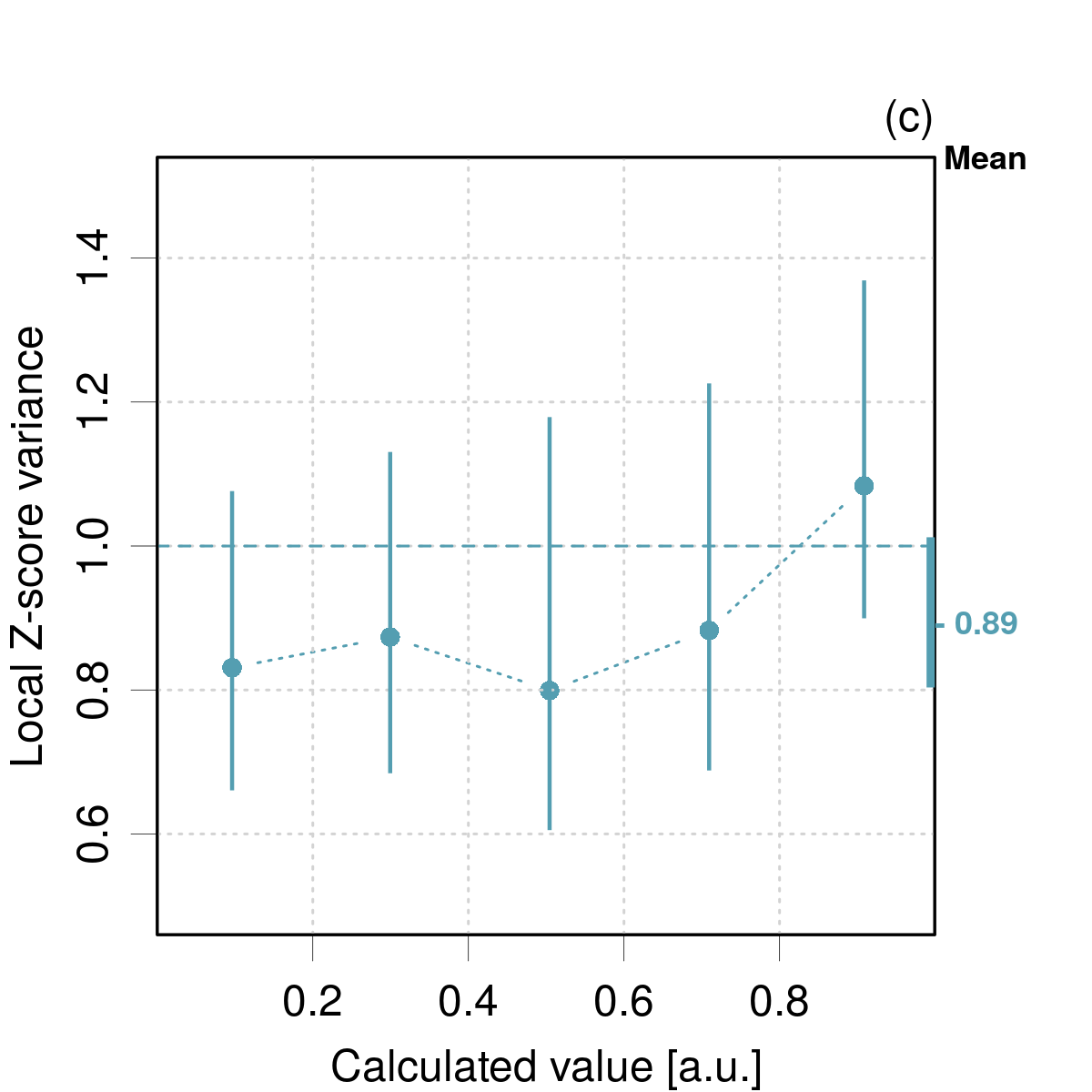} & \includegraphics[height=6cm]{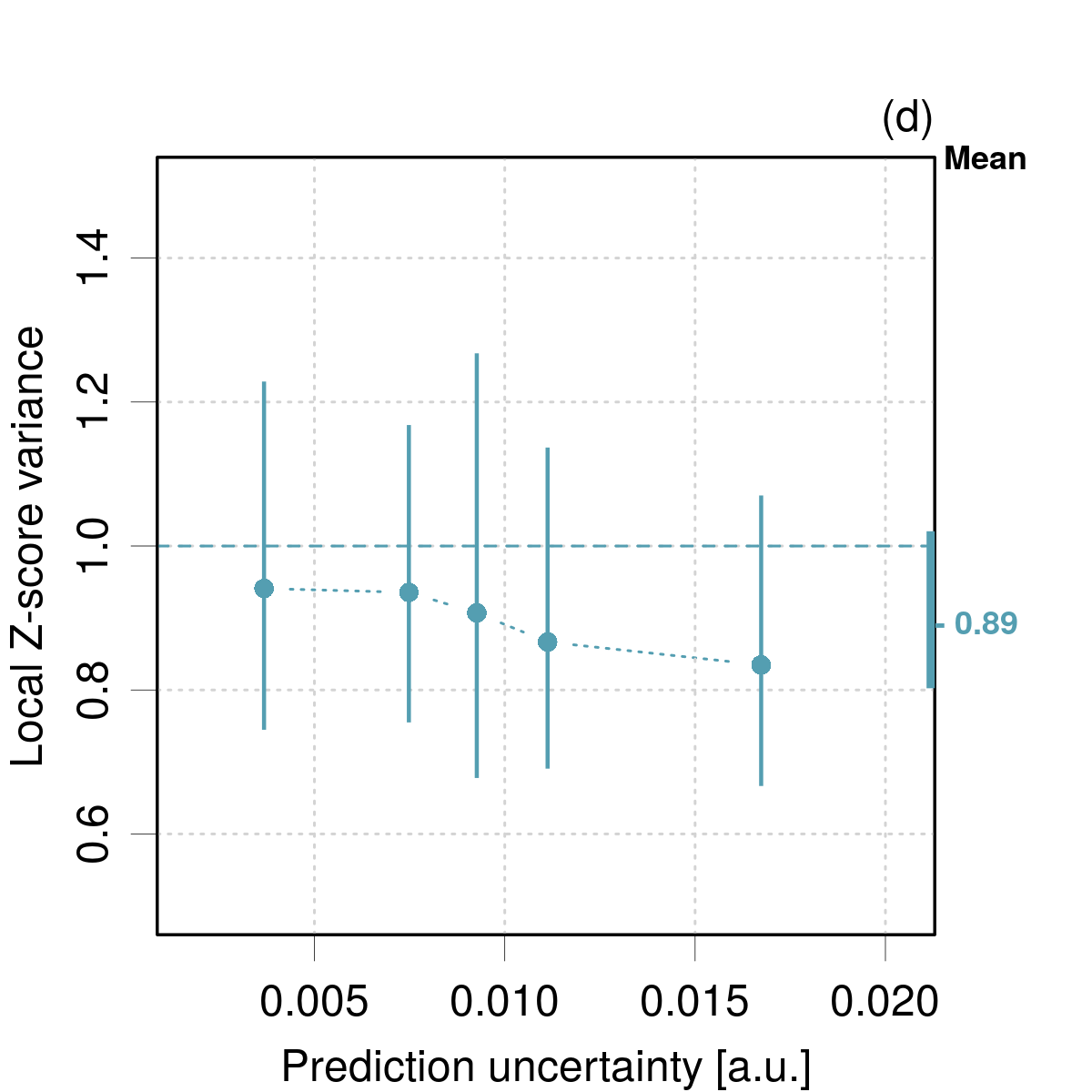}\tabularnewline
\end{tabular}
\par\end{centering}
\caption{\label{fig:example-zscores}Toy model error distribution generated
from Eqs.\,\ref{eq:genz-1}-\ref{eq:genz-2} with $N=1000$: (a)
thinned subset ($N=250$) and $1.96*u_{E}$ for the error bars; (b)
full set of $z$-scores; (c) and (d) LZV analysis of the $z$-scores
with respect to the calculated value (c) and the prediction uncertainty
(d). }
\end{figure}

The LZV analysis has been performed along the calculated values (Fig.\,\ref{fig:example-zscores}(c))
and along the prediction uncertainty (Fig.\,\ref{fig:example-zscores}(d)),
using five areas of 200 points. Both graphs show the compatibility
of the $z$-scores with the unity variance requirement, and no remarkable
trend is observed. 
\end{quote}

\subsection{Implementation}

\subsubsection{Binomial proportions confidence intervals\label{subsec:Binomial-proportions-confidence}}

As presented in Sec. I of the Supplementary Material about PICP testing,
the discreteness of binomial proportions $\nu_{p}$ and the asymmetry
of the associated confidence intervals make the estimation of binomial
proportions confidence intervals a complex problem. Numerous methods
are available, from which I retained the continuity corrected Wilson
method for this study \citep{Newcombe1998}. 

Sec. I of the Supplementary Material also reports considerations about
the power of PICP testing. For instance, to reject the hypothesis
that a PICP value $\nu=0.99$ is equal to the target coverage $p=0.95$,
one needs at least $M=150$ points to achieve a power of 0.8. This
has to be taken into account for the LCP analysis, where I adopt a
systematic strategy to find a balance between testing power and resolution:
the number of subsets is taken as $n=\left\lfloor M/150\right\rfloor $
and constrained to lie between 2 and 15. If the number of subsets
is smaller than 5, a sliding window of size $M/n$ is used. 

\subsubsection{Testing $z$-scores variance\label{subsec:Testing--scores-variance}}

In order to test the variance of $z$-scores with respect to the target
value (1), Sec. II of the Supplementary Material concludes on the
use of a bootstrapping method, with the limits of a 95\,\% confidence
intervals estimated by the BC$_{a}$ method \citep{DiCiccio1996}.
The design parameters for the LZV analysis are the same as for the
LCP analysis. 

\subsubsection{Code availability}

Graphical functions \texttt{plotPIT}, \texttt{plotCalCurve, plotLZV}
and \texttt{plotLCP} have been included in \texttt{ErrViewLib-v1.4}
\url{https://github.com/ppernot/ErrViewLib/releases/tag/v1.4}, also
available in Zenodo at \url{https://doi.org/10.5281/zenodo.5817888}.

\section{Applications\label{sec:Results}}

The tools and statistics presented above for the validation of prediction
uncertainty estimates are applied below to several datasets extracted
from the literature or provided by kind colleagues. The choice of
datasets is subjective, mostly guided by availability and complementarity
in methods and properties.\textcolor{orange}{{} }They cover the major
embedded and a-posteriori CC-UQ methods. These datasets are available
online at \url{https://github.com/ppernot/PU2022}, or in Zenodo at
\url{https://doi.org/10.5281/zenodo.5818026}.

\subsection{Datasets with expanded prediction uncertainty}

Despite the recommendations of Ruscic\citep{Ruscic2014} for the systematic
use of $U_{95}$ expanded uncertainty in reference databases\citep{Klippenstein2017a},
I did not find many CC-UQ studies providing them explicitly for a
reasonable dataset size. I selected two recent studies in which I
found considerations about prediction sharpness.

\subsubsection{Expanded uncertainty of errors}

The ideal scenario is to have sets of predictions and reference data
with their expanded uncertainties: $V_{i}$, $U_{95;i}^{(V)}$, $R_{i}$
and $U_{95,i}^{(R)}$. The estimation of the errors and their associated
expanded uncertainty using the combination of variances requires the
hypothesis that the expansion factors for predictions and reference
values have similar values ($f_{95,i}^{(R)}\simeq f_{95,i}^{(V)}$),
leading to
\begin{equation}
U_{95,i}=\sqrt{(U_{95,i}^{(R)})^{2}+(U_{95,i}^{(V)})^{2}}\label{eq:comb-u95-1-1}
\end{equation}
 It is then directly possible to test the PICP value
\begin{equation}
\nu_{0.95}=\frac{1}{M}\sum_{i=1}^{M}\boldsymbol{1}\left(|E_{i}|/U_{95,i}\le1\right)
\end{equation}
leading to a single assessment of calibration at the $p=0.95$ level. 

When the uncertainties of the reference data are negligible, this
approach directly tests the quality of the expanded uncertainties
for the predicted values $U_{95}^{(V)}$. Otherwise, the quality of
the reference set uncertainties $U_{95}^{(R)}$ will also affect the
results. 

\subsubsection{The BAK2021 dataset\label{subsec:The-BAK2021-dataset}}

Bakowies and von\,Lilienfeld \citep{Bakowies2021} proposed a new
method to estimate zero-point energies (ZPE) and the corresponding
expanded $U_{95}$ uncertainty in the framework of the composite ATOMIC
method. Interestingly, they observed a quadratic dependence of the
ZPE scaling factors and their dispersion with the fraction of heteroatoms
in a set of 279 molecules. From this, they built a statistical model
for 95\,\% prediction intervals, and they validated this heteroscedastic
expanded uncertainty on a set of 99 molecules against CCSD(T) ZPE
values, by looking for outlying points. They did not identify any
concern about the predicted prediction intervals and conclude that
their method ``provides a fair estimate of 95\,\% confidence''
\citep{Bakowies2021}. 

\begin{figure}[t]
\begin{centering}
\includegraphics[height=6cm]{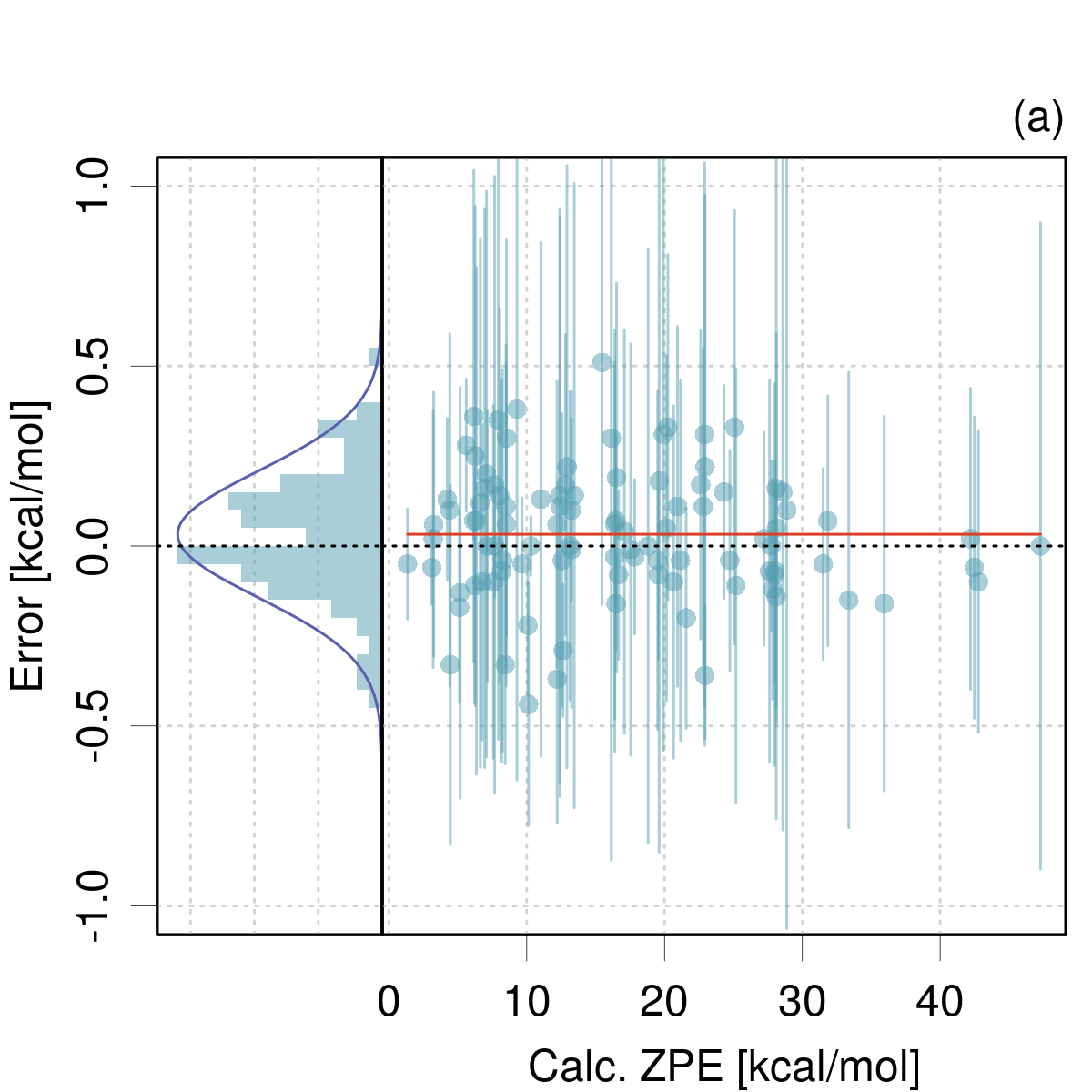}\includegraphics[height=6cm]{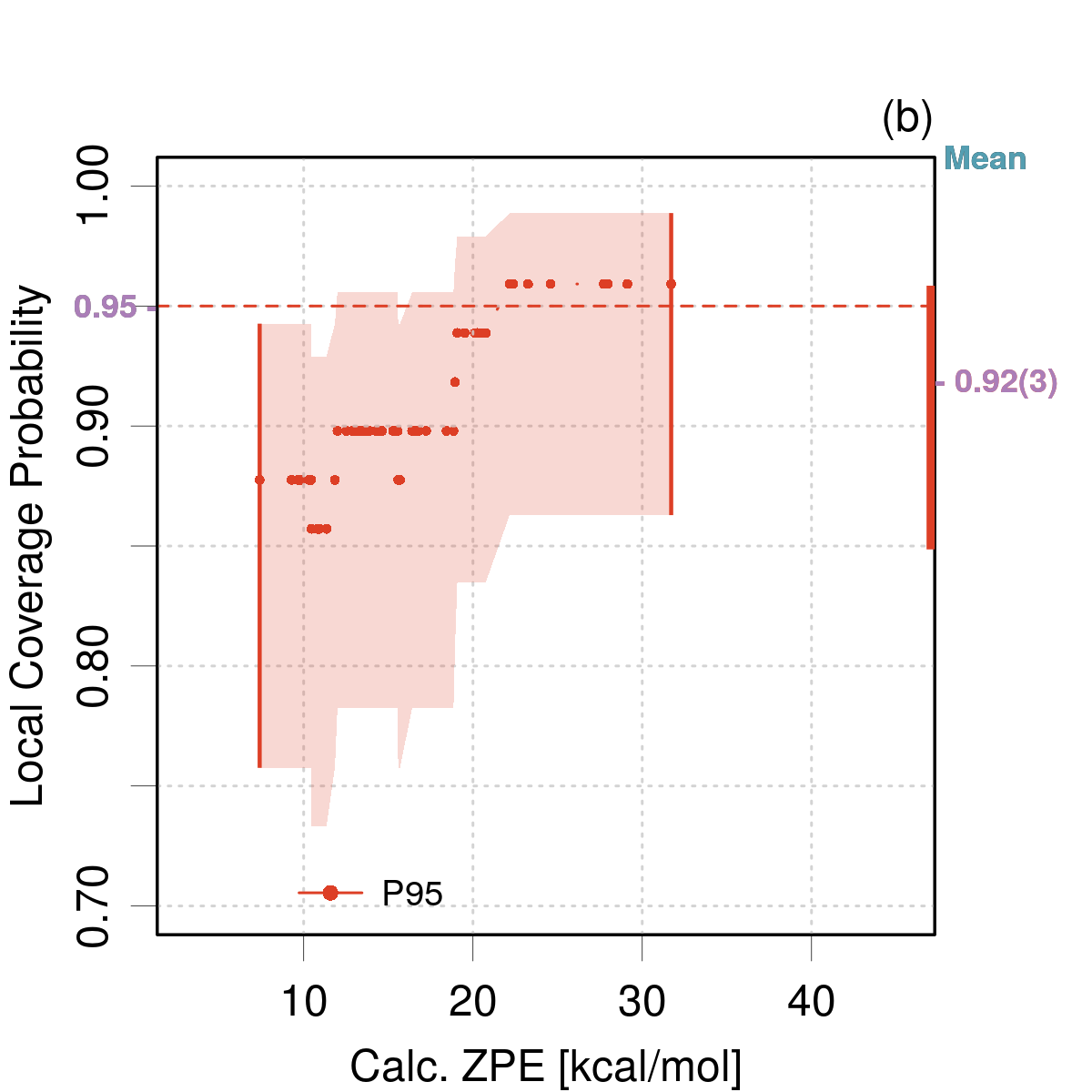}
\par\end{centering}
\begin{centering}
\includegraphics[height=6cm]{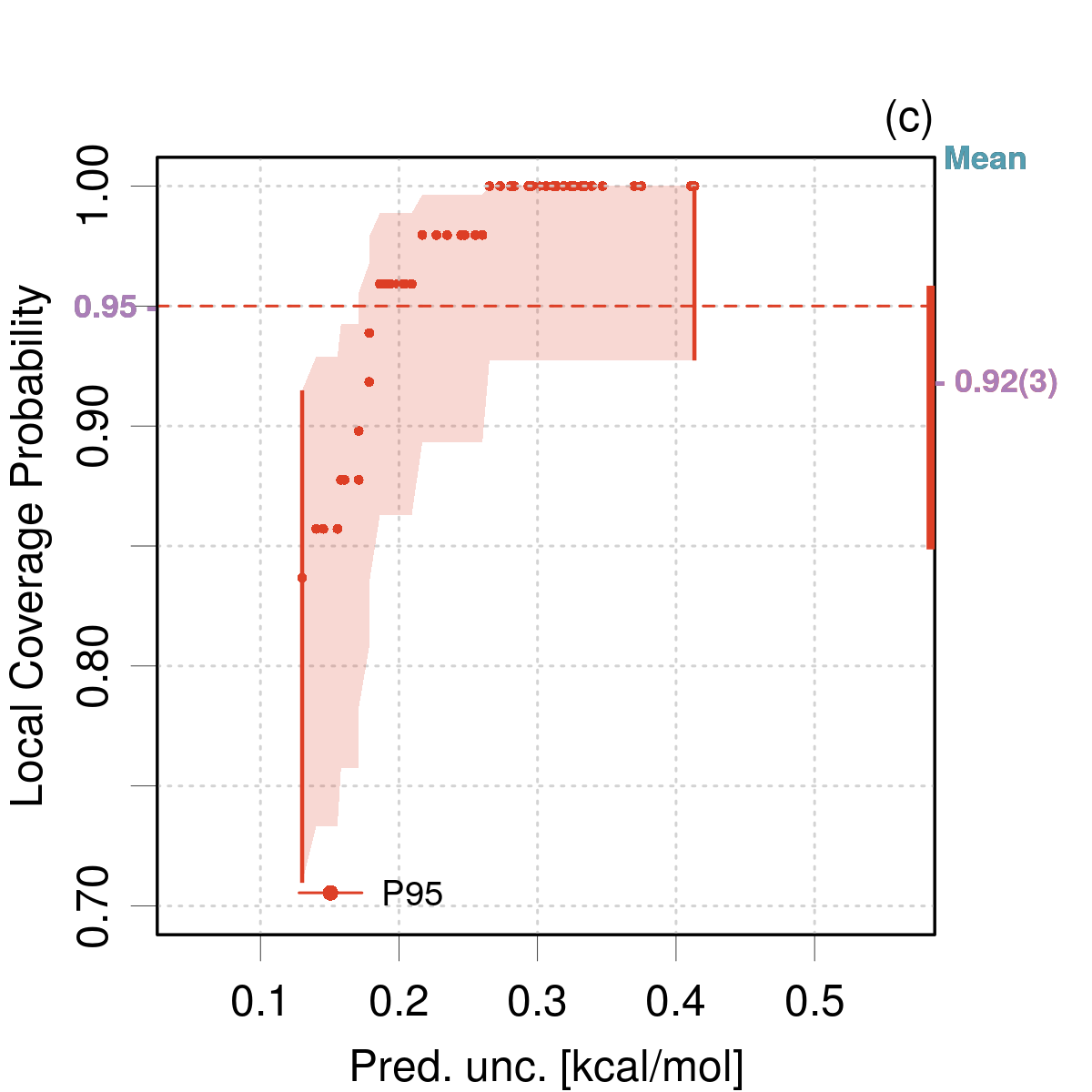}\includegraphics[height=6cm]{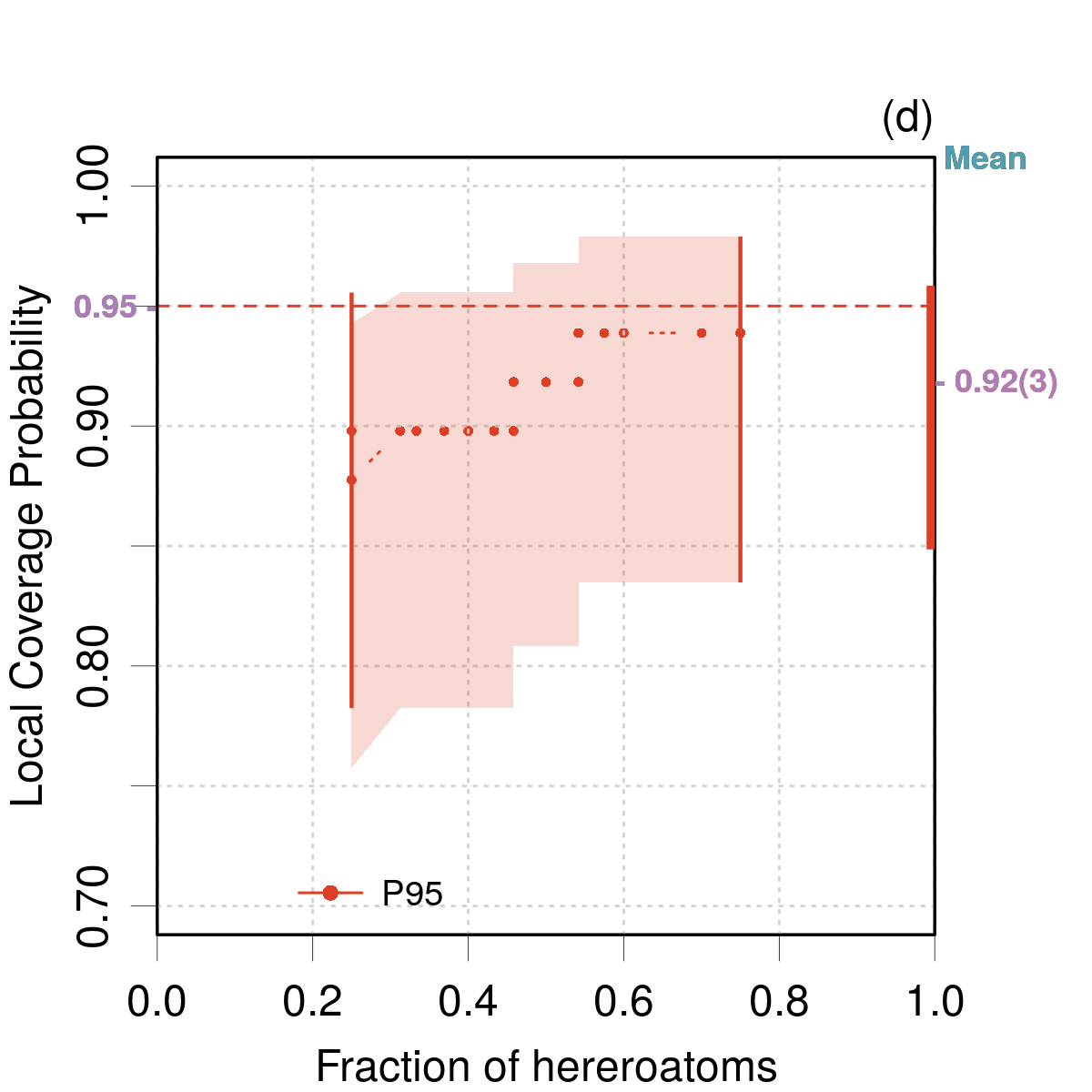}
\par\end{centering}
\caption{\label{fig:BAK2021}Distribution of errors (a) and alternative versions
of LCP analysis (b)-(d) for the BAK2021 dataset. The LCP analysis
has been done using (b) the calculated value, (c) the prediction uncertainty
and (d) the heteroatoms content, as predictor variables.}
\end{figure}

This is a small dataset ($N=99$), and I find interesting to check
how the validation tools perform in this context. The values for the
ATOMIC-2(um) and CCSD(T) ZPE data were manually (and painstakingly)
extracted from Table S6 of the Supporting Information of the original
article \citep{Bakowies2021}. There is no uncertainty on the reference
values for this dataset. The errors distribution is shown in Fig.\,\ref{fig:BAK2021}(a).

Considering that expanded $U_{95}$ uncertainties are reported, I
tested only this probability level. The global PICP $\nu_{0.95}=0.92(3)$
is not significantly different from the target value. 

Three alternative versions of the LCP analysis are tested. The LCP
areas have been chosen along the calculated ZPE values (Fig.\,\ref{fig:BAK2021}(b)),
the estimated prediction uncertainty (Fig.\,\ref{fig:BAK2021}(c))
and the heteroatoms content of the molecules (Fig.\,\ref{fig:BAK2021}(d)).
Because of the small dataset, a sliding window of width $M=N/2$ was
used in all cases. Despite the large error bars, a few points indicate
a statistical rejection of local PICP values. In ZPE space (Fig.\,\ref{fig:BAK2021}(b)),
one notices a global trend, with PICP values increasing from 0.89
to 0.96 as the ZPE value increases. This might reveal an underestimation
of the uncertainties for small ZPE values. An analogous trend is observed
in prediction uncertainty space (Fig.\,\ref{fig:BAK2021}(c)), with
PICP values increasing from 0.84 to 1. This would indicate that small
prediction uncertainties are too small, and large ones too large.
The third representation is along the property that was used to design
the prediction uncertainty model (Fig.\,\ref{fig:BAK2021}(d)), and
the positive trend, although still present, is much weaker than in
the ZPE or prediction uncertainty space. 

Overall, one can agree with Bakowies and von\,Lilienfeld that their
UQ method globally provides ``fair'' $U_{95}$ values, but with
a caveat about a local calibration issue, as revealed in the PU-space
LCP analysis.

\subsubsection{The PRO2021 dataset\label{subsec:The-PRO2021-dataset}}

In a recent study, Proppe and Kircher \citep{Proppe2021} estimated
prediction uncertainty for reaction rates derived by the Mayr-Patz
equation. Their uncertainty model accounts for parametric uncertainty
and model discrepancy, for which they compared two versions. In the
first version (a), the\textcolor{violet}{{} }model dispersion is uniform
(homoscedastic),\textcolor{violet}{{} }whereas in the second version
(b), it has a polynomial dependence on the parametric uncertainty.
Using graphs showing the correlation between prediction uncertainty
and dispersion of the residuals (their Fig. 5), Proppe and Kircher
conclude that the second version offers a much better fit to the residuals
distribution (better sharpness). However, both models provide 95\,\%
prediction intervals with an excessive 99\,\% coverage. 

\begin{figure}[t]
\begin{centering}
\includegraphics[height=6cm]{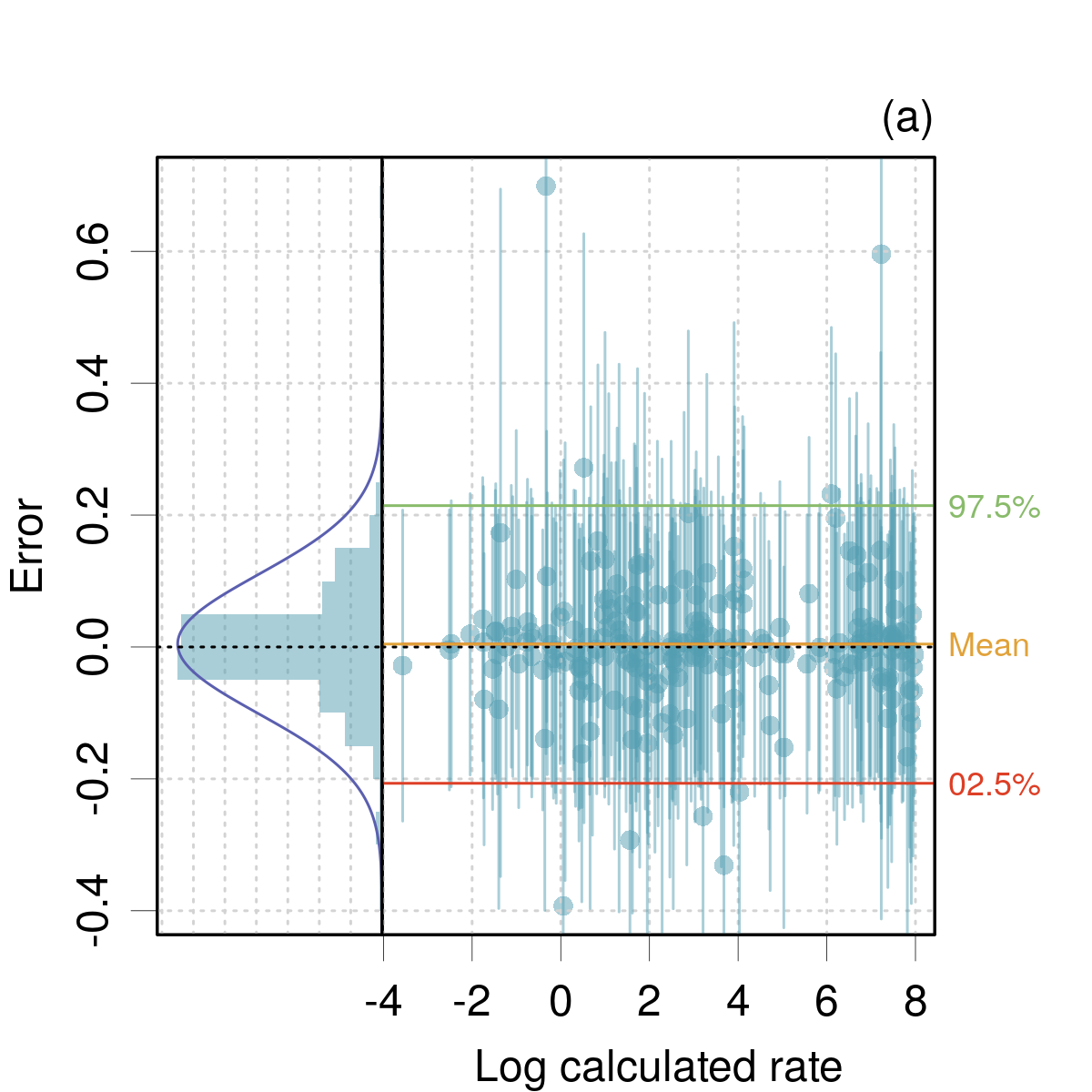}\includegraphics[height=6cm]{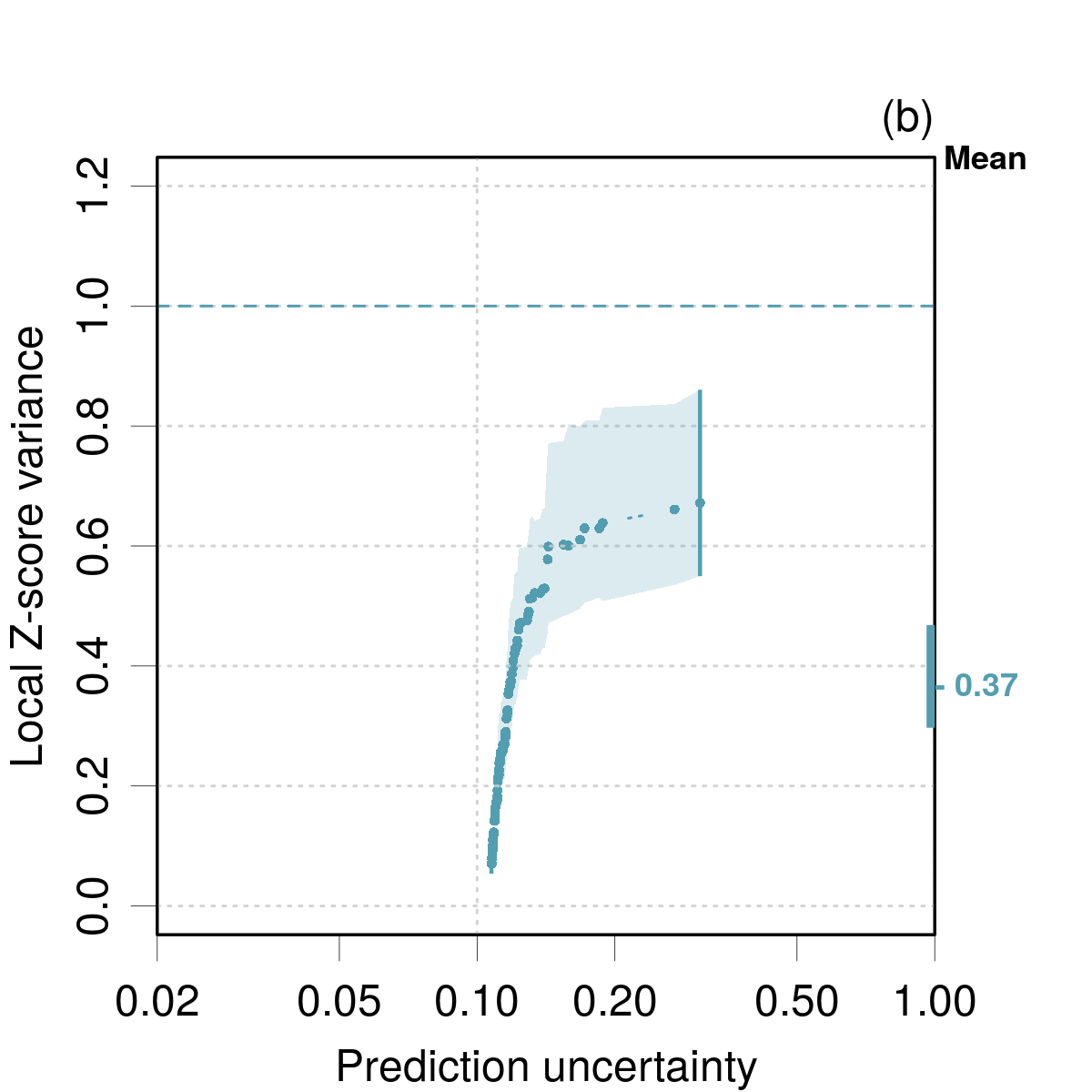}\includegraphics[height=6cm]{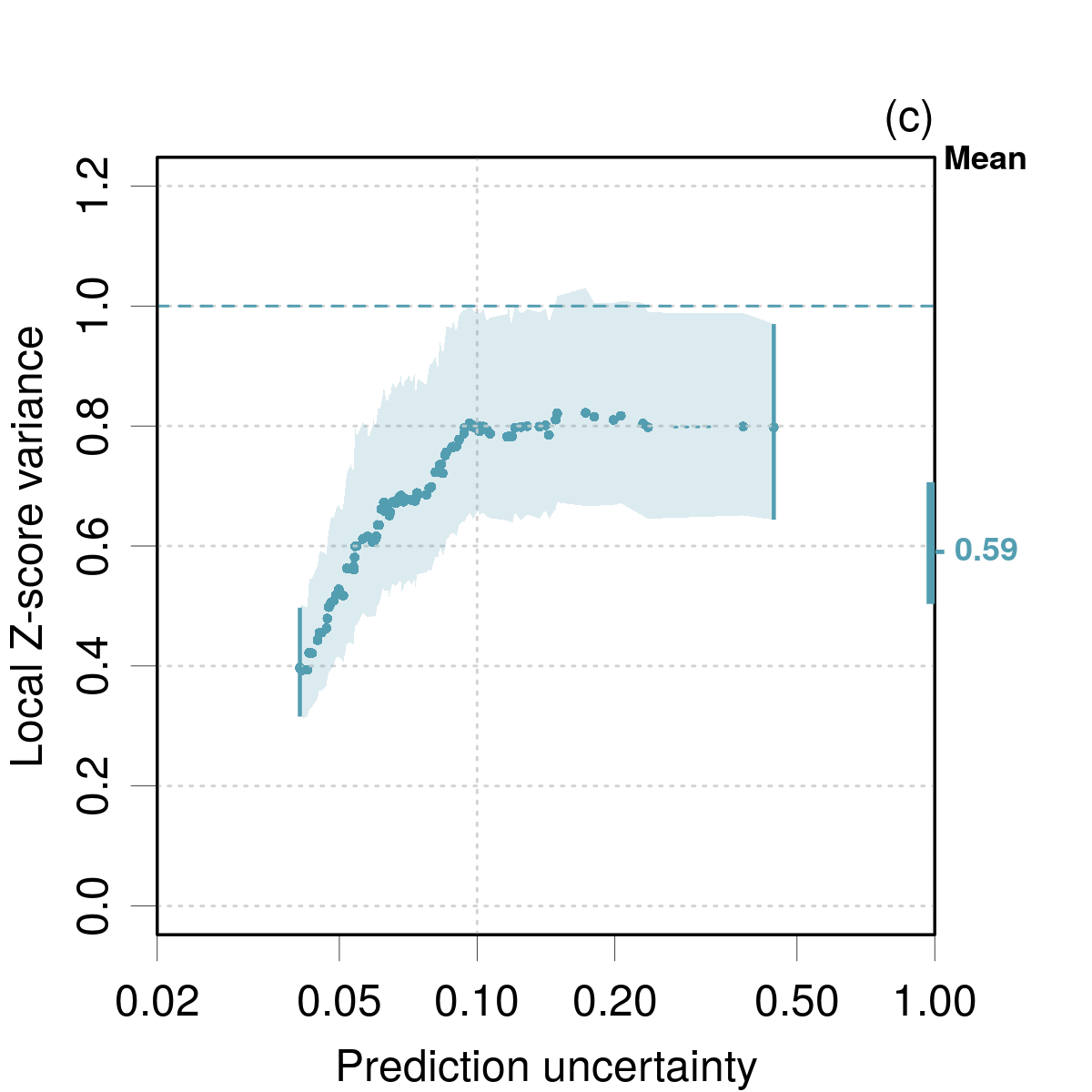}
\par\end{centering}
\caption{\label{fig:PRO2021} Errors distribution (a) and LZV analysis (b)
and (c) for the PRO2021 dataset. (a) and (b) Uniform model dispersion.
(c) Polynomial model dispersion. }
\end{figure}

The authors kindly provided me the corresponding datasets containing
$M=212$ points with a reference value (log of the experimental reaction
rate), a calculated value (log of the calculated reaction rate) and
an expanded uncertainty on the error, for both model versions (a)
and (b). No further data treatment was necessary. The distribution
of errors and their expanded uncertainties is shown in Fig.\,\ref{fig:PRO2021}(a)
for method (a). There is no easily perceptible difference on the same
plot for method (b), which is not shown. 

The PICP is $\nu_{0.95}=0.995(5)$ for both models, \emph{i.e., }211
points out of 212 are included in the predicted $I_{95}$ intervals,
and the lists of included systems for both models differ only by two
points. In such conditions, the coverage statistics cannot be used
to differentiate the models, and due to the saturation of the PICP
values (very close to the upper limit), the LCP analysis does not
provide useful information (not shown).

Considering that the expanded uncertainties were derived by Proppe
and Kircher from standard uncertainties by an enlargement factor of
1.96, I derived the standard uncertainties and estimated the $z$-scores
statistics. Both sets are unbiased but present variances significantly
different from 1: 0.370(43) for model (a) and 0.595(49) for model
(b), corresponding to an overestimation of uncertainty. The LZV analysis
with a sliding interval of width $M/2$ was performed against prediction
uncertainty. The results are shown in Figs.\,\ref{fig:PRO2021}(b,c).
Three features call for comments: (1) model (b) has a wider range
of PUs than model (a), which might help it to better fit the data;
(2) model (b) is much closer to the target, notably for small PUs,
and (3) both models present a similar trend showing that small PUs
are more overestimated than the large ones.

These tests bring evidence that the polynomial model (b) is a notable
improvement over the uniform model (a). However there remains a global
overestimation of the uncertainties, and the lack of local calibration
is not fully resolved. 

\subsection{Datasets with standard prediction uncertainty}

I consider here datasets where the coverage level of prediction uncertainty
is not explicitly provided. 

\subsubsection{Standard uncertainty of errors}

The most common scenario in the CC-UQ literature is based on standard
uncertainties of the predictions and reference data: $V_{i}$, $u_{V,i}$,
$R_{i}$ and $u_{R,i}$. The uncertainty on the errors is directly
accessible through the combination of variances
\begin{equation}
u_{E,i}=\sqrt{u_{R,i}^{2}+u_{V,i}^{2}}
\end{equation}
enabling the computation of $z$-scores without further hypothesis,
to be tested by their mean and variance. 

Again, non-negligible uncertainties on the reference data might affect
the results. If the reference uncertainty is missing, it can be taken
as null (for instance in the case of high-accuracy calculated values)
or as constant, if a typical value can be found.

\subsubsection{The FEL2008 dataset\label{subsec:The-FEL2008-dataset}}

This dataset has been extracted manually from Table VII of a 2008
article by D. Feller \emph{et al.} \citep{Feller2008}, reporting
atomization energies estimated by the FPD method for 106 small molecules.
After removing data with missing uncertainty, one is left with a set
of 102 systems with a predicted value, a prediction uncertainty, and
a reference value with its uncertainty.\textcolor{orange}{{} }The prediction
and reference uncertainties have to be combined to estimate the errors
uncertainty, by a rule that depends on their nature (standard, expanded
or other...).

For the reference data, in the absence of specific information, I
assumed that expanded uncertainties were used ($U_{95}^{(R)}$). The
process of prediction uncertainty estimation in the FPD method has
been summarized above (Section\,\ref{subsec:Bottom-up-correction-methods}),
and it is difficult to infer its nature, beyond a possible over-estimation
implied by the worst-case scenario strategy.

As an initial assessment, I estimated that FPD PUs are close to standard
uncertainties and computed $u_{E}$ values as $u_{E,i}=\sqrt{u_{V,i}^{2}+\left(U_{95,i}^{(R)}/1.96\right)^{2}}$and
derived the corresponding $z$-scores.

\begin{figure}[t]
\begin{centering}
\includegraphics[height=6cm]{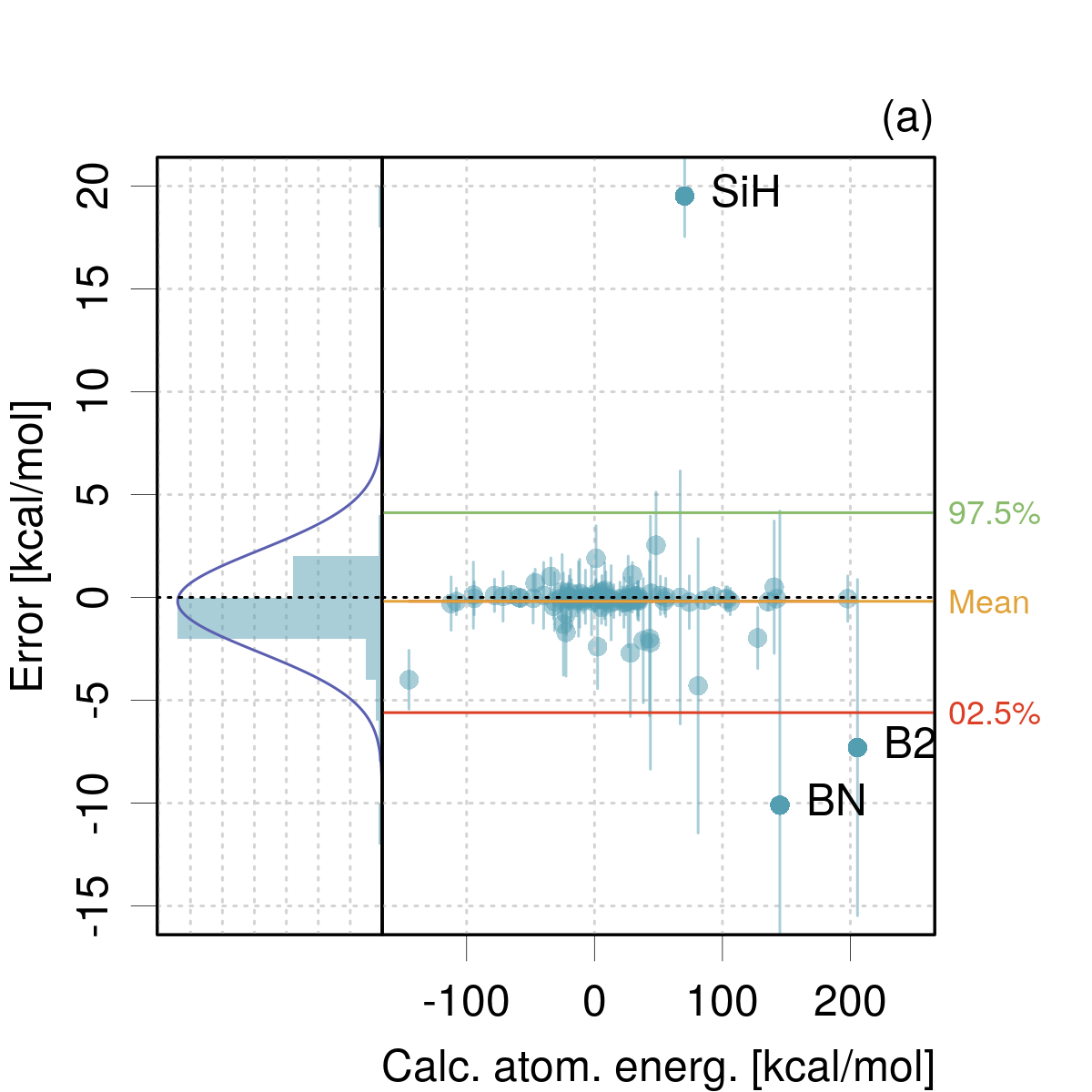}\includegraphics[height=6cm]{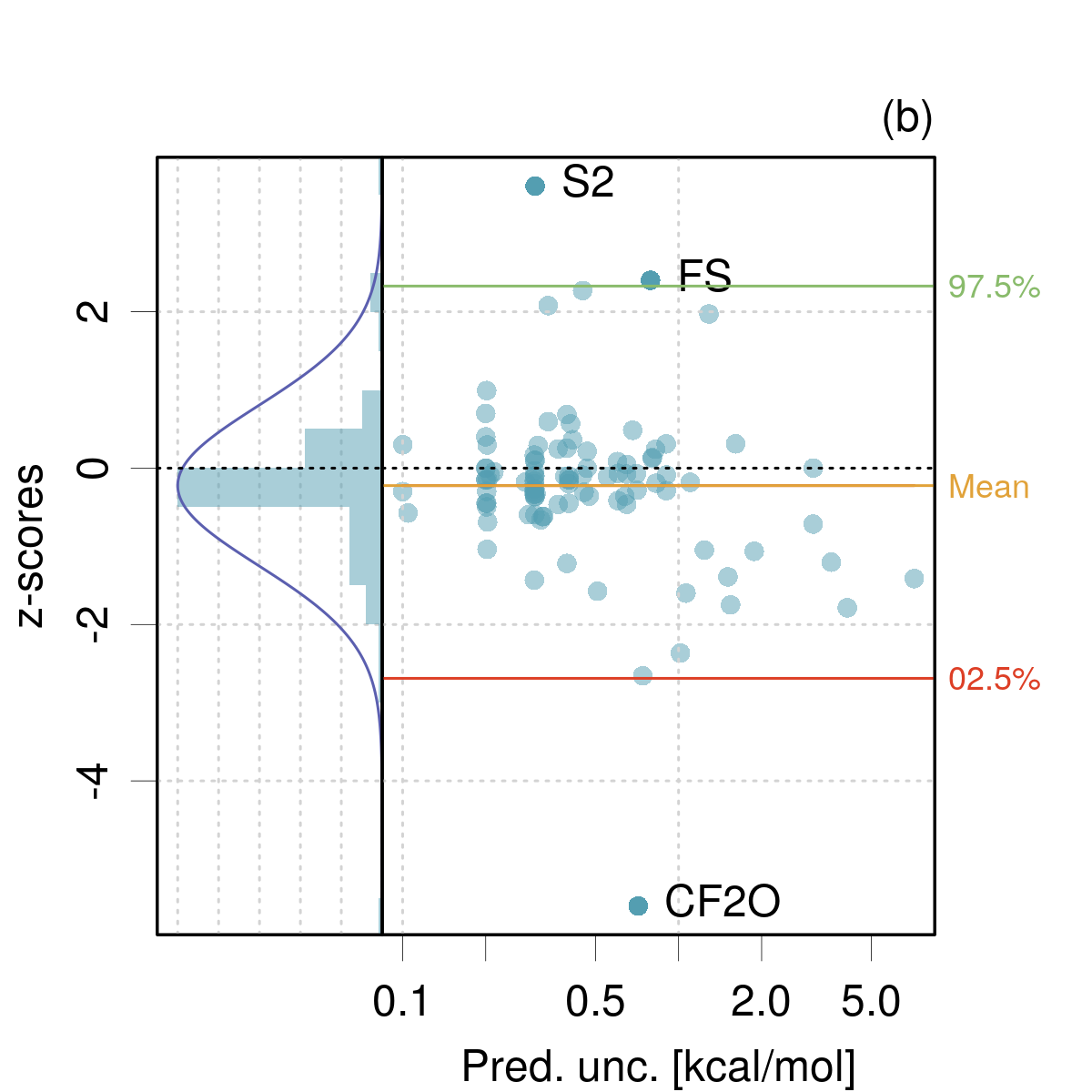}\includegraphics[height=6cm]{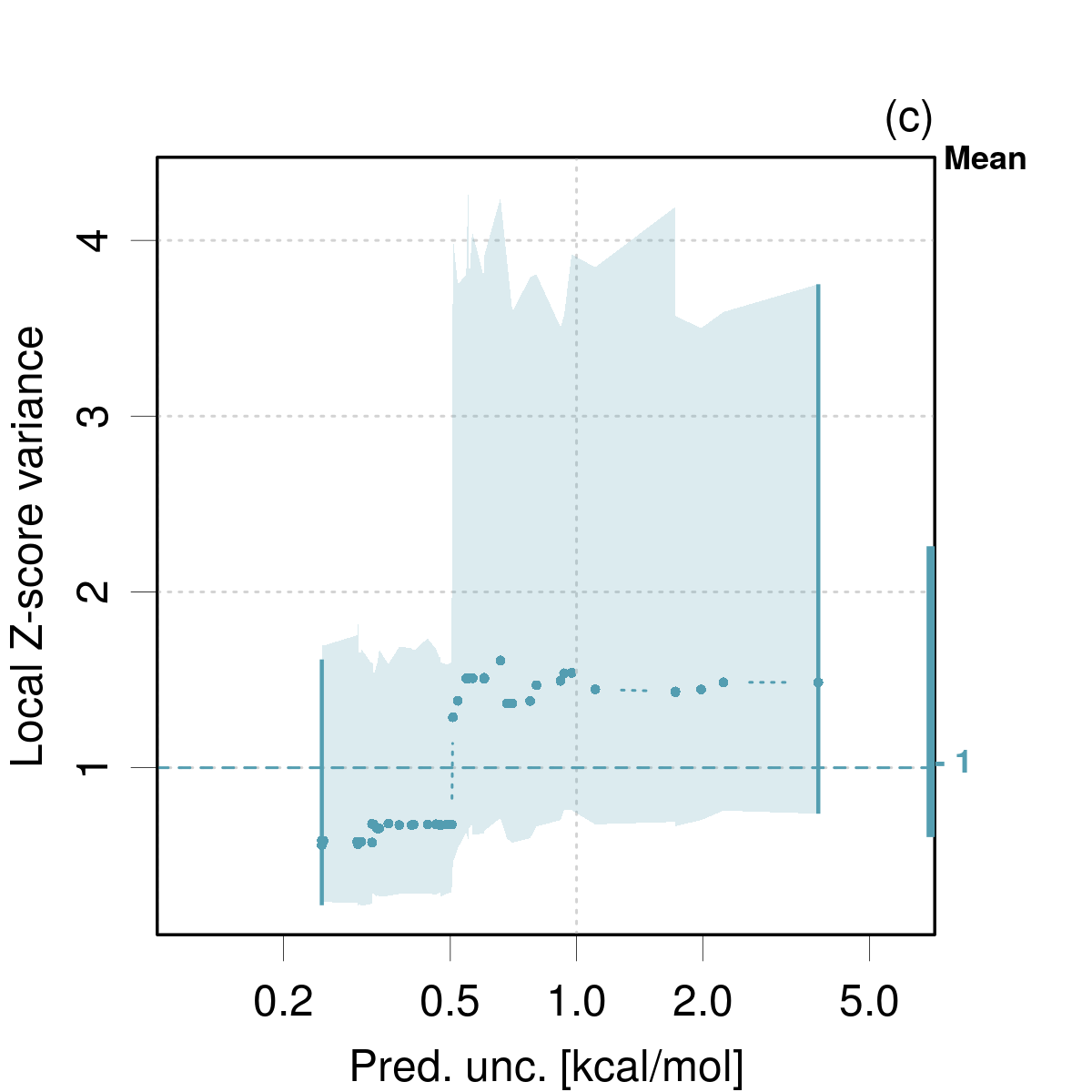}
\par\end{centering}
\caption{\label{fig:dist-FEL2008}Distribution of errors (a), z-scores (b)
and LZV analysis (c) for the FEL2008 dataset. In (a) and (b), the
left panel shows an histogram of the data with its best amplitude-normalized
Gaussian fit. The SiH system has been removed for the $z$-score plot
and LZV analysis.}
\end{figure}

The distribution of errors with their uncertainty is shown in Fig.\,\ref{fig:dist-FEL2008}(a).
SiH stands as a strong outlier, with an uncertainty too small to cover
its large error level. The case is not discussed in the original article,
although it is stated that the heat of formation of silicon was not
well established at the time of publication. However, there does not
seem to be a systematic error, as other Si-containing systems are
not visibly affected. Other outliers such as BN or B$_{2}$ have error
bars large enough to cover their deviation. 

A plot of the $z$-score distribution after the removal of the SiH
system (Fig.\,\ref{fig:dist-FEL2008}(b)) reveals two points with
outstanding values: the atomization energies of S$_{2}$ and CF$_{2}$O
have apparently underestimated uncertainties. The histogram of the
$z$-scores (Fig.\,\ref{fig:dist-FEL2008}(b), left panel) shows
a strong concentration of small $z$-scores values and a strong departure
from a normal shape. Nevertheless, the variance of the $z$-scores
is close to 1: $\mathrm{Var}(Z)=1.04(33)$, which does not invalidate
the derivation of $u_{E}$. The LZV analysis against prediction uncertainty
is shown in Fig.\,\ref{fig:dist-FEL2008}(c) for a sliding window
of width $M/2$, showing no significant deviation from the target,
considering the large uncertainties. There is a slight positive trend,
with a step around 0.5\,kcal/mol, which is linked to the prevalence
of negative $z$-score values for larger PUs (Fig.\,\ref{fig:dist-FEL2008}(b)).

Unless my treatment of the original data is unduly lucky, it appears
that the ``crude and hopefully conservative'' appreciation of the
original authors \citep{Feller2017} might be self-deprecating, but
more data would be necessary to conclude.

\subsubsection{The PAN2015 dataset\label{subsec:PAN2015}}

A dataset of 257 formation heats predicted by the mBEEF method, and
their standard uncertainties, have been extracted from a 2015 article
Pandey and Jacobsen \citep{Pandey2015}. The reference data have no
uncertainty reported. I previously analyzed this dataset (PAN2015)
\citep{Pernot2017b}, showing an inconsistency between the prediction
uncertainties and the errors amplitudes. For instance, the mean prediction
uncertainty (Eq.\,\ref{eq:MPU-1}; 0.18~kcal/mol) significantly
exceeds~the standard deviation of the errors ($\mathrm{sd}(E)=0.13$~kcal/mol).
I propose here to check how the proposed validation methods perform
with this dataset. 
\begin{figure}[t]
\begin{centering}
\includegraphics[viewport=0bp 0bp 1200bp 1200bp,clip,height=6cm]{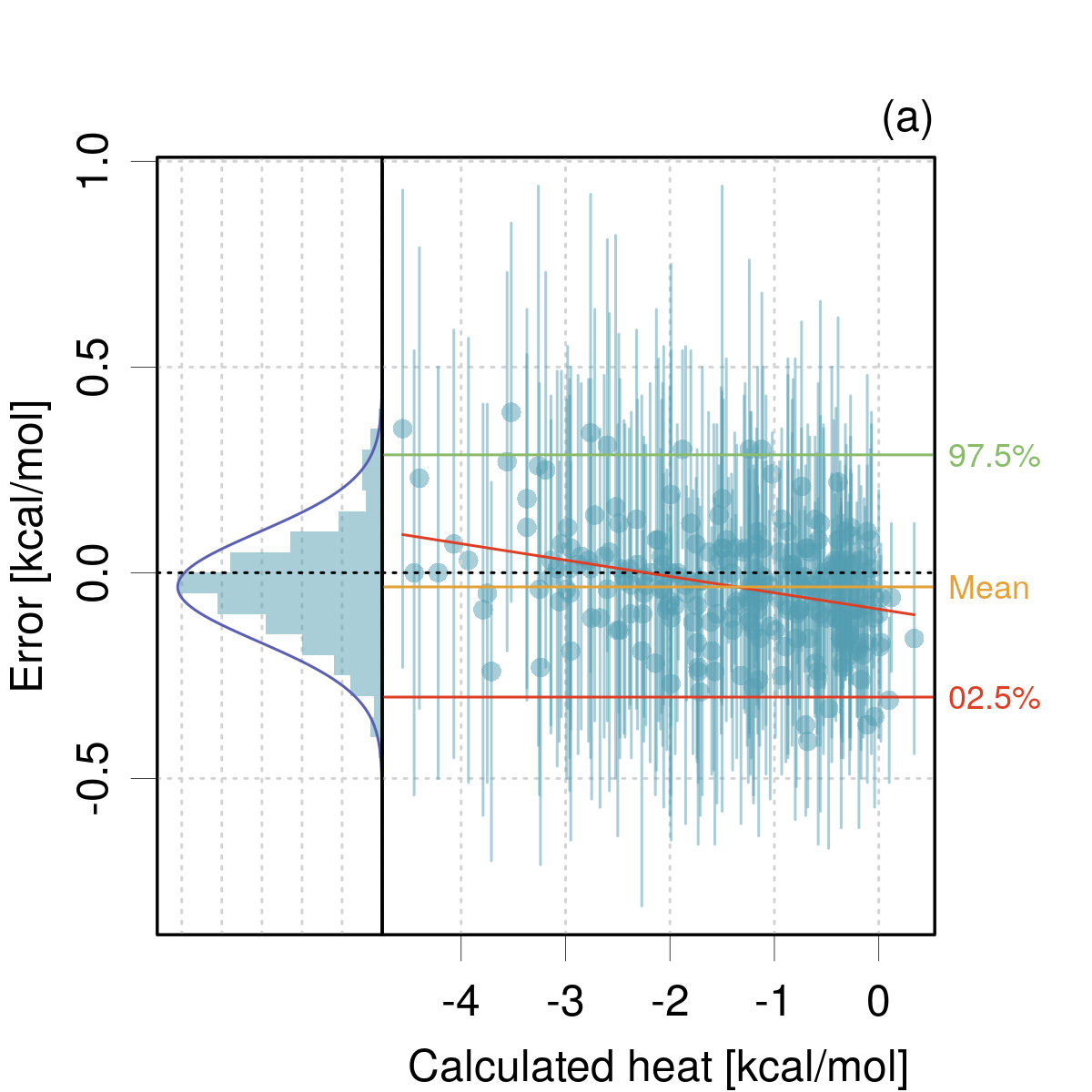}\includegraphics[clip,height=6cm]{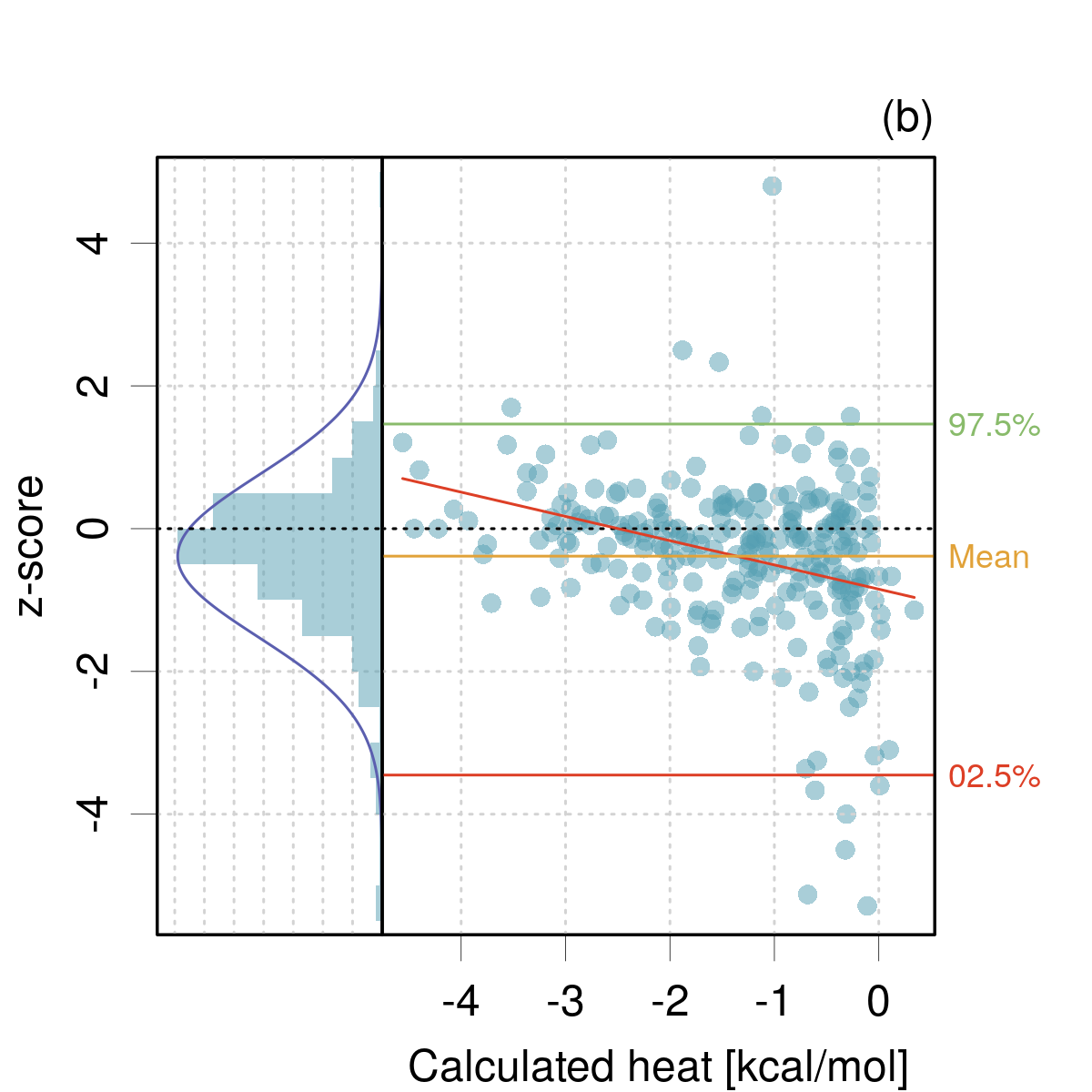}
\par\end{centering}
\begin{centering}
\includegraphics[height=6cm]{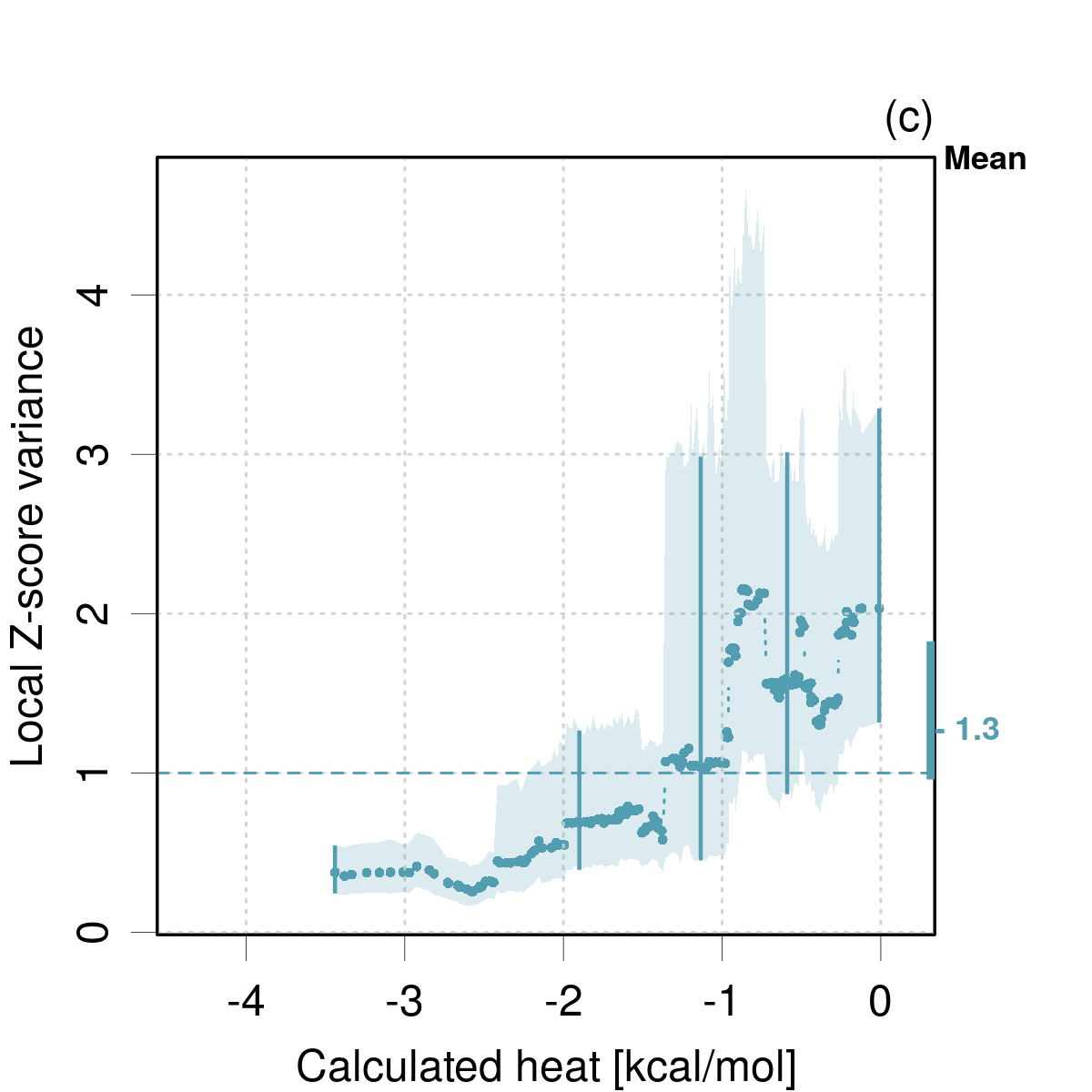}\includegraphics[clip,height=6cm]{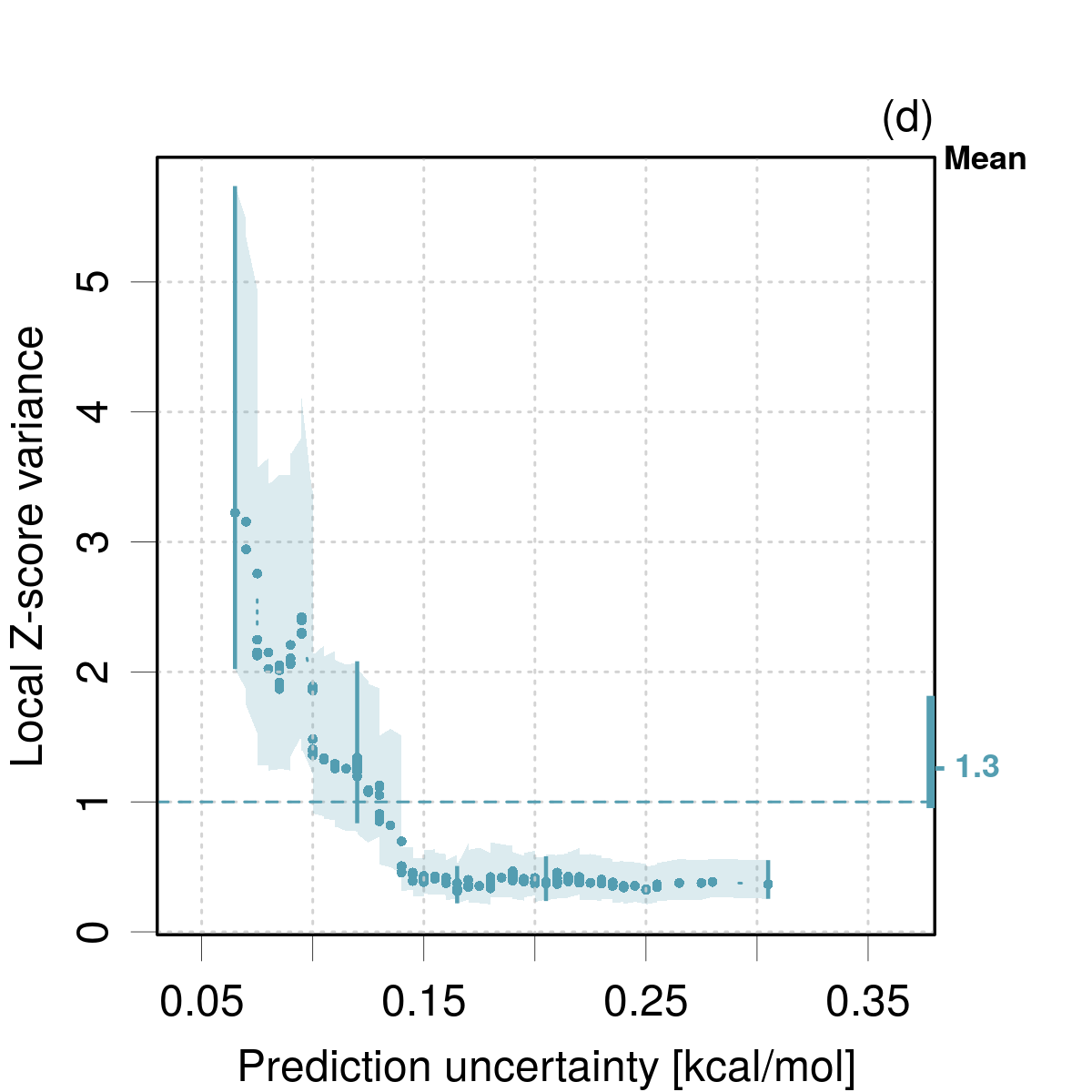}
\par\end{centering}
\caption{\label{fig:PAN2015}Distribution of errors (a) and $z$-scores (b)
for the PAN2015 dataset. LZV analysis of $z$-scores along the predicted
values (c) and the predicted uncertainties (d), using a sliding window
of width $M/5$. }
\end{figure}

The error set and the corresponding $z$-scores are plotted in Fig.\,\ref{fig:PAN2015}.
Both distributions deviate from normality. Besides a small trend in
both quantities, the main feature in these plots is the heterogeneity
of the $z$-scores distribution along the predicted values, with a
noticeable negative tail for heats above -1\,kcal/mol. Some uncertainties
in this area seem to be underestimated. 

The global $z$-scores statistics are not ideal: $\mathrm{E}(Z)=-0.385(72)$
and $\mathrm{Var}(Z)=1.28(20)$ with a 95\,\% confidence interval
$[0.96,\thinspace1.80]$. The variance is not incompatible with its
target (1), but considering the heteroscedasticity of the $z$-scores,
it increases from about 0.5 for the first half of the dataset to about
2 for the second half, with confidence intervals excluding 1 (LZV
analysis, Fig.\,\ref{fig:PAN2015}(c)). The LZV analysis with respect
to the prediction uncertainty shows the opposite trend, decreasing
from a value above 3 to about 0.5 for the larger PUs (Fig.\,\ref{fig:PAN2015}(d)).
Briefly, small uncertainties are underestimated, while large ones
are overestimated. 

\subsubsection{The PAR2019 dataset\label{subsec:PAR2019}}

\begin{figure}[t]
\begin{centering}
\includegraphics[viewport=0bp 0bp 1440bp 1200bp,clip,height=6cm]{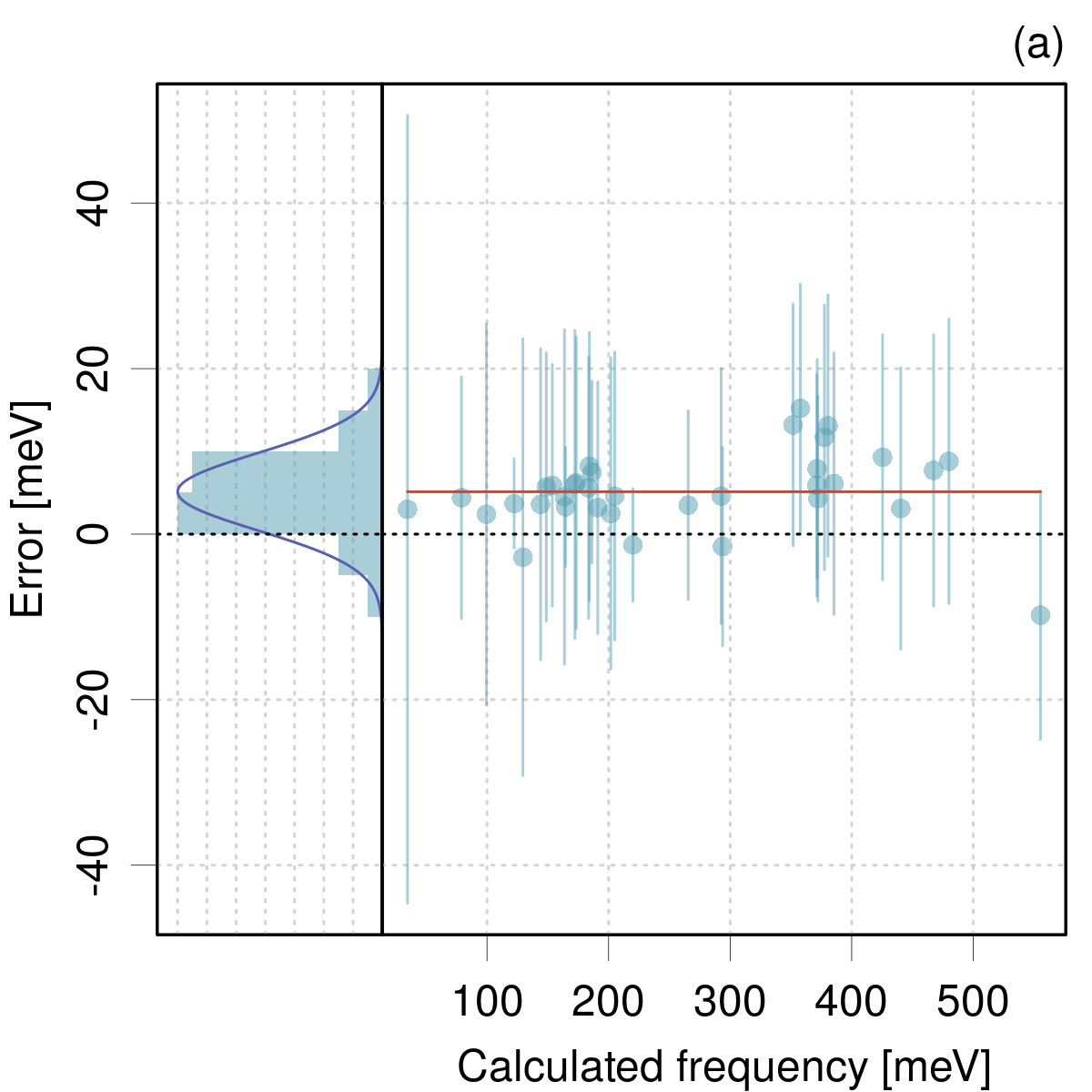}\includegraphics[clip,height=6cm]{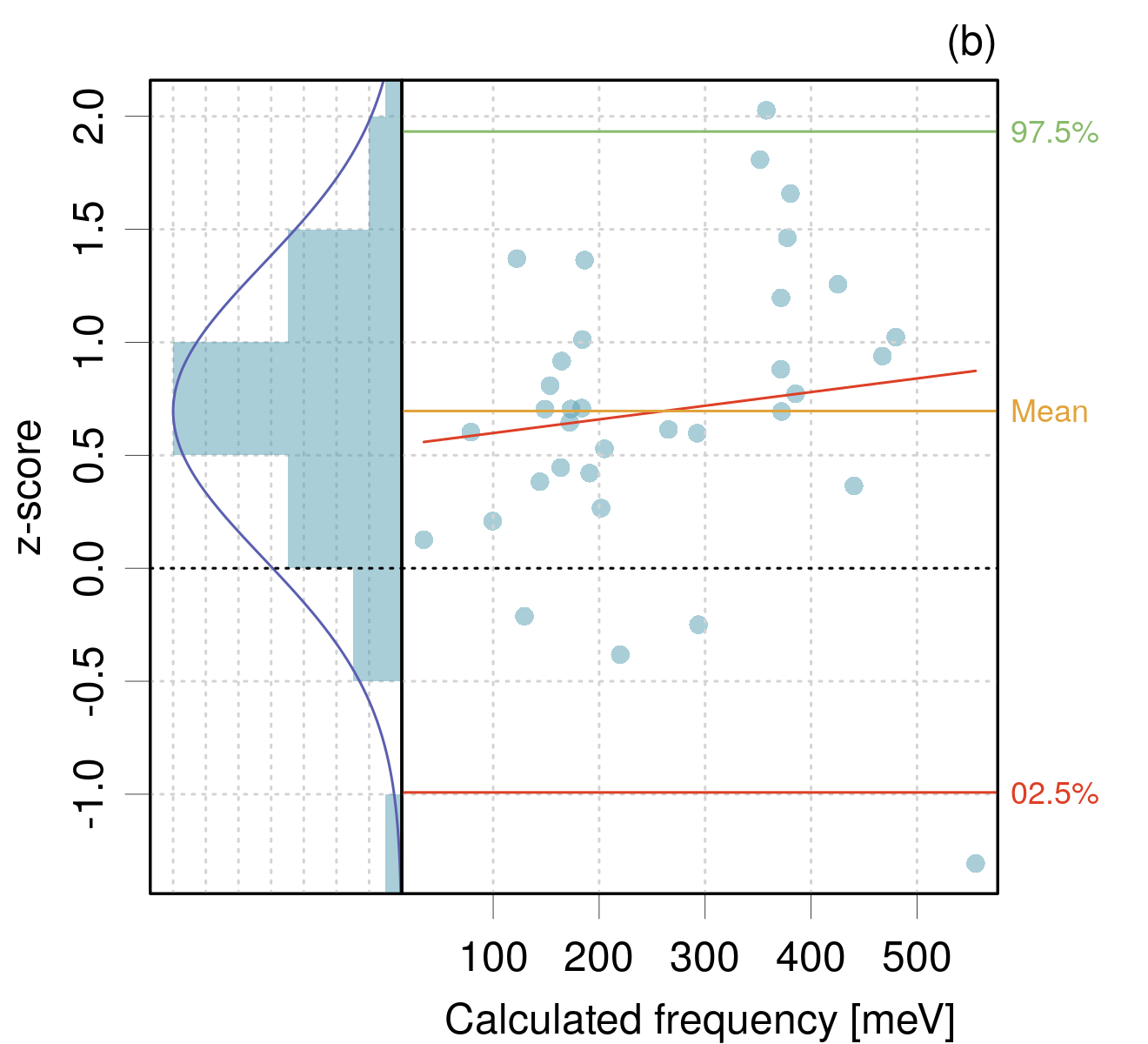}
\par\end{centering}
\caption{\label{fig:PAR2019}Distribution of errors (a) and $z$-scores (b)
for the PAR2019 dataset.}
\end{figure}

A dataset of 35 harmonic vibrational frequencies (PAR2019) has been
extracted from an article by Parks \emph{et al. }\citep{Parks2019}
(Table I, columns $\mu$ and $\sigma$), as another example of BEEF-generated
uncertainties. This is a small dataset, which puts the validation
methods to their limits. The errors are plotted in Fig.\,\ref{fig:PAR2019}(a),
and the corresponding $z$-scores are plotted in Fig.\,\ref{fig:PAR2019}(b).

In Fig.\,\ref{fig:PAR2019}(a), many error bars appear too large
with respect to the error amplitudes and there is a non-negligible
bias. The $z$-scores are also notably biased, while their variance
is 0.42(13) with a 95\,\% confidence interval of $\left[0.23,\thinspace0.81\right]$,
excluding the target value. The data are too sparse to attempt a LZV
analysis. 

BEEF-based CC-UQ methods seem to enjoy some popularity, but we saw
on two examples that they have to be used with care. Pernot and Cailliez
\citep{Pernot2017} showed that the capture of model errors into the
variance-covariance matrix of a model's parameters (or into their
Bayesian posterior pdf) might be problematic. As the calibration is
quantified by the mean prediction variance, there is no guarantee
that the prediction uncertainty is reliable for any single prediction.
In fact, this parameters uncertainty inflation (PUI) scheme \citep{Pernot2017b}
implies strong functional constraints which play against its local
calibration \citep{Simm2016,Reiher2022}. 

\subsection{A-posteriori UQ methods}

A-posteriori prediction uncertainty has to be estimated by a statistical
model using the differences between model predictions and reference
data, and complementary uncertainty information, when available. The
main target is to provide reliable prediction uncertainty in the shape
of standard and/or expanded uncertainty.

The first step in prediction uncertainty estimation is the correction
of trends in the errors dataset. The estimation of trends typically
relies on the calculated value, $V$, as a predictor variable \citep{Pernot2011,Lejaeghere2014,Pernot2015,Proppe2017},
but more complex scenarios can be considered, as might be done in
QSAR \citep{Cronin2019} and ML \citep{Tran2020,Venkatraman2021}
methods. 

The workhorse model of trend correction is the low-degree polynomial.
I note below PTC$n$ (Polynomial Trend Correction of order $n)$ the
correction of the error trend vs. $V$ by a polynomial of degree $n$.
For this study, the PTC$n$ model's parameters and their uncertainty
are estimated by standard least-squares, which is known to be robust
to non-normal error distributions \citep{Knief2021}. More complex
prediction uncertainty models could be used for datasets presenting
complex trends, requiring weighting schemes (\emph{e.g.}, in the case
of heterogeneous reference values uncertainties \citep{Pernot2015})
or the consideration of heteroscedasticity \citep{Bakowies2021,Proppe2021}.
More specific trend correction models are defined in their application
case. Once trends have been corrected, one is left with a set of residual
errors which can be considered as unpredictable and can be treated
as random variables. 

I have considered in this study three methods (DIST, PRED and EQ)
to estimate either prediction uncertainty or the limits of 100$p$\,\%
prediction intervals. All these methods assume the homoscedasticity
of the errors:
\begin{itemize}
\item DIST: based on mean and standard deviation of the error set that provide
a standard uncertainty,
\begin{align}
\mu & =\mathrm{mean}(E)\\
u_{E} & =\mathrm{sd}(E)
\end{align}
The interest of this model is to enable the user to infer a prediction
uncertainty from statistics often reported in benchmark tables. Note
that for well corrected trends, one should have $\mu\simeq0$. In
order to avoid hypotheses on the errors distribution, the use of $z$-scores
based validation methods is best suited to this case. Note that, by
construction, one will get $\mathrm{Var}(Z)=1$, and the focus should
be on the sharpness assessment by LZV analysis.
\item PRED: based on the least-squares based statistical predictions of
the trend correction model \citep{Pernot2015,DeWaele2016,Proppe2017}.
The least-squares formalism provides standard prediction uncertainty
or expanded uncertainty at any probability level. The prediction intervals
account for the parametric uncertainty of the correction polynomial,
but are symmetrical and derive from the expansion of the prediction
uncertainty by Student's-$t$ factors. This approach can be validated
by PICP-based and $z$-score-based methods.
\item EQ: based on empirical quantiles $q_{p}$ estimated from the errors
set. This model makes no distribution hypothesis, except that the
errors empirical cumulative distribution function is a good proxy
for the CDF of prediction errors $F$. It ignores the parametric uncertainty
due to the trend correction model, but it enables to treat non-normal
errors distributions and to estimate 100$p$\,\% prediction intervals
from quantiles of the errors distribution
\begin{align}
I_{p} & =\left[q_{(1-p)/2}(E),q_{(1+p)/2}(E)\right]
\end{align}
This approach guarantees a good average calibration, but not local
calibration, as the global distribution might not be locally optimal
in the presence of heteroscedasticity. The EQ method can be validated
by the full arsenal based on PICP estimation. A symmetrized expanded
uncertainty, noted SEQ$_{p}$ can be defined as the half range of
$I_{p}$. 
\end{itemize}

\paragraph{Cross-validation.}

For a-posteriori methods, one can benefit from the ability to estimate
a prediction uncertainty repeatedly to enrich the LCP analysis with
cross-validation. Schematically, one proceeds as follows:
\begin{enumerate}
\item Cross-validation (repeat until all points are tested)
\begin{enumerate}
\item Split randomly the data into learning and test sets.
\item Use the learning set to estimate prediction intervals.
\item Test these intervals on the test set (Does the interval contain the
test value or not ?).
\end{enumerate}
\item Dispatch the test results into the chosen areas of the predictor variable,
and estimate the PICPs as the percentage of positive tests in each
interval $\nu_{i}$.
\end{enumerate}
Leave-One-Out (LOO) or k-fold cross-validation can be used in step
1. To improve the estimation of $\nu$, k-fold cross-validation can
also be repeated several times on randomly reordered datasets.

\subsubsection{The PER2017 dataset\label{subsec:The-PER2017-dataset}}

The most famous a-posteriori method might be the scaling of harmonic
frequencies. A few years ago, I showed that it had the notable interest
to enable the estimation of a prediction uncertainty \citep{Pernot2017b}.
As an example, I used a set of 2278 frequencies calculated at the
CCD/6-31G{*} level, extracted from the CCCBDB \citep{cccbdb_6_1}.
To check calibration, I then estimated the 2-$\sigma$ PICP, which
ideally was 95\,\%, and much better than for a concurrent method.
I propose here to revisit this dataset.

\begin{figure}[t]
\begin{centering}
\includegraphics[viewport=0bp 0bp 1200bp 1200bp,clip,height=6cm]{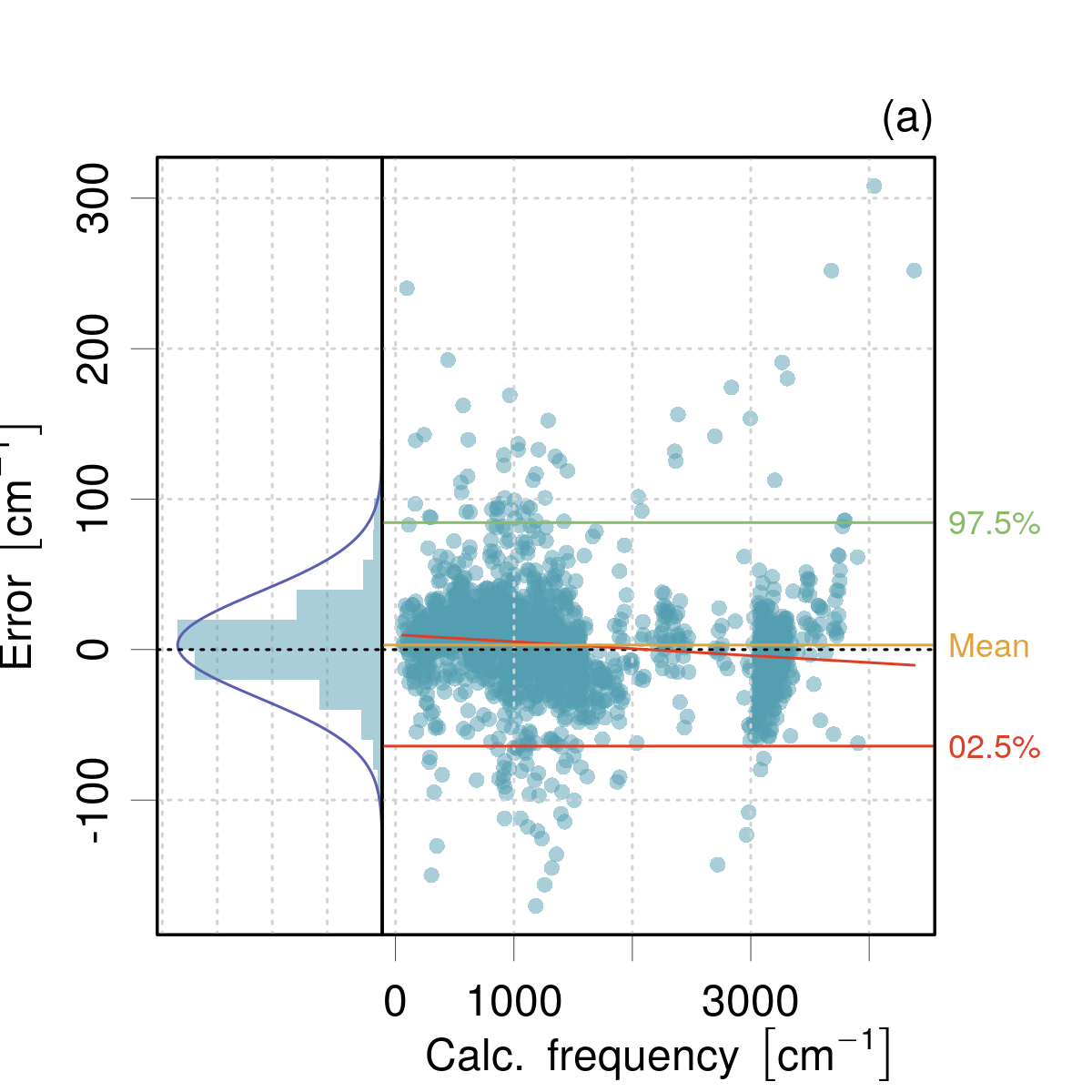}\includegraphics[viewport=0bp 0bp 1200bp 1200bp,clip,height=6cm]{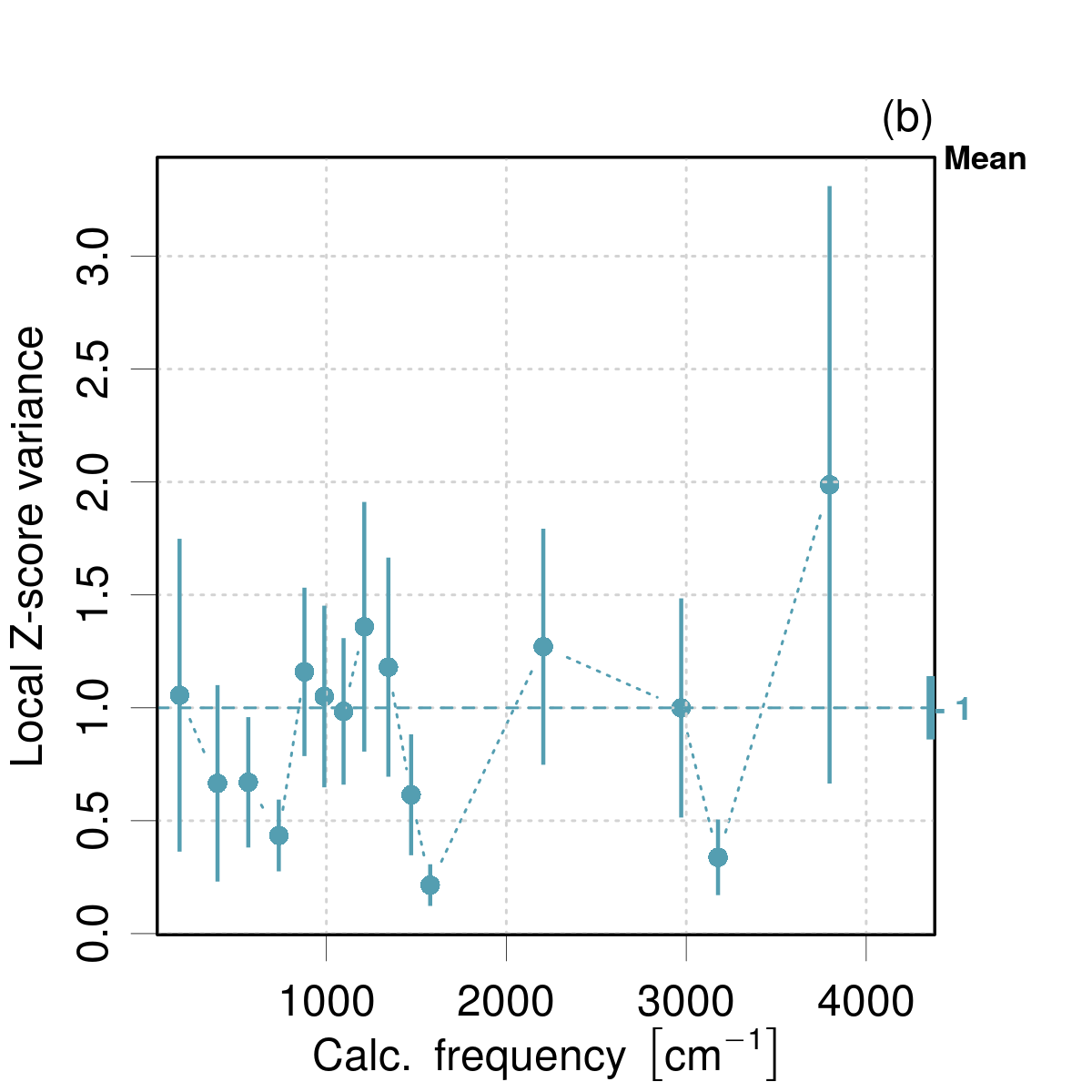}\includegraphics[viewport=0bp 0bp 1200bp 1200bp,clip,height=6cm]{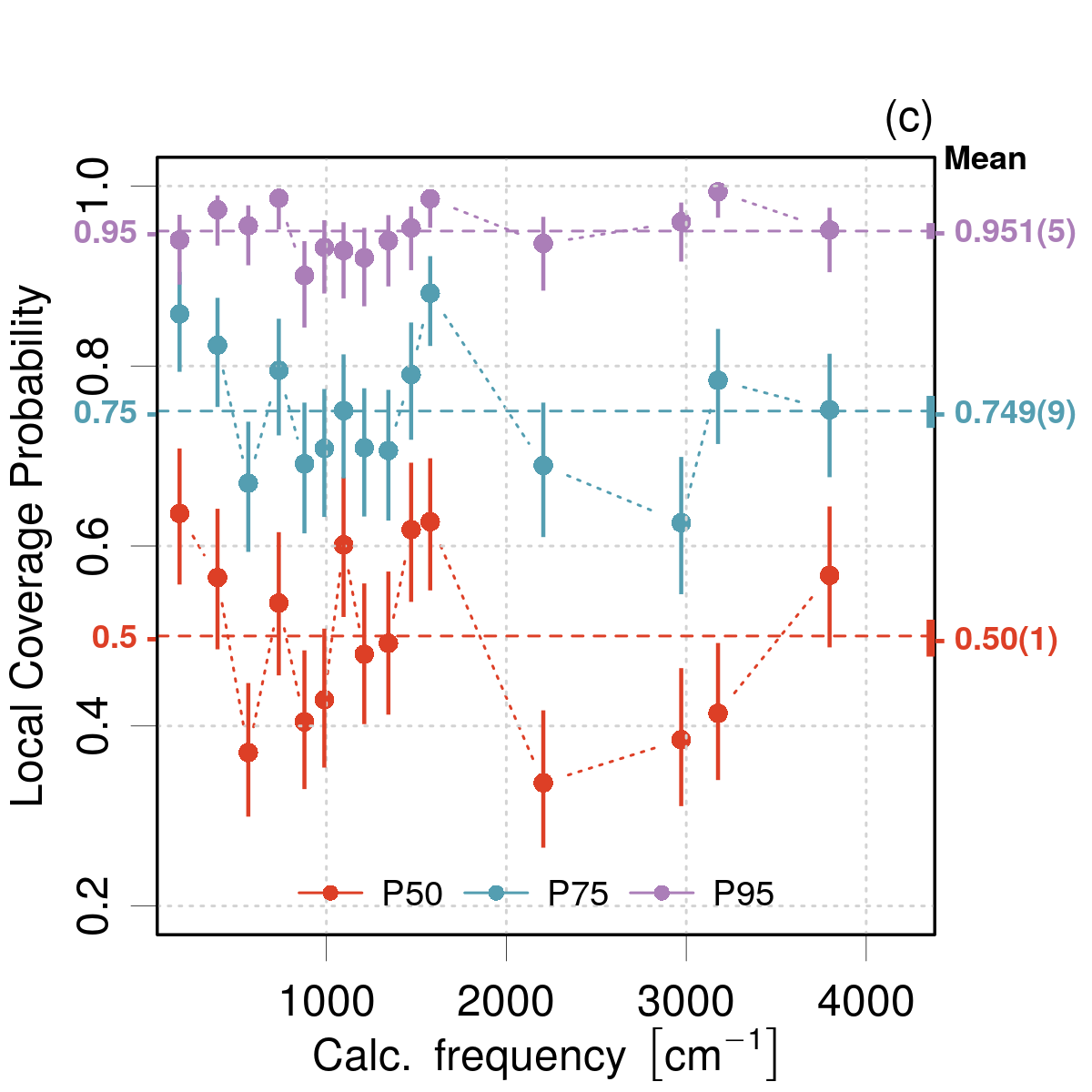}
\par\end{centering}
\caption{\label{fig:PER2017}(a) Distribution of errors after scaling of harmonic
frequencies obtained at the CCD/6-31G{*} level of theory (set PER2017);
(b) LZV analysis; and (c) LCP analysis of PICPs estimated by the EQ
method.}
\end{figure}

The distribution of errors after scaling is presented in Fig.\,\ref{fig:PER2017}(a),
from which one can make two observations: (1) in the absence of the
constant term in the correction model, the scaling does not fully
correct the bias (there is still a very small trend in the errors,
which is probably irrelevant); and (2) the histogram is far from being
normal. More problematic, the dispersion of the errors does not seem
to be uniform along the predictor axis. One might thus expect less
than optimal local calibration. 

Let us consider first the $z$-scores analysis with a prediction uncertainty
derived by the DIST approach (Fig.\,\ref{fig:PER2017}(b)). By construction,
the variance of the $z$-scores is 1, but the LZV analysis reveals
a local calibration issue, with several areas with small variance
values deviating significantly from the target. 

Not surprisingly, considering the shape of the errors distribution
in Fig.\,\ref{fig:PER2017}(a), the PICP values returned by the PRED
approach are off, 0.70(1) vs 0.50 and 0.864(7) vs 0.75, except at
the 0.95 level, where the PICP is 0.944(5). The latter value was the
one I used to validate the prediction uncertainty estimation in my
earlier study.

A much better calibration is automatically obtained by using prediction
intervals based on the EQ approach (Fig.\,\ref{fig:PER2017}(c)),
but the issue of sharpness becomes prominent, as some local PICP values
deviate significantly from their targets. Note however that the situation
is not so bad at the 0.95 level, and one would still get a reasonable
estimation of a uniform $U_{95}$ value (70\,cm$^{-1}$) depending
on the intended application. 

Nevertheless, this questions the reliability of the prediction uncertainty
derived by the scaling procedure for this dataset, which features
multiple underlying trends. More elaborate, bond-based, scaling methods
\citep{Pulay1979,Legler2015} might enable to derive better predictions.

\subsubsection{The PAN2015 dataset: An alternative view}

\begin{figure}[t]
\begin{centering}
\includegraphics[viewport=0bp 0bp 1200bp 1200bp,clip,height=6cm]{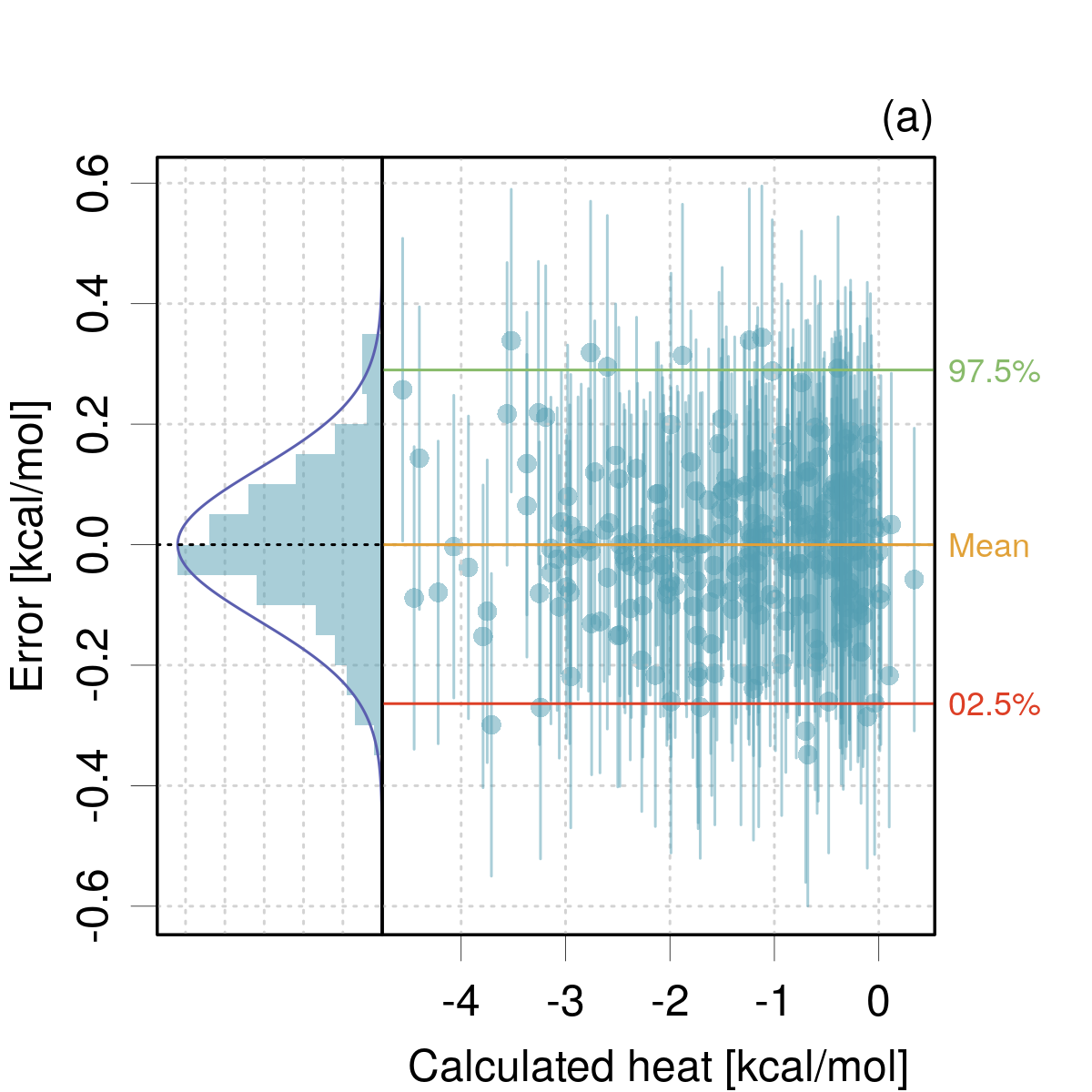}\includegraphics[viewport=0bp 0bp 1200bp 1200bp,clip,height=6cm]{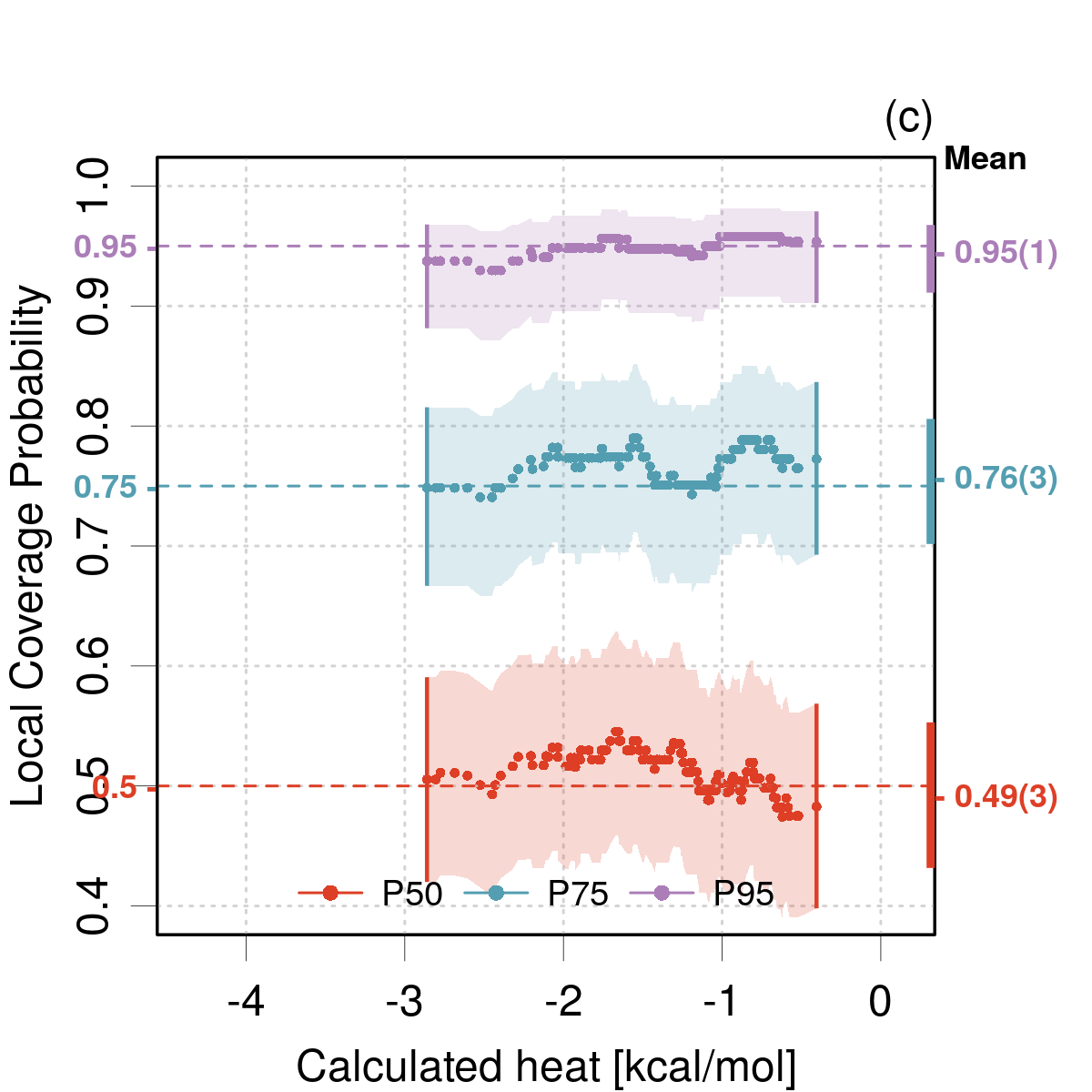}\includegraphics[viewport=0bp 0bp 1200bp 1200bp,clip,height=6cm]{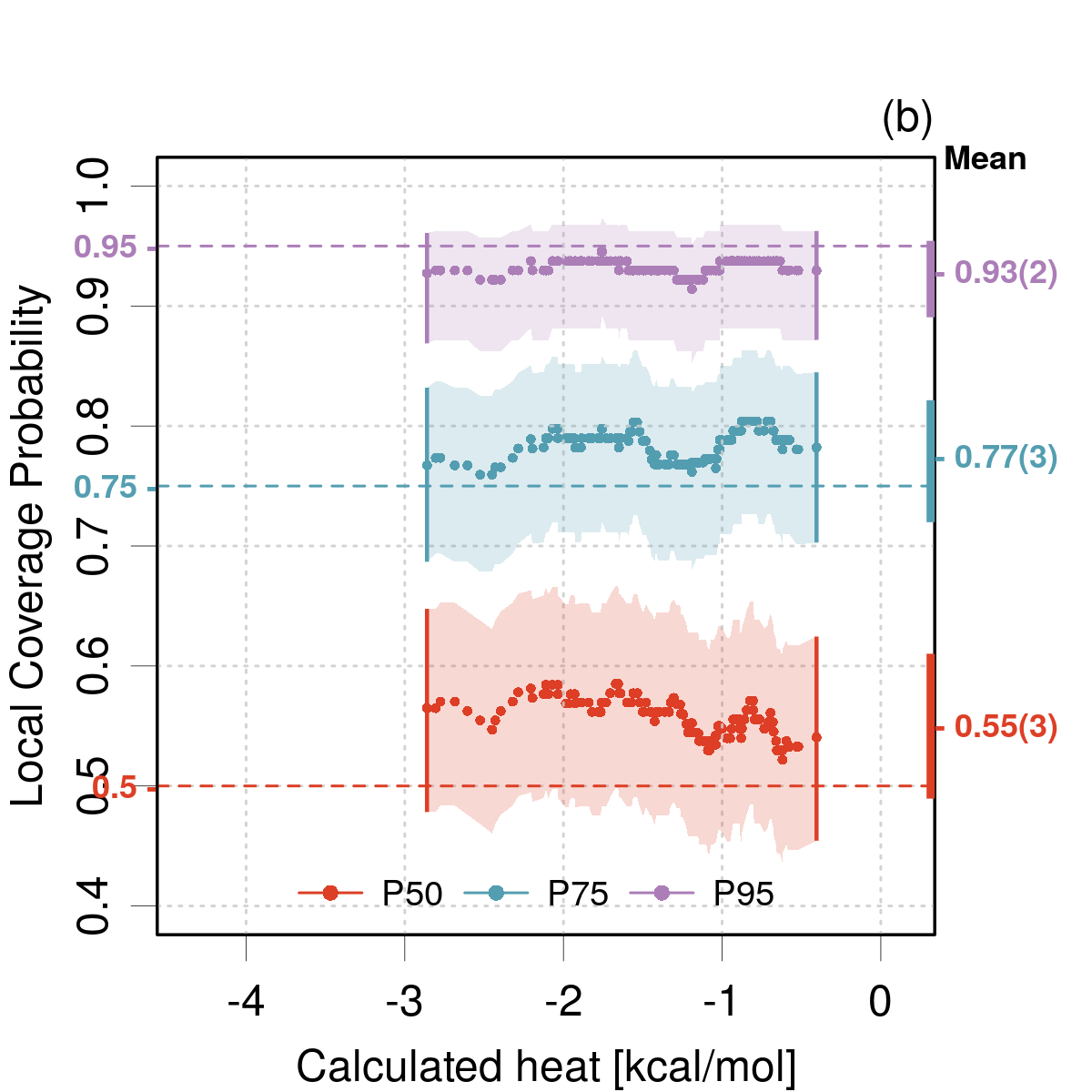}
\par\end{centering}
\caption{\label{fig:PAN2015-1}Distribution of errors (a) and LCP analysis
for an \emph{a-posteriori} reanalysis of the PAN2015 errors with a
PTC1 correction and PRED model (b) or EQ model (c).}
\end{figure}

For comparison with the original prediction uncertainty estimation,
I applied an \emph{a-posteriori} analysis to the PAN2015 dataset analyzed
above (Sections\,\ref{subsec:PAN2015}). Considering the trend observed
in the errors, I used a linear trend correction (PTC1). The distribution
of residual errors is shown in Fig.\,\ref{fig:PAN2015-1}(a). The
distribution is still not perfectly normal. 

Using the PRED method, the PICP values 0.55(3)/0.77(3)/0.93(2) are
nevertheless consistent with their targets Fig.\,\ref{fig:PAN2015-1}(b).
Although the LCP analysis does not provide evidence of local calibration
issues, there seems to be a small systematic bias on the local PICP
values that could be compensated for by using the EQ approach Fig.\,\ref{fig:PAN2015-1}(c). 

The mean prediction uncertainty estimated by the PRED model is about
0.13\,kcal/mol. By comparison, the BEEF-based PUs range from 1/4th
to 3 times this value, with about 67\,\% of them in excess. 

\subsubsection{ML and $\Delta$-ML datasets\label{subsec:ML-and-delta-ML}}

The data are issued from a study by Zaspel \emph{et al.} \citep{Zaspel2019},
as used and analyzed by Pernot \emph{et al.} \citep{Pernot2020b}.
They contain the effective atomization energies (EAEs) for the QM7b
dataset \citep{Montavon2013}, for molecules up to seven heavy atoms
(C, N, O, S or Cl). I consider here values for the cc-pVDZ basis set,
the MP2 and CCSD(T) \emph{ab initio} methods, and two machine learning
algorithms (CM-L1 and SLATM-L2). The ML methods have been trained
over a random sample of 1000 CCSD(T) energies completed by a set of
350 outliers for the SLATM-L2 method identified by Pernot \emph{et
al.} \citep{Pernot2020b}. The final dataset for this study contains
the prediction errors for the 5861 remaining systems by the MP2 and
SLATM-L2 methods. In complement to the previous data, the errors of
MP2 with respect to CCSD(T) have been learned by the CM-L1 and SLATM-L2
methods, providing new $\Delta$-ML/CM-L1 and $\Delta$-ML/SLATM-L2
datasets (unpublished, kindly provided by B. Huang). 

\begin{figure}[t]
\begin{centering}
\includegraphics[viewport=0bp 0bp 1200bp 1200bp,clip,height=6cm]{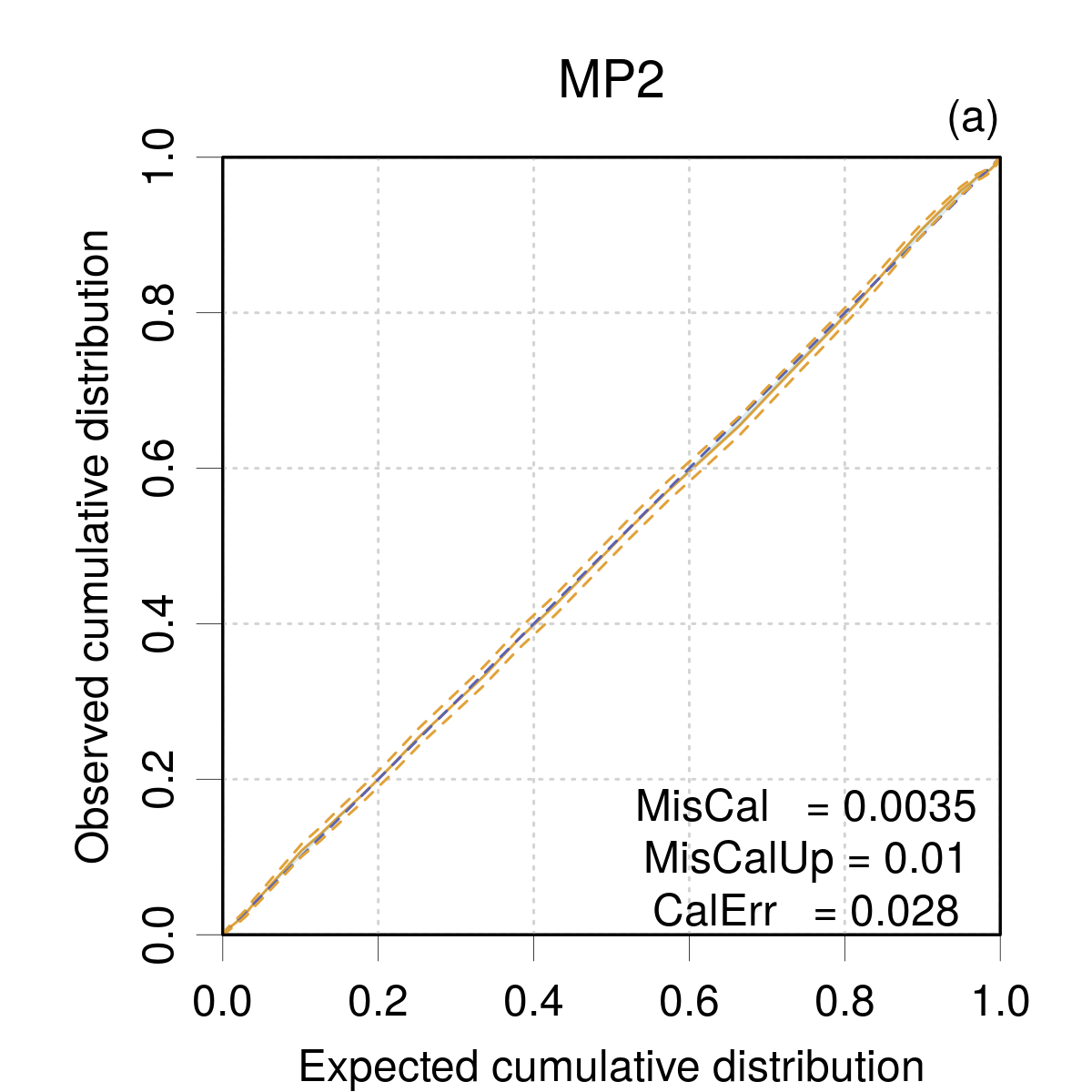}\includegraphics[viewport=0bp 0bp 1200bp 1200bp,clip,height=6cm]{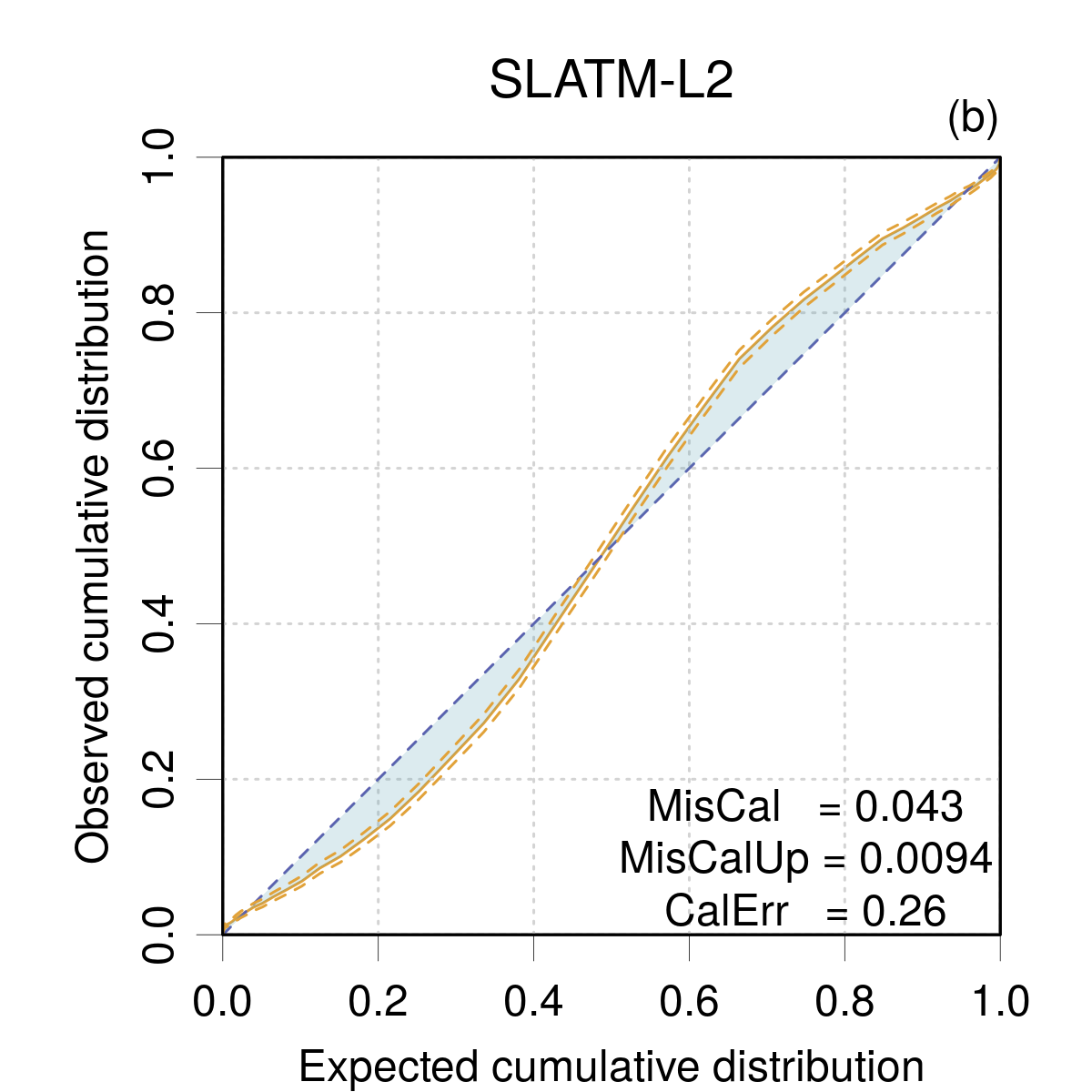}\includegraphics[viewport=0bp 0bp 1200bp 1200bp,clip,height=6cm]{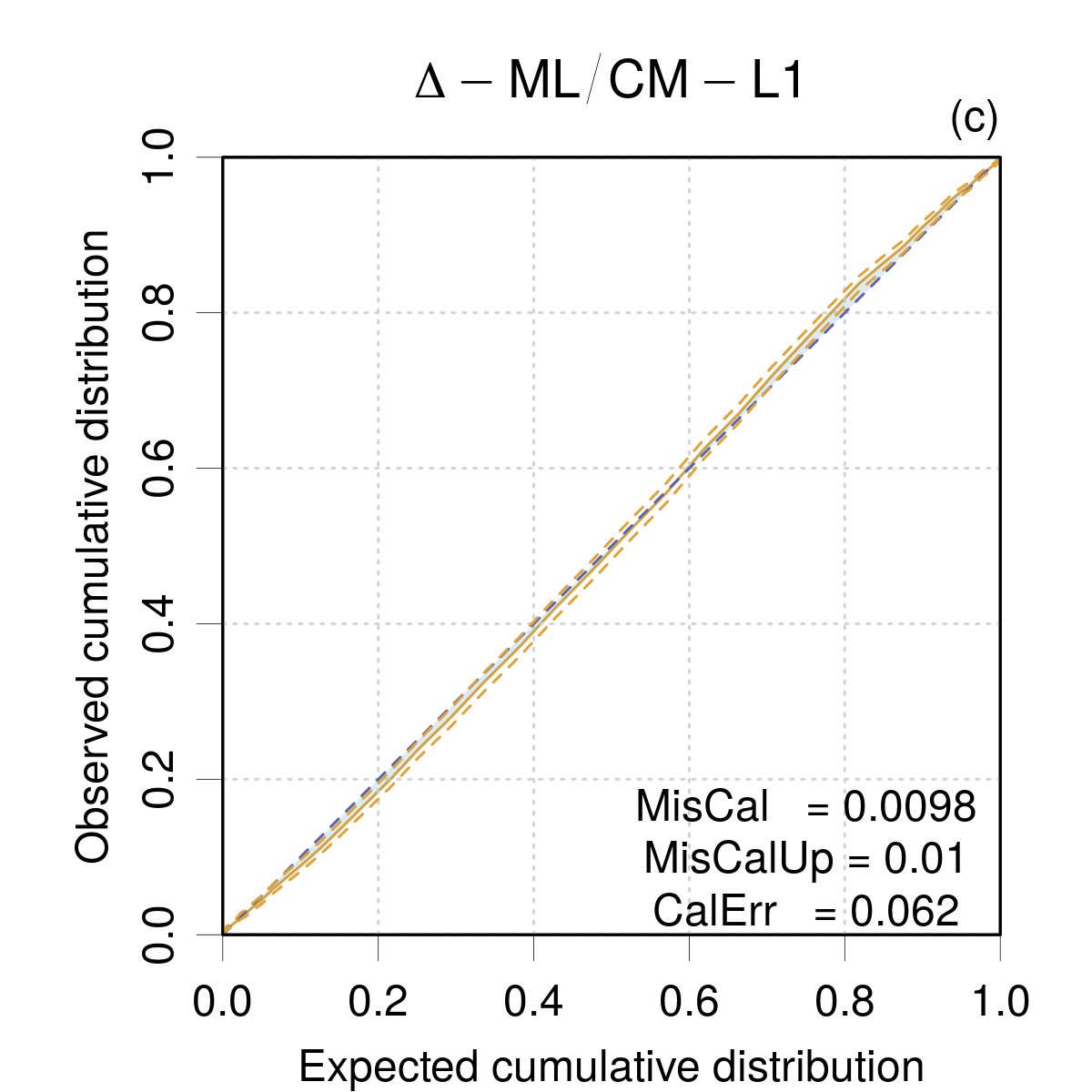}
\par\end{centering}
\caption{\label{fig: ML-Calib}Calibration curves for the prediction uncertainty
of effective atomization energies by (a) MP2, (b) SLATM-L2 and (c)
$\Delta$-ML/CM-L1 methods with a PTC1/PRED calibration scenario.}
\end{figure}

It has been observed previously that the MP2 errors have a quasi-normal
distribution, which was not the case for the SLATM-L2 dataset. Assuming
a normal CDF of prediction errors through the use of a PTC1/PRED scenario,
I built calibration curves to illustrate their diagnostic features.
Note that, considering the large number of points, the parametric
uncertainty for the linear correction is negligible, and one is basically
testing here the normality of the corrected error sets. The plots
in Fig.\,\ref{fig: ML-Calib}, confirm the calibration for MP2 PUs,
while the notable heavy tails of the SLATM-L2 errors do not benefit
from a trend correction, leading to a distorted calibration curve.
The $\Delta$-ML/CM-L1 dataset presents a slight distortion of the
calibration curve. In contrast, the $\Delta$-ML/SLATM-L2 dataset
(not shown) presents the same profile as SLATM-L2. Considering the
MisCal statistic, the MP2/PTC1/PRED and $\Delta$-ML/CM-L1/PTC1/PRED
methods are validated, but MP2/PTC1/PRED is the one with the smallest
calibration error (CalErr\,=\,0.028). 

\begin{figure}[t]
\begin{centering}
\includegraphics[viewport=0bp 0bp 1200bp 1200bp,clip,height=6cm]{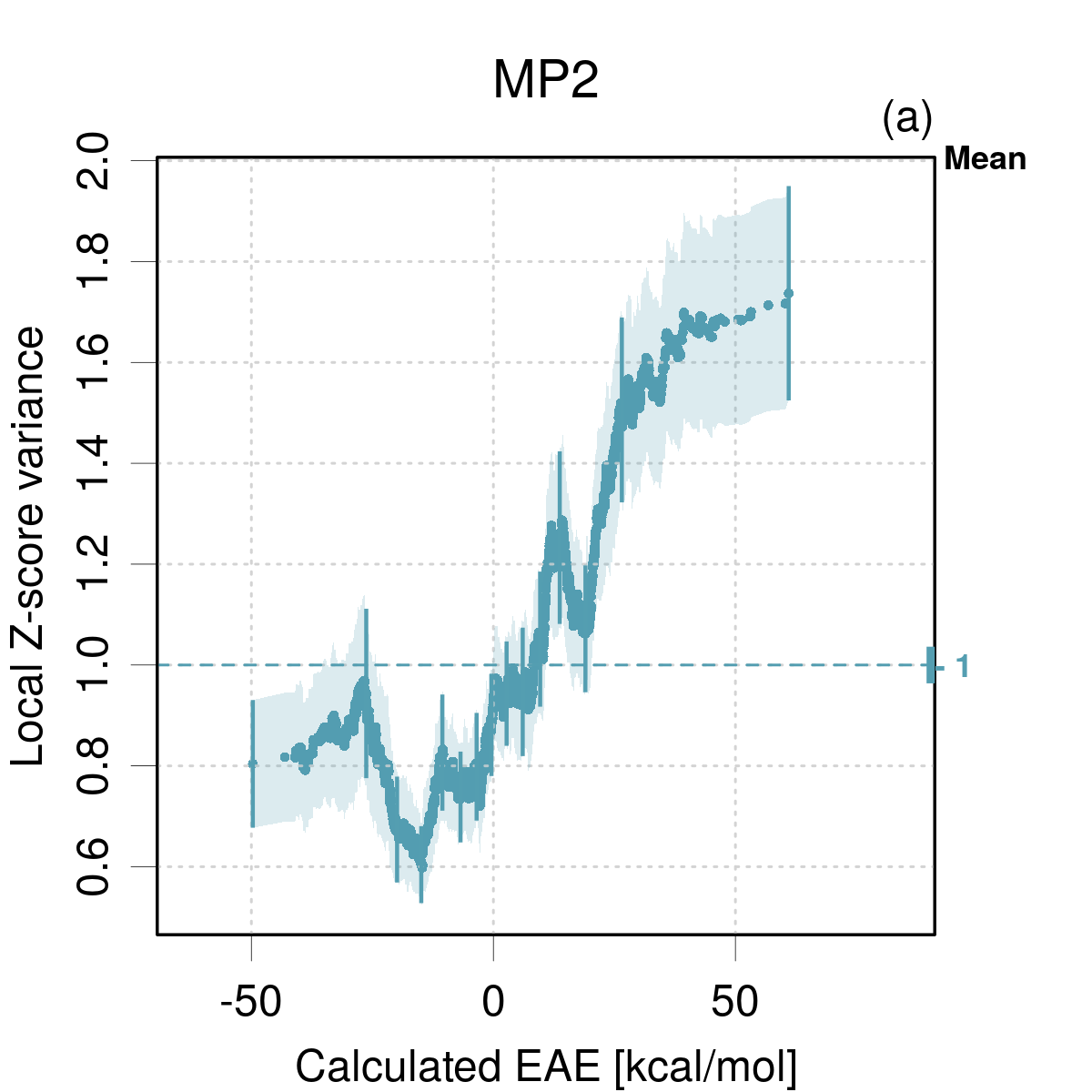}\includegraphics[viewport=0bp 0bp 1200bp 1200bp,clip,height=6cm]{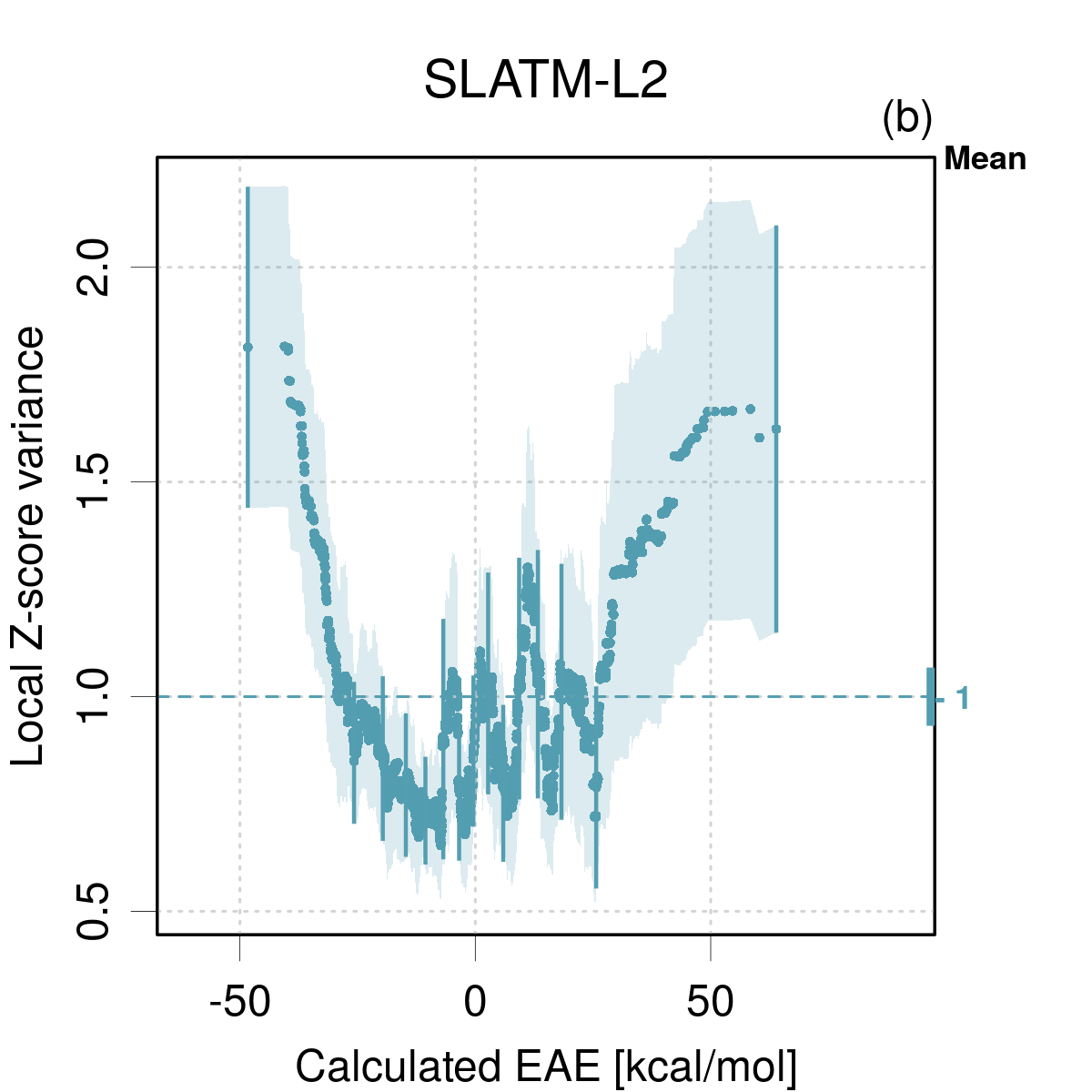}\includegraphics[viewport=0bp 0bp 1200bp 1200bp,clip,height=6cm]{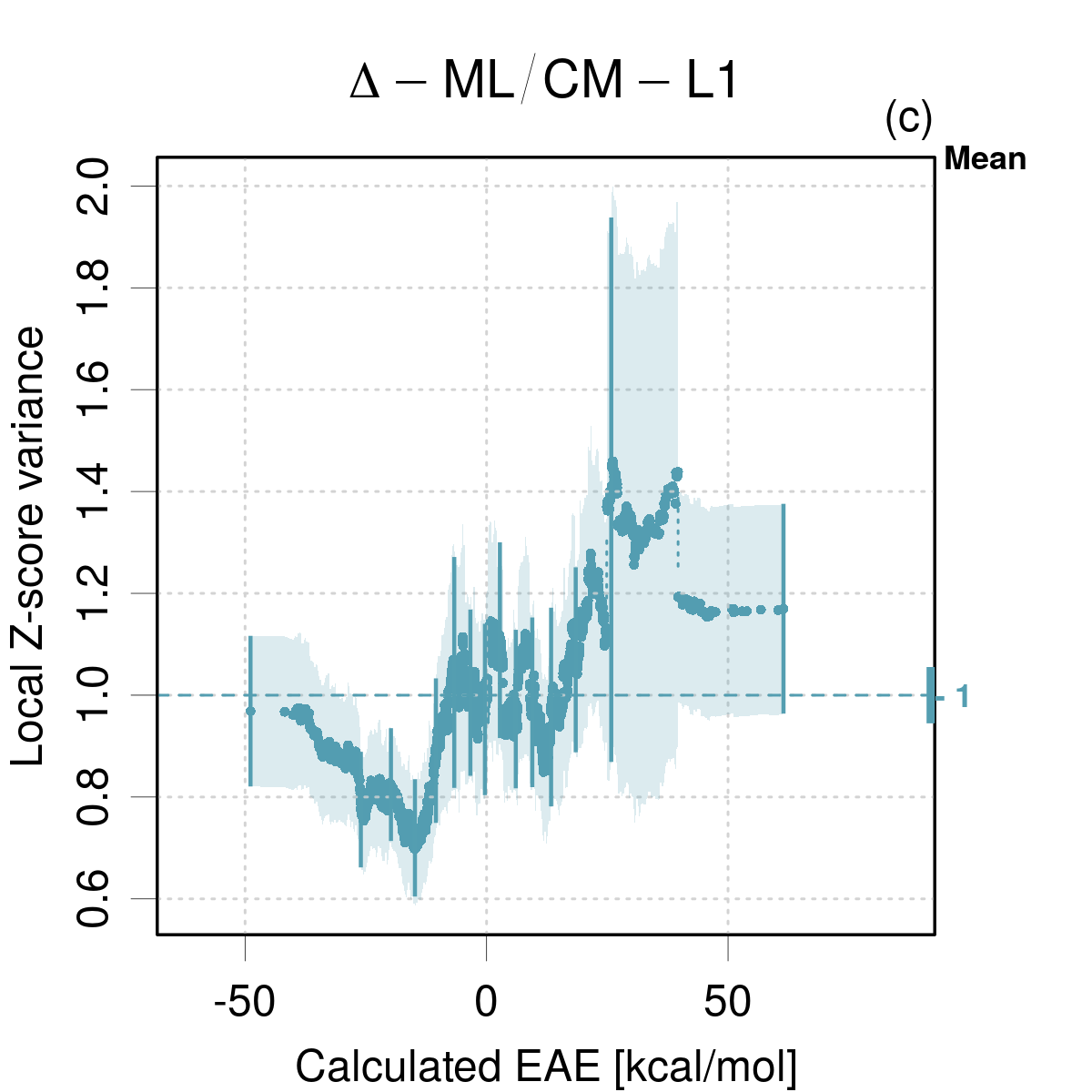}
\par\end{centering}
\caption{\label{fig: ML-LZV}LZV analysis for the prediction uncertainty of
effective atomization energies (EAEs) by (a) MP2, (b) SLATM-L2 and
(c) $\Delta$-ML/CM-L1 methods with a PTC1/PRED calibration scenario.}
\end{figure}

\begin{figure}[t]
\begin{centering}
\includegraphics[viewport=0bp 0bp 1200bp 1200bp,clip,height=6cm]{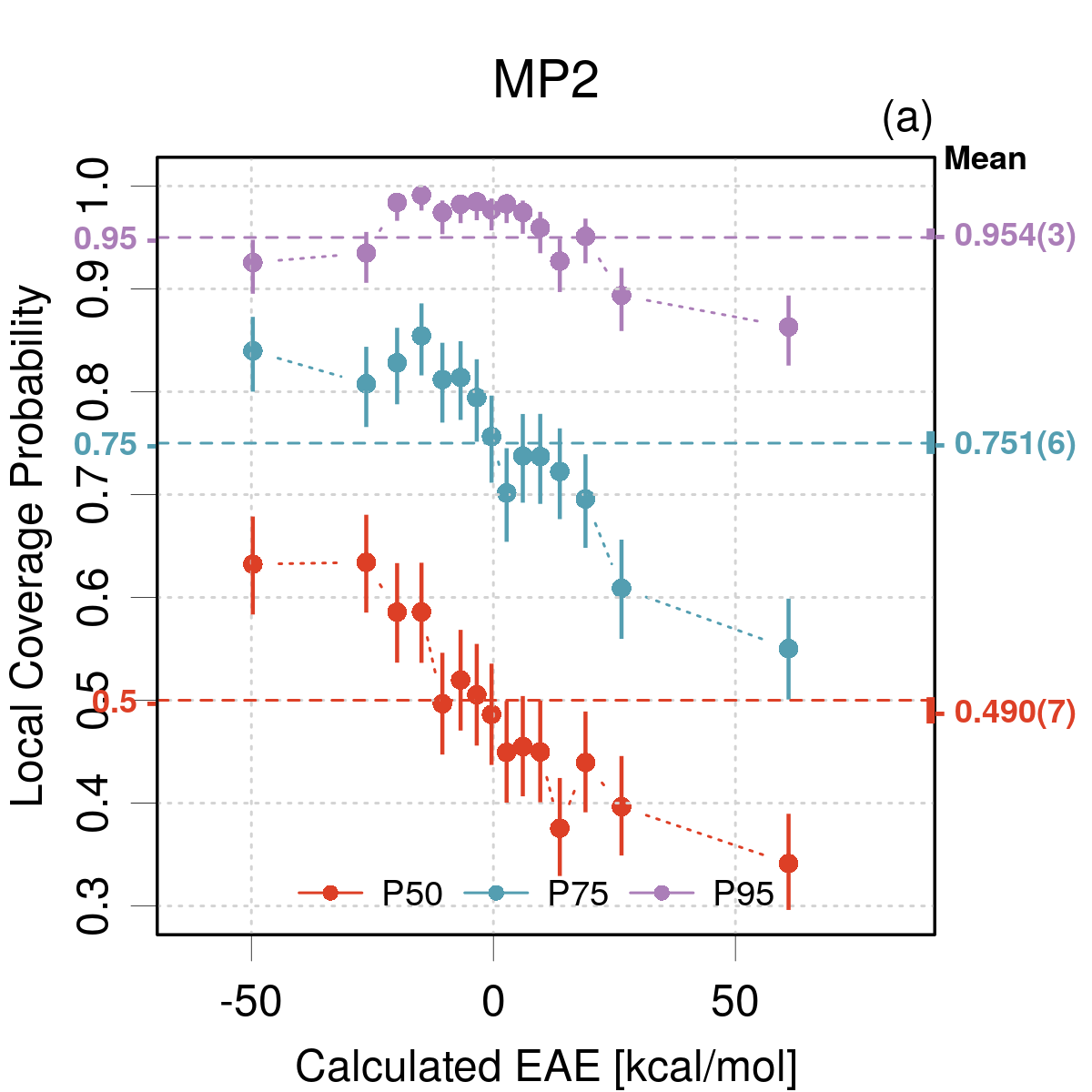}\includegraphics[viewport=0bp 0bp 1200bp 1200bp,clip,height=6cm]{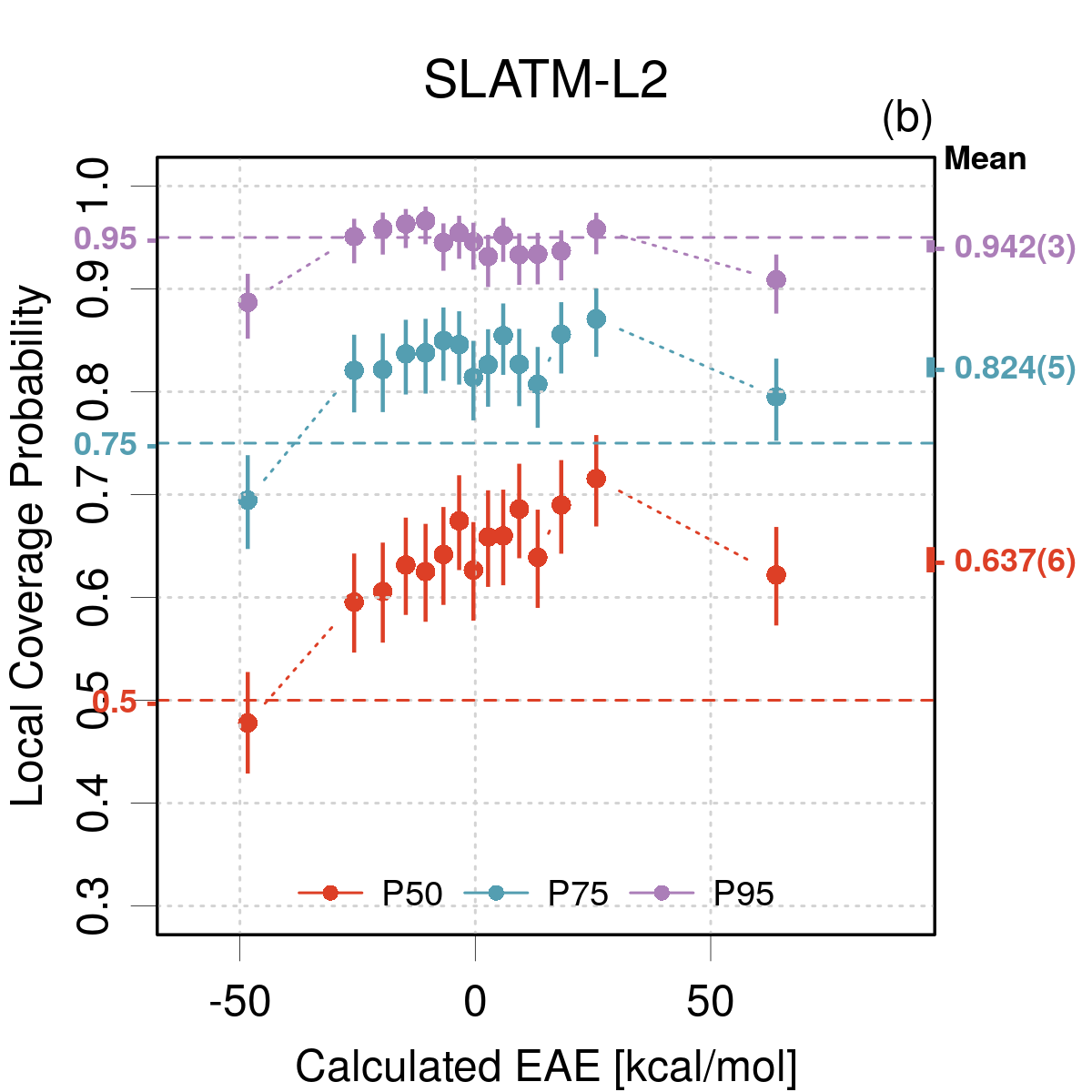}\includegraphics[viewport=0bp 0bp 1200bp 1200bp,clip,height=6cm]{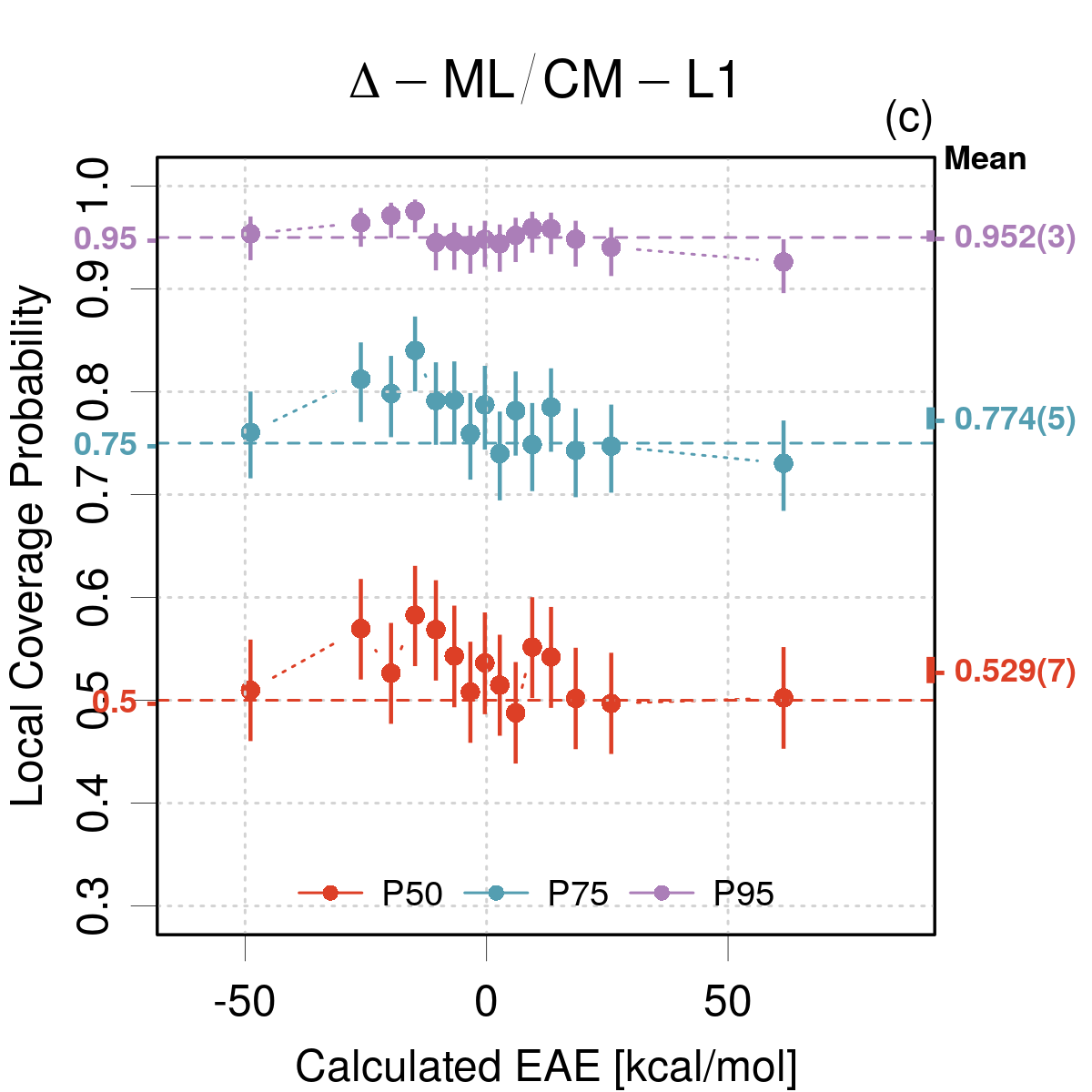}
\par\end{centering}
\caption{\label{fig: ML-LCP}LCP analysis for the prediction uncertainty of
effective atomization energies (EAEs) by (a) MP2, (b) SLATM-L2 and
(c) $\Delta$-ML/CM-L1 methods with PTC1/PRED calibration scenario.}
\end{figure}

The PTC1/PRED method provides us with standard PUs and confidence
intervals. Their sharpness can be tested by LZV and LCP analysis,
respectively. The LZV plots are presented in Fig.\,\ref{fig: ML-LZV},
and the LCP plots are given in Fig.\,\ref{fig: ML-LCP}. 

The uniform standard prediction uncertainty estimated by the PTC1/PRED
method seems to have severe local calibration issues, notably for
MP2 and SLATM-L2 (Fig.\,\ref{fig: ML-LZV}). For MP2, the prediction
uncertainty overestimates the dispersion of the errors for negative
energies, while it underestimates it for the positive ones. For SLATM-L2,
the underestimation occurs a both extremities. The situation is better
for $\Delta$-ML/CM-L1, which deviates significantly from the target
for small negative values. A similar dip in the $z$-score variance
is visible for the other methods, which might correspond to a set
of molecules for which the CCSD(T) energies are more closely approximated
by all three methods. We can consider that the prediction uncertainty
extracted from $\Delta$-ML/CM-L1 is rather reliable. Note that its
largest significant deviation of $\mathrm{Var}(Z)$ (about 0.8) corresponds
to a local excess of about 10\,\% on the prediction uncertainty ($1/\sqrt{0.8}\simeq1.1$).
By comparison, the largest variance deviation for MP2 (about 1.7)
corresponds to an underestimation of the local prediction uncertainty
by a factor $1/\sqrt{1.7}\simeq2.6$. 

The PRED method also enables to generate prediction intervals for
a LCP analysis. Local calibration of the PCT1/PRED scenario is evaluated
in Fig.\,\ref{fig: ML-LCP}. One sees that despite its quasi-normal
errors distribution and good calibration statistics, MP2 presents
a sharpness problem at all coverage levels (Fig.\,\ref{fig: ML-LCP}(a)).
At the 0.5 and 0.75 levels, the LCP curves confirm the observations
made through the LZV analysis (prediction uncertainty overestimation
for the negative energy values and underestimation for the positive
ones). This effect is partially lost at the 0.95 level. For SLATM-L2
(Fig.\,\ref{fig: ML-LCP}(b)), the poor calibration of the PRED intervals
is clearly visible, notably at the 0.5 and 0.75 levels, and local
calibration problems are also detected, notably at the smaller coverage
probabilities. Using a quantile-based method (EQ) would improve average
calibration but not locally. In the $\Delta$-ML/CM-L1 case (Fig.\,\ref{fig: ML-LCP}(c))
calibration is globally very good, but one still observes a deviation
for the small negative energy values, as seen in the LZV analysis. 

Considering the acceptable-to-good local calibration of the ML methods
at the 95\,\% level, one can consider the possibility to define a
uniform prediction uncertainty. Table~\ref{tab:cal_DML} reports
the values of the bias (mean error), 95\,\% prediction intervals
($I_{95}$), prediction uncertainty ($u_{p}$, estimated as the standard
deviation of the errors), expanded prediction uncertainty ($U_{95}$,
as the half range of $I_{95}$), and expansion factor $(f_{95}=U_{95}/u_{p})$.
MP2 is added for comparison although it is not well calibrated enough
to provide reliable uncertainty statistics. 
\begin{table}[ht]
\noindent \centering{}%
\begin{tabular}{lr@{\extracolsep{0pt}.}lr@{\extracolsep{0pt}.}lr@{\extracolsep{0pt}.}lr@{\extracolsep{0pt}.}lr@{\extracolsep{0pt}.}l}
\hline 
Method & \multicolumn{2}{c}{Bias } & \multicolumn{2}{c}{$I_{95}$} & \multicolumn{2}{c}{$u_{p}$} & \multicolumn{2}{c}{$U_{95}$} & \multicolumn{2}{c}{$f_{95}$}\tabularnewline
 & \multicolumn{2}{c}{(kcal/mol)} & \multicolumn{2}{c}{(kcal/mol)} & \multicolumn{2}{c}{(kcal/mol)} & \multicolumn{2}{c}{(kcal/mol)} & \multicolumn{2}{c}{}\tabularnewline
\hline 
MP2  & -0&04(2)  & {[}-3&39(6),~2.84(8){]} & 1&61(2)  & 3&12(5)  & 1&94(2) \tabularnewline
SLATM-L2 & 0&05(2)  & {[}-2&37(7),~2.63(8){]}  & 1&20(2)  & 2&50(5)  & 2&08(3) \tabularnewline
 $\Delta$-ML/SLATM-L2 & 0&003(2)  & {[}-0&35(1),~0.32(1){]}  & 0&169(3)  & 0&336(7)  & 1&99(3) \tabularnewline
 $\Delta$-ML/CM-L1 & 0&03(1)  & {[}-1&49(3),~1.50(3){]}  & 0&77(1)  & 1&49(2)  & 1&93(2) \tabularnewline
\hline 
\end{tabular}\caption{\label{tab:cal_DML}Bias and prediction uncertainties for effective
atomization energies by MP2 and ML methods.}
\end{table}

The bias is negligible in all cases, smaller or equal to the uncertainty
on the limits of $I_{95}$. For simplicity, the other statistics have
been obtained without correction of the error sets. The $I_{95}$
limits present a good symmetry, with a slight defect for MP2 and SLATM-L2.
The LCP analysis showed that $U_{95}$ provides a prediction interval
with a reliable 95\,\% coverage for all ML methods, except for extreme
energy values for SLATM-L2 and $\Delta$-ML/SLATM-L2. As the errors
distributions are not normal in these cases, one should also provide
a standard prediction uncertainty for further uncertainty propagation.
It is remarkable that the expansion factor between $u_{p}$ and $U_{95}$
is close to the value expected for a normal distribution (1.96). The
largest difference is for SLATM-L2, with $f_{95}=2.08(3)$. For all
practical purposes, a factor two might be acceptable. As a byproduct
of this analysis, one can note the excellent performance of the $\Delta$-ML/SLATM-L2
method to predict CCSD(T) values, with a prediction uncertainty of
0.17\,kcal/mol, about 1/10th of the MP2 error standard deviation,
albeit with large local calibration problems.

In a recent article comparing MP2 (normal error distribution) with
SLATM-L2 (non-normal) \citep{Pernot2020b}, the authors argued that
the former would probably be a wiser choice for reliable predictions.
This suggestion does not stand against the present results, as MP2
is worse than SLATM-L2 at the LCP test. In fact, it would even be
reasonable to use SLATM-L2 to define a constant expanded uncertainty
$U_{95}$, which is not the case for MP2. 

\section{Conclusion\label{sec:Conclusion}}

The use of computational chemistry in practical applications is a
strong incentive to establish levels of confidence on the calculated
properties. In this paper, I considered the concepts of calibration
and sharpness, and the associated metrics and graphical checks, developed
for probabilistic forecasters, notably in meteorology and more recently
in machine learning. I adapted these methods to computational chemistry
uncertainty quantification scenarios and I applied them to a series
of datasets covering both embedded and a-posteriori CC-UQ methods. 

The validation methods were adapted to the level of available information,
notably to avoid uncontrolled hypotheses on the error distributions.
In particular, while the calibration of expanded uncertainties can
be directly tested using prediction interval coverage probabilities
(PICP), validation of prediction uncertainty sets was based on the
analysis of $z$-scores variance. Instead of sharpness assessment,
I focused here on local calibration along a predictor quantity, leading
to local versions of the PICP (LCP analysis) and $z$-scores variance
(LZV analysis) tests, which proved to be very convenient and useful
in our context.

The main conclusions arising from the application cases are the following: 
\begin{enumerate}
\item Calibration validation should preferably be performed by the UQ providers.
There is a formidable loss of information when summarizing UQ to prediction
uncertainty. Validation of prediction uncertainty is strongly dependent
on the hypothesis of errors distributions.\textcolor{violet}{{} }We
met several instances where uncertainties reported in the literature
had ambiguous meanings. They might be standard or expanded uncertainties
or provide only a part of the error budget, for instance the variability
of stochastic systems \citep{Bergmann2020,Lin2021}.
\item Establishing reliable PUs is not an easy task. In fact, very few of
the considered examples provide acceptable levels of average and local
calibration. The most satisfying example was provided by a $\Delta$-ML
method, which is in line with observations by Tran \emph{et al.} \citep{Tran2020}
who observe that ``\emph{methods that use one model to make value
predictions and then a subsequent model to make uncertainty estimates
were more calibrated than models that attempted to make value and
uncertainty predictions simultaneously}''. We saw that a-posteriori
methods can be used to ensure average calibration, however, they cannot
always provide local calibration.
\item The sample sizes required to test confidently either PICP values or
$z$-scores variances often exceed the size of computational chemistry
benchmark datasets. This is even more limiting when testing local
calibration. In such cases, the trend in curves of the LCP/LZV analysis
provide an interesting diagnostic, even when statistical uncertainties
are large.
\item Reliably testing PICP or variance values requires some care, as the
standard tests often make hypotheses on the sample distribution which
are not met by computational chemistry errors.
\item Considering that most prediction uncertainty estimates analyzed in
this study were rejected on the basis of rigorous statistical criteria,
one might want to loosen the acceptance thresholds. The pending question
is thus how much of miscalibration is acceptable for specific applications.
The community has to seize this problem.
\item One has always to keep in mind that error statistics are affected
by the quality of reference data. In the present validation framework,
an additional factor to consider is the quality of reference data
uncertainty. In the limit of perfect correction of model predictions,
and therefore of tiny prediction uncertainty, reference data uncertainty
would become the main contribution to, and subject of, calibration
statistics. 
\end{enumerate}
I hope that the concepts and tools presented here will enable a unified
and more rigorous testing framework for my colleagues interested in
virtual measurements and more generally in prediction uncertainty
of computational chemistry methods. 

\section*{Data availability statement }

The data and codes that enable to reproduce the figures and tables
of this study are openly available at the following URL: \url{https://github.com/ppernot/PU2022},
or in Zenodo at \url{https://doi.org/10.5281/zenodo.5818026}.

\section*{Supplementary Material}

See \href{https://aip.scitation.org/doi/suppl/10.1063/5.0084302/suppl_file/suppinf.pdf}{supplementary material}
for details on PICP testing, $z$-scores variance testing and an atlas
of calibration checks.

\section*{Acknowledgments}

I would like to warmly thank A. Savin for enlightening and constructive
discussions, B. Huang for providing the $\Delta$-ML dataset and J.
Proppe for the PRO2021 dataset \citep{Proppe2021}. 

\bibliographystyle{unsrturlPP}
\bibliography{NN}

\end{document}